%% file: AJRDR12obssysrev.tex
\documentclass[useAMS,usenatbib,breaklinks=true]{mnras}
\usepackage{graphicx}
\usepackage{times}
\usepackage{mn2e-breakabs}
\usepackage{amssymb}
\usepackage{amsmath}
\begin{document}

% --- title --- %
\title[BOSS galaxy correlation functions and BAO Measurements] 
{The clustering of galaxies in the completed SDSS-III Baryon Oscillation Spectroscopic Survey: Observational systematics and baryon acoustic oscillations in the correlation function}

\author[A. J. Ross et al.]{\parbox{\textwidth}{
Ashley J. Ross\thanks{Email: ross.1333@osu.edu; Ashley.Jacob.Ross@gmail.com}$^{1,2}$, 
Florian Beutler$^{3,2}$,
Chia-Hsun Chuang$^{4,5}$,
Marcos Pellejero-Ibanez$^{6,7}$,
Hee-Jong Seo$^{8}$,
Mariana Vargas-Maga\~na$^{9}$,
Antonio J. Cuesta$^{10}$,
Will J. Percival$^{2}$,
Angela Burden$^{11,2}$,
Ariel G. S\'{a}nchez$^{12}$,
Jan Niklas Grieb$^{13,12}$,
Beth Reid$^{3}$,
Joel R. Brownstein$^{14}$,
Kyle S. Dawson$^{14}$,
Daniel J. Eisenstein$^{15}$,
Shirley Ho$^{16,3,17}$,
Francisco-Shu Kitaura$^{5}$,
Robert C. Nichol$^{2}$,
Matthew D. Olmstead$^{18}$,
Francisco Prada$^{4,19,20}$,
Sergio A. Rodr{\'i}guez-Torres$^{4,19,21}$,
Shun Saito$^{22,23}$,
Salvador Salazar-Albornoz$^{13,12}$,
Donald P. Schneider$^{24,25}$,
Daniel Thomas$^{2}$,
Jeremy Tinker$^{26}$,
Rita Tojeiro$^{27}$,
Yuting Wang$^{28,2}$,
Martin White$^{3,17,29}$,
Gong-bo Zhao$^{28,2}$
}
  \vspace*{4pt} \\ 
$^{1}$Center for Cosmology and AstroParticle Physics, The Ohio State University, Columbus, OH 43210, USA\\
$^{2}$Institute of Cosmology \& Gravitation, Dennis Sciama Building, University of Portsmouth, Portsmouth, PO1 3FX, UK\\
$^{3}$Lawrence Berkeley National Lab, 1 Cyclotron Rd, Berkeley CA 94720, USA\\
$^{4}$ Instituto de F\'{\i}sica Te\'orica, (UAM/CSIC), Universidad Aut\'onoma de Madrid, Cantoblanco, E-28049 Madrid, Spain \\
$^{5}$ Leibniz-Institut f\"ur Astrophysik Potsdam (AIP), An der Sternwarte 16, 14482 Potsdam, Germany\\
$^{6}$ Instituto de Astrof\'isica de Canarias (IAC), C/V\'ia L\'actea, s/n, E-38205, La Laguna, Tenerife, Spain\\
$^{7}$ Departamento Astrof\'isica, Universidad de La Laguna (ULL), E-38206 La Laguna, Tenerife, Spain\\
$^{8}$Department of Physics and Astronomy, Ohio University, 251B Clippinger Labs, Athens, OH 45701, USA\\
$^{9}$Instituto de Fisica, Universidad Nacional Autonoma de Mexico, Apdo. Postal 20-364, Mexico\\
$^{10}$Institut de Ci{\`e}ncies del Cosmos (ICCUB), Universitat de Barcelona (IEEC-UB), Mart{\'\i} i Franqu{\`e}s 1, E08028 Barcelona, Spain\\
$^{11}$Department of Physics, Yale University, 260 Whitney Ave, New Haven, CT 06520, USA \\
$^{12}$Max-Planck-Institut f\"ur extraterrestrische Physik, Postfach 1312, Giessenbachstr., 85741 Garching, Germany\\
$^{13}$Universit\"ats-Sternwarte M\"unchen, Ludwig-Maximilians-Universit\"at M\"unchen, Scheinerstra\ss{}e 1, 81679 M\"unchen, Germany\\
$^{14}$Department of Physics and Astronomy, University of Utah, 115 S 1400 E, Salt Lake City, UT 84112, USA\\
$^{15}$Harvard-Smithsonian Center for Astrophysics, 60 Garden St., Cambridge, MA 02138, USA\\
$^{16}$McWilliams Center for Cosmology, Department of Physics, Carnegie Mellon University, 5000 Forbes Ave., Pittsburgh, PA 15213\\
$^{17}$Department of Physics, University of California, Berkeley, CA 94720, USA\\
$^{18}$Department of Chemistry and Physics, King's College, 133 North River St, Wilkes Barre, PA 18711, USA\\
$^{19}$Campus of International Excellence UAM+CSIC, Cantoblanco, E-28049 Madrid, Spain\\
$^{20}$Instituto de Astrof\'isica de Andaluc\'ia (CSIC), E-18080 Granada, Spain\\
$^{21}$Departamento de F\'isica Te\'orica M8, Universidad Auton\'oma de Madrid (UAM), Cantoblanco, E-28049, Madrid, Spain\\
$^{22}$Max-Planck-Institut f\"{u}r Astrophysik, Karl-Schwarzschild-Star{\ss}e 1, D-85740 Garching bei M\"{u}nchen, Germany\\
$^{23}$Kavli Institute for the Physics and Mathematics of the Universe (WPI),\\ The University of Tokyo Institutes for Advanced Study, The University of Tokyo, Kashiwa, Chiba 277-8583, Japan\\
$^{24}$Department of Astronomy and Astrophysics, The Pennsylvania State University, University Park, PA 16802, USA\\
$^{25}$Institute for Gravitation and the Cosmos, The Pennsylvania State University, University Park, PA 16802, USA\\
$^{26}$Center for Cosmology and Particle Physics, Department of Physics, New York University, 4 Washington Place, New York, NY 10003, USA\\
$^{27}$School of Physics and Astronomy, University of St Andrews, North Haugh, St Andrews KY16 9SS, UK\\
$^{28}$National Astronomy Observatories, Chinese Academy of Science, Beijing, 100012, P.R. China\\
$^{29}$Department of Astronomy, University of California, Berkeley, CA 94720, USA
}
\date{To be submitted to MNRAS} 

\pagerange{\pageref{firstpage}--\pageref{lastpage}} \pubyear{2014}
\maketitle
\label{firstpage}

\begin{abstract}
We present baryon acoustic oscillation (BAO) scale measurements determined from the clustering of 1.2 million massive galaxies with redshifts $0.2 < z < 0.75$ distributed over 9300 square degrees, as quantified by their redshift-space correlation function. In order to facilitate these measurements, we define, describe, and motivate the selection function for galaxies in the final data release (DR12) of the SDSS III Baryon Oscillation Spectroscopic Survey (BOSS). This includes the observational footprint, masks for image quality and Galactic extinction, and weights to account for density relationships intrinsic to the imaging and spectroscopic portions of the survey. We simulate the observed systematic trends in mock galaxy samples and demonstrate that they impart no bias on baryon acoustic oscillation (BAO) scale measurements and have a minor impact on the recovered statistical uncertainty. We measure transverse and radial BAO distance measurements in $0.2 < z < 0.5$, $0.5 < z < 0.75$, and (overlapping) $0.4 < z < 0.6$ redshift bins. In each redshift bin, we obtain a precision that is 2.7 per cent or better on the radial distance and 1.6 per cent or better on the transverse distance. The combination of the redshift bins represents 1.8 per cent precision on the radial distance and 1.1 per cent precision on the transverse distance. This paper is part of a set that analyses the final galaxy clustering dataset from BOSS. The measurements and likelihoods presented here are combined with others in \cite{Acacia} to produce the final cosmological constraints from BOSS. 
%A companion paper tests cosmological models using these measurements, combined with BAO scale measurements presented in other companion papers that use alternative methods (e.g., the power spectrum and modeling the full shape of the galaxy clustering).  
\end{abstract}


\section{Introduction}
The Baryon Oscillation Spectroscopic Survey (BOSS) has built on the legacy of previous wide-field surveys such as Two Degree Field Galaxy Redshift Survey (2dFGRS; \citealt{2df}) and the Sloan Digital Sky Survey I-II (SDSS; \citealt{York00}) to amass a sample (\citealt{DR12,Reid15}) of more than 1 million spectroscopic redshifts of the galaxies with the greatest stellar mass to $z < 0.75$. This final BOSS data set represents the premier large-scale structure catalog for use in measuring cosmologic distances based on the baryon acoustic oscillation (BAO) feature and the rate of structure growth via the signature of redshift-space distortions (RSD).

Previous results have demonstrated that the current and previous BOSS data sets produce precise and robust BAO and RSD measurements (c.f., \citealt{Reid12,alphDR9,Chuang13,Kazin13,Sanchez13,Anderson14DR9,alph,Sanchez14,Samushia14,CuestaDR12,Gil15BAO,Gil15RSD}). The results of \cite{Ross12,Ross14,Alam15,Osumi15} have demonstrated that the BOSS results are robust to observational systematic concerns and details of sample selection related to galaxy evolution. This paper represents a final, detailed, investigation of observational systematic concerns in the BOSS sample. We detail how the angular selection functions of the BOSS galaxy samples are defined and test for any systematic uncertainty that is imparted into BAO measurements based on this process. The work we present details how BOSS galaxy data can be combined into one BOSS galaxy catalog, and that robust BAO distance and RSD growth measurements can be obtained from the data set. 

This work uses the `combined' BOSS galaxy catalog to determine BAO scale distance measurements, making use of density field `reconstruction' (c.f., \citealt{Pad12}). Following \cite{Xu13,Anderson14DR9,alph,Ross152D,CuestaDR12}, we use the monopole and quadrupole of the correlation function to measure the expansion rate, $H(z)$, and the angular diameter distance, $D_A(z)$, at the redshift of BOSS galaxies. BAO measurements obtained using the monopole and quadrupole of the power spectrum are presented in \cite{BeutlerDR12BAO}, while \cite{VargasDR12BAO} diagnoses the level of  theoretical systematic uncertainty in the BOSS BAO measurements. Measurements of the rate of structure growth from the RSD signal are presented in \cite{BeutlerDR12RSD,GriebDR12RSD,SanchezDR12RSD,SatpathyDR12RSD}. \cite{Acacia} combines the results of these seven (including this work) results together into a single likelihood that can be used to test cosmological models.

The paper is outlined as follows: In Section \ref{sec:analysis} we describe how clustering measurements and their covariance are determined, and how these measurements are used to determine the distance to BOSS galaxies using the BAO feature; in Section \ref{sec:data}, we describe how BOSS galaxies are selected, masked, and simulated. In section \ref{sec:weights}, we describe how weights that correct for observational systematic relationships with galaxy density are determined and applied to clustering measurements. In Section \ref{sec:clus}, we present the configuration-space clustering of BOSS galaxies, demonstrating the effect of systematic weights, comparing the clustering of different BOSS selections and showing that the clustering in the independent NGC and SGC hemispheres is consistent and that the separate BOSS selections can be combined into one BOSS sample to be used for clustering measurements. In Section \ref{sec:BAOrob}, we show that the BOSS BAO measurements are robust to observational systematics (both for data and mock samples). In Section \ref{sec:BAOres}, we present the BAO measurements of the BOSS combined sample; these measurements are used in \cite{Acacia}, combined with the BAO distance measurements and RSD growth measurements of \cite{BeutlerDR12BAO,BeutlerDR12RSD,GriebDR12RSD,SanchezDR12RSD,SatpathyDR12RSD,VargasDR12BAO} and using the methods described in \cite{SanchezDR12comb} to constrain cosmological models. In Section \ref{sec:disc}, we compare our BAO results with those obtained from other BOSS studies and make general recommendations for how to consider any residual observation systematic uncertainty when using BOSS clustering results. 

Unless otherwise noted, we use a flat $\Lambda$CDM cosmology given by $\Omega_m = 0.31$, $\Omega_bh^2 = 0.0220$, $h=0.676$. This is consistent with \cite{Planck2015} and is the same as used in the companion papers studying the BOSS combined sample.

\section{Analysis Tools}
\label{sec:analysis}
\subsection{Clustering statistics}
We work in configuration space. The procedure we use is the same as in \cite{alph}, except that our fiducial bin-size is 5 $h^{-1}$Mpc (as justified in Appendix \ref{app:binsize}). We repeat some of the details here. We determine the multipoles of the correlation function, $\xi_{\ell}(s)$, by finding the redshift-space separation, $s$, of pairs of galaxies and randoms, in units $h^{-1}$Mpc assuming our fiducial cosmology, and cosine of the angle of the pair to the line-of-sight, $\mu$, and employing the standard \cite{LS} method 
\begin{equation}
\xi(s,\mu) =\frac{DD(s,\mu)-2DR(s,\mu)+RR(s,\mu)}{RR(s,\mu)}, 
\label{eq:xicalc}
\end{equation}
where $D$ represents the galaxy sample and $R$ represents the uniform random sample that simulates the selection function of the galaxies. $DD(s,\mu)$ thus represent the number of pairs of galaxies with separation $s$ and orientation $\mu$. 

When counting, each pair is summed as the multiplication of the weights associated with the pair galaxy/random points. For galaxies, the total weight corrects for systematic dependencies in the imaging and spectroscopic data (see Section \ref{sec:weights}) multiplied by a weight, $w_{\rm FKP}$, that is meant to optimally weight the contribution of galaxies based on their number density at different redshifts. The random points are weighted only by $w_{\rm FKP}$. The $w_{\rm FKP}$ weight is based on \cite{FKP} and defined as
\begin{equation}
w_{\rm FKP} = 1/(1+n(z)P_0).
\label{eq:wfkp}
\end{equation}
In this analysis (and other companion DR12 papers), we use $P_0 = 10^4h^{3}$Mpc$^{-3}$, while previous BOSS analyses have used $P_0 = 2\times10^4h^{3}$Mpc$^{-3}$. The choice of $P_0 = 10^4h^{3}$Mpc$^{-3}$ is motivated by the fact that this is close to the value of the BOSS power spectrum at $k= 0.14h$Mpc$^{-1}$ and \cite{FB14} suggest this scale is the effective scale to use for BOSS BAO measurements.  

We calculate $\xi(s,|\mu|)$ in evenly-spaced bins\footnote{The pair-counts are tabulated using a bin width of 1 $h^{-1}$Mpc and then summed into 5 $h^{-1}$Mpc bins, allowing different choices for bin centres.} of width 5 $h^{-1}$Mpc in $s$ and 0.01 in $|\mu|$. We then determine the first two even moments of the redshift-space correlation function via 
\begin{equation}
\frac{2\xi_{\ell}(s)}{2\ell+1} = \sum^{100}_{i=1} 0.01\xi(s,\mu_i)L_{\ell}(\mu_i), 
\label{eq:xiell}
\end{equation}
where $\mu_i = 0.01i-0.005$ and $L_\ell$ is a Legendre polynomial of order $\ell$.

We will also use data that has had the ``reconstruction'' process applied \citep{Eis07rec,Pad12}. In this case, there is a shifted random field, denoted S, and the original random field, and equation (\ref{eq:xicalc}) becomes
\begin{equation}
\xi(s,\mu) =\frac{DD(s,\mu)-2DS(s,\mu)+SS(s,\mu)}{RR(s,\mu)}, 
\label{eq:xicalcrec}
\end{equation}

\subsection{Likelihood analysis/parameter inference}

We assume the likelihood distribution, ${\cal L}$, of any parameter (or vector of parameters), $p$, of interest is a multi-variate Gaussian:  
\begin{equation}
{\cal L}(p) \propto e^{-\chi^2(p)/2}.
\end{equation}
The $\chi^2$ is given by the standard definition
\begin{equation}
\chi^2 = {\bf D}{\sf C}^{-1}{\bf D}^{T},
\end{equation}
where ${\sf C}$ represents the covariance matrix of a data vector and ${\bf D}$ is the difference between the data and model vectors, when model parameter $p$ is used. We assume flat priors on all model parameters, unless otherwise noted.

In order to estimate covariance matrices, we use a large number of mock galaxy samples (see Section \ref{sec:mocks}), unless otherwise noted. The noise from the finite number of mock realizations requires some corrections to the $\chi^2$ values, the width of the likelihood distribution, and the standard deviation of any parameter determined from the same set of mocks used to define the covariance matrix. These factors are defined in \cite{Hartlap07,Dod13,Per14} and we apply them in the same way as in, e.g., \cite{alph}. We use 996 mocks and thus the factors end up being only 3 per cent.

\subsection{Fitting the BAO Scale}

The fundamental aim of BAO measurements is to measure the angular diameter distance, $D_A(z)$ and the expansion rate, $H(z)$. We do so by measuring how different the BAO scale is in our clustering measurements compared to its location in a template constructed using our fiducial cosmology. There are two effects that determine the difference between the observed BAO position and that in the template. The first is the difference between the BAO position in the true intrinsic primordial power spectrum, and that in the model, with the multiplicative shift depending on the ratio $r_{\rm d}/r^{\rm fid}_{\rm d}$, where $r_{\rm d}$ is the sound horizon at the drag epoch (and thus represents the expected location of the BAO feature in co-moving distance units, due to the physics of the early Universe) . The second is the difference in projection. The data is measured using a fiducial distance-redshift relation, matching that of the template: if this is wrong we will see a shift that depends on $H(z)$ in the radial direction, and $D_A(z)$ in the angular direction. The combination of these effects means that our comparison of BAO positions measures:
\begin{equation}
\alpha_{||} = \frac{\left(H(z)r_{\rm d}\right)^{\rm fid}}{H(z)r_{\rm d}},~~\alpha_{\perp} = \frac{D_A(z)r^{\rm fid}_{\rm d}}{D^{\rm fid}_A(z)r_{\rm d}}.
\end{equation}
 It is often convenient for the purposes of comparison to translate these to
\begin{equation}
\alpha = \alpha_{||}^{1/3}\alpha_{\perp}^{2/3}, ~~ 1+\epsilon = \left(\frac{\alpha_{||}}{\alpha_{\perp}}\right)^{1/3},
\end{equation}
here $\alpha$ is the BAO measurement expected from spherically averaged clustering measurements and $\epsilon$ the significance of the BAO feature introduced into the quadrupole by assuming a fiducial cosmology that does not match the true cosmology. 

The methodology we use to measure $\alpha_{||}, \alpha_{\perp}$ is based on that used in \cite{Xu13,alph,Ross152D}, but we employ improved modeling of the post-reconstruction quadrupole based on the results of \cite{Seo15}, which are similar to \cite{White15} and \cite{Cohn16}. We present the relevant details here.

We generate a template $\xi(s)$ using the linear power spectrum, $P_{\rm lin}(k)$, obtained from {\sc Camb}\footnote{camb.info} \citep{camb} and a `no-wiggle' $P_{\rm nw}(k)$ obtained from the \cite{EH98} fitting formulae, both using our fiducial cosmology (except where otherwise noted). We account for redshift-space distortion (RSD) and non-linear effects via
\begin{equation}
P(k,\mu) = C^2(k,\mu,\Sigma_s)\left((P_{\rm lin}-P_{\rm nw})e^{-k^2\sigma_v^2}+P_{\rm nw}\right),
\end{equation}
where
\begin{equation}
\sigma^2_v = (1-\mu^2)\Sigma^2_{\perp}/2+\mu^2\Sigma^2_{||}/2,
\end{equation}

\begin{equation}
C(k,\mu,\Sigma_s) = \frac{1+\mu^2\beta(1-S(k))}{(1+k^2\mu^2\Sigma^2_s/2)},
\label{eq:Csk}
\end{equation}
$S(k)$ is the smoothing applied in reconstruction; $S(k) = e^{-k^2\Sigma_r^2/2}$ and $\Sigma_r = 15 h^{-1}$Mpc for the reconstruction applied to the BOSS DR12 sample. Finally, 
we fix $\beta=0.4$ and $\Sigma_s = 4 h^{-1}$Mpc and use $\Sigma_{\perp} = 2.5 h^{-1}$Mpc and $\Sigma_{||} = 4 h^{-1}$Mpc for post-reconstruction results and $\Sigma_{||}= 10 h^{-1}$Mpc and $\Sigma_{\perp}= 6 h^{-1}$Mpc pre-reconstruction. The choices to the damping scales are similar to those of \cite{BeutlerDR12BAO,VargasDR12BAO} and the values found in \cite{Seo15}. We show in Appendix \ref{app:rob} that the specific choices have little impact on our results. Note, the bias priors we define below effectively allow the amplitude of $\xi_2$ to vary.

Given $P(k,\mu)$, we determine the multipole moments
\begin{equation}
P_{\ell}(k) = \frac{2\ell+1}{2}\int_{-1}^1 P(k,\mu)L_{\ell}(\mu)d\mu,
\end{equation}
where $L_{\ell}(\mu)$ are Legendre polynomials. These are transformed to $\xi_{\ell}$ via
\begin{equation}
\xi_{\ell}(s) = \frac{i^{\ell}}{2\pi^2}\int dk k^2P_{\ell}(k)j_{\ell}(ks)
\end{equation}
We then use 
\begin{equation}
\xi(s,\mu) = \sum_{\ell}\xi_{\ell}(s)L_{\ell}(\mu) 
\end{equation}
(summing to $\ell = 4$) and take averages over any given $\mu$ window to create any particular template:
\begin{equation}
\xi(s,\alpha_{\perp},\alpha_{||})_{F, {\rm mod}}(s) = \int_0^1d\mu F(\mu^{\prime})\xi(s^{\prime},\mu^{\prime}),
\end{equation}
where\footnote{This is essentially the \cite{AP} effect on the BAO feature.} $\mu^{\prime} =  \mu\alpha_{||}/\sqrt{\mu^2\alpha_{||}^2+(1-\mu^2)\alpha_{\perp}^2}$ and $s^{\prime} = s\sqrt{\mu^2\alpha_{||}^2+(1-\mu^2)\alpha_{\perp}^2}$ and the specific $F(\mu^{\prime})$ are defined below.

In practice, we fit for $\alpha_{\perp},\alpha_{||}$ using $\xi_0,\xi_2$. To fit $\xi_0,\xi_2$, we recognize $\xi_2 = 5\int_0^1d\mu\left(1.5\mu^2\xi(\mu)-0.5\xi(\mu)\right)$ and, denoting $3\int_0^1d\mu\mu^2\xi(\mu)$ as $\xi_{\mu2}$ (so here $F(\mu) = 3\mu^2$), we fit to the data using the model
\begin{equation}
\xi_{0, {\rm mod}}(s) = B_0\xi_{0}(s,\alpha_{\perp},\alpha_{||}) + A_{0}(s)  
\label{eq:xi0mod}
\end{equation}
\begin{equation}
\xi_{2, {\rm mod}}(s) = \frac{5}{2}\left(B_2\xi_{\mu2}(s,\alpha_{\perp},\alpha_{||}) - B_0\xi_0(s,\alpha_{\perp},\alpha_{||})\right) + A_{2}(s), 
\label{eq:xi2mod}
\end{equation}
where $A_x(s) = a_{x,1}/s^2+a_{x,2}/s+a_{x,3}$. In each case, the parameter $B_x$ essentially sets the size of the BAO feature in the template. We apply a Gaussian prior of width ${\rm log}(B_x) = 0.4$ around the best-fit $B_0$ in the range $50 < s < 80h^{-1}$Mpc with $A_x = 0$. We have fixed $\beta = 0.4$ in the fiducial template and the $1-S(k)$ term in Equation (\ref{eq:Csk}) forces its effective value to zero at large scales (in the post-reconstruction case). However, note that the greater the difference there is between $B_2$ and $B_0$, the greater the amplitude of $\xi_{2,{\rm mod}}$ will be. Thus, $B_2$ plays essentially the same role in our analysis as $\beta$ has in previous analyses (e.g., \citealt{alph}).

Modeling $\xi_{0,2}$ in the manner described above isolates the anisotropic BAO scale information, while marginalizing over broad-band shape and amplitude information. The pair of moments $\xi_{0,2}$ represent an optimal and complete pair in the case where BAO scale information is spherically distributed \citep{Ross152D}.

\section{Data}
\label{sec:data}
\subsection{The BOSS DR12 Galaxy Sample}
The SDSS-III \citep{Eis11} BOSS \citep{Dawson12} targeted galaxies for spectroscopy using SDSS imaging data, as described in \cite{Reid15}. The SDSS-I, II, and III surveys obtained wide-field CCD photometry \citep{C,Gunn06} in five passbands ($u,g,r,i,z$; \citealt{F}), amassing a total footprint of 14,455 deg$^2$. From this data, BOSS targeted and subsequently observed spectra for 1.4 million galaxies \citep{DR12}, using the BOSS spectrograph \citep{Smee13} and the SDSS telescope \citep{Gunn06}. Observations were performed in a series of 15-minute exposures and integrated until a fiducial minimum signal-to-noise ratio, chosen to ensure a high redshift success rate, was reached. Redshifts were determined as described in \cite{Bolton12}.

The full details of the BOSS galaxy samples are given in \cite{Reid15}\footnote{Code to produce the BOSS catalogs, {\sc MKSAMPLE}, is available from \ the main SDSS web site http://www.sdss.org/surveys/boss}. Here, we summarise the most relevant details in order to provide the background required to understand the analysis of observational effects presented in Section \ref{sec:weights}.

The CMASS sample is designed to be approximately stellar mass limited above $z = 0.45$. Such galaxies are selected from the SDSS DR8 \citep{DR8} imaging via
\begin{eqnarray}
 17.5 < i_{\rm cmod} & < &19.9\\
r_{\rm mod} - i_{\rm mod} &  < & 2 \\
d_{\perp} & > & 0.55 \label{eq:hcut}\\
i_{\rm fib2} & <  &21.5\\
i_{\rm cmod} & < &19.86 + 1.6(d_{\perp} - 0.8) \label{eq:slide}
\end{eqnarray}
where all magnitudes are corrected for Galactic extinction (via the \citealt{SFD} dust maps), $i_{\rm fib2}$ is the $i$-band magnitude within a $2^{\prime \prime}$ aperture, the subscript $_{\rm mod}$ denotes `model' magnitudes \citep{EDR}, the subscript $_{\rm cmod}$ denotes `cmodel' magnitudes \citep{DR2}, and 
\begin{equation}
d_{\perp} = r_{\rm mod} - i_{\rm mod} - (g_{\rm mod} - r_{\rm mod})/8.0.
\label{eq:dp}
\end{equation}

For CMASS targets, stars are further separated from galaxies by only keeping objects with
\begin{eqnarray}
i_{\rm psf} - i_{\rm mod} &>& 0.2 + 0.2(20.0-i_{\rm mod})  \label{eq:sgsep1}\\
z_{\rm psf}-z_{\rm mod} &>& 9.125 -0.46z_{\rm mod} \label{eq:sgsep2}
\end{eqnarray}

\noindent unless the object also passes the LOWZ cuts. 

The LOWZ sample is selected based on the following
\begin{eqnarray}
r_{\rm cmod} < 13.5 + c_{\parallel}/0.3 \label{eq:lzslide}\\
|c_{\perp}| < 0.2 \\
16 < r_{\rm cmod} < 19.6 \label{eq:lzrc}\\
r_{\rm psf}-r_{\rm mod} > 0.3 
\end{eqnarray}
where
\begin{equation}
c_{\parallel} = 0.7(g_{\rm mod}-r_{\rm mod})+1.2(r_{\rm mod}-i_{\rm mod}-0.18)
\end{equation}
and
\begin{equation}
c_{\perp} = r_{\rm mod}-i_{\rm mod}-(g_{\rm mod}-r_{\rm mod})/4.0 -0.18 .
\end{equation}

\begin{figure}
\includegraphics[width=84mm]{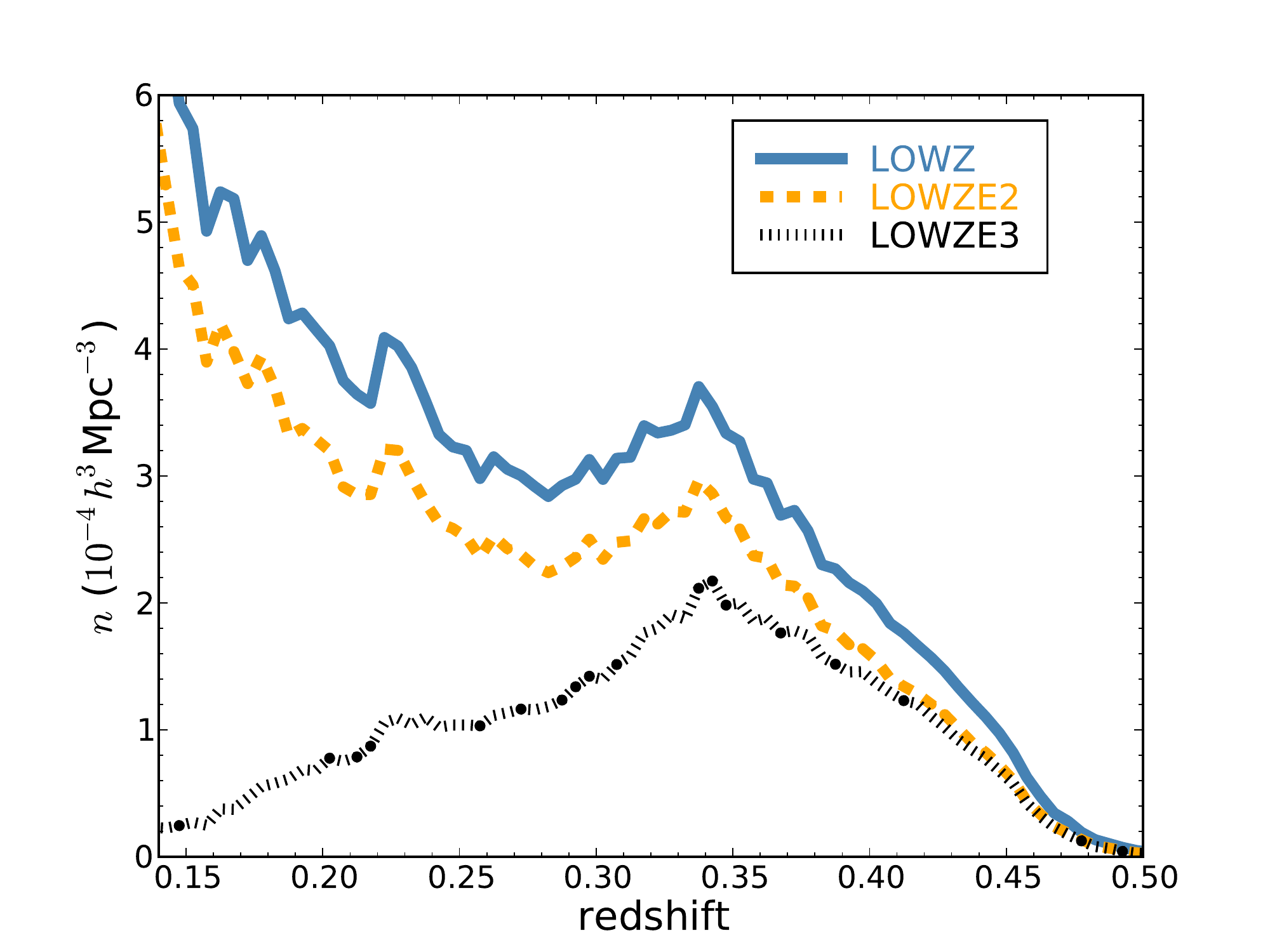}
  \caption{The number density as a function of redshift for the three different LOWZ selections, in the North Galactic Cap (NGC). The LOWZE2 and LOWZE3 selections were applied to early BOSS observations.}
  \label{fig:nzlowz}
\end{figure}

As detailed in \cite{Reid15}, approximately 900 deg$^2$ of the LOWZ sample was targeted with more restrictive cuts than the nominal LOWZ selection. This 900 deg$^2$ area is divided into two separate selections. Covering 130 deg$^2$, the `LOWZE2' selection applies the CMASS $i$-band star/galaxy separation cut (equation \ref{eq:sgsep1}) and had an $r_{\rm cmod}$ limit that was 0.1 magnitudes brighter for both equation (\ref{eq:lzslide}) (13.4) and equation (\ref{eq:lzrc}) (19.5). These bright limits reduce the density of the sample by 16 per cent (as can be seen in Fig. \ref{fig:nzlowz}). Covering 760 deg$^2$, the `LOWZE3' sample is the same as the LOWZE2 selection, except that the $z$-band star/galaxy selection (equation \ref{eq:sgsep2}) is also applied and the bright limit is $r_{\rm cmod} > 17$. The $z$-band star/galaxy separation cut reduces the density of the sample by an additional 39 per cent, in a manner that depends strongly on the size of the PSF, as detailed in Section \ref{sec:see}. This gives the LOWZE3 sample approximately half the number density of the LOWZ sample. 

Given that each sample is a subset of the nominal LOWZ sample, we are able to apply the respective cuts to reproduce LOWZE2 and LOWZE3 samples over the full BOSS footprint. Thus, unless explicitly stated otherwise, when studying each respective sample, we will do so over the full BOSS NGC footprint in order to obtain the best statistical understanding of the samples. Doing so allows us to test the properties of these samples and thereby combine them into one full BOSS galaxy sample. The number density as a function of redshift is displayed in Fig. \ref{fig:nzlowz} for each of the LOWZ selections. Compared to the nominal LOWZ selection, the reduction in number density is approximately constant as a function of redshift for the LOWZE2, while for LOWZE3 the difference grows greater at lower redshifts.

In addition to the color cuts applied to targeting, we apply cuts in redshift of $0.43 < z < 0.7$ to CMASS and $0.15 < z < 0.43$ to the LOWZ, LOWZE2, and LOWZE3 samples when measuring their individual clustering signals. These samples are combined into one full BOSS sample, applying no redshift cuts on the individual samples. We do not expect the galaxies that are removed to have a statistically significant effect on the trends observed, and thus we consider the effect of this to be negligible.

\subsection{Mask}
The BOSS mask is described in detail in section 5.1 of \cite{Reid15}. The most basic mask to be applied to BOSS is defined by the coverage of the spectroscopic tiles, i.e., the survey footprint; this is shown in figure 1 of \cite{Acacia}. On top of the survey footprint, a series of veto masks are applied. These include masks for bright stars, bright objects \citep{Rykoff14}, and non-photometric conditions.

We define additional veto masks based on the seeing at the time the imaging data was observed and the Galactic extinction. Survey area is discarded where the $i$-band seeing, given in terms of the full-width-half-maximum of the point spread function (`PSF\_FWHM') is greater than $2^{\prime\prime}$. This is due to the $i_{\rm fib2}$ selection, as these magnitudes are convolved with $2^{\prime\prime}$ seeing and are therefore ill-defined where the seeing is worse. We additionally remove areas where the $g$- and $r$-band PSF\_FWHM are greater than $2^{\prime\prime}.3$ and $2^{\prime\prime}.1$; these values are roughly equivalent to the $i$-band value of $2^{\prime\prime}.0$, given the optics of the SDSS telescope. These cuts on seeing remove 0.5 and 1.7 per cent of the area in NGC and SGC footprints.

We cut areas where the Galactic extinction, as given by the \cite{SFD} $E(B-V)$ value, is greater than 0.15. A negligible amount of area in the NGC (0.06 per cent) has worse extinction than this. This cut removes 2.2 per cent of the area in the SGC. We find a correlation between the projected density of LOWZ galaxies and $E(B-V)$ at high extinction values (see Fig. \ref{fig:sea}), and thus cut at $E(B-V) = 0.15$ to remove this trend and make the data quality more similar between the NGC and SGC.

\subsection{Galactic Hemisphere}
\label{sec:NSdata}
\begin{figure}
\includegraphics[width=84mm]{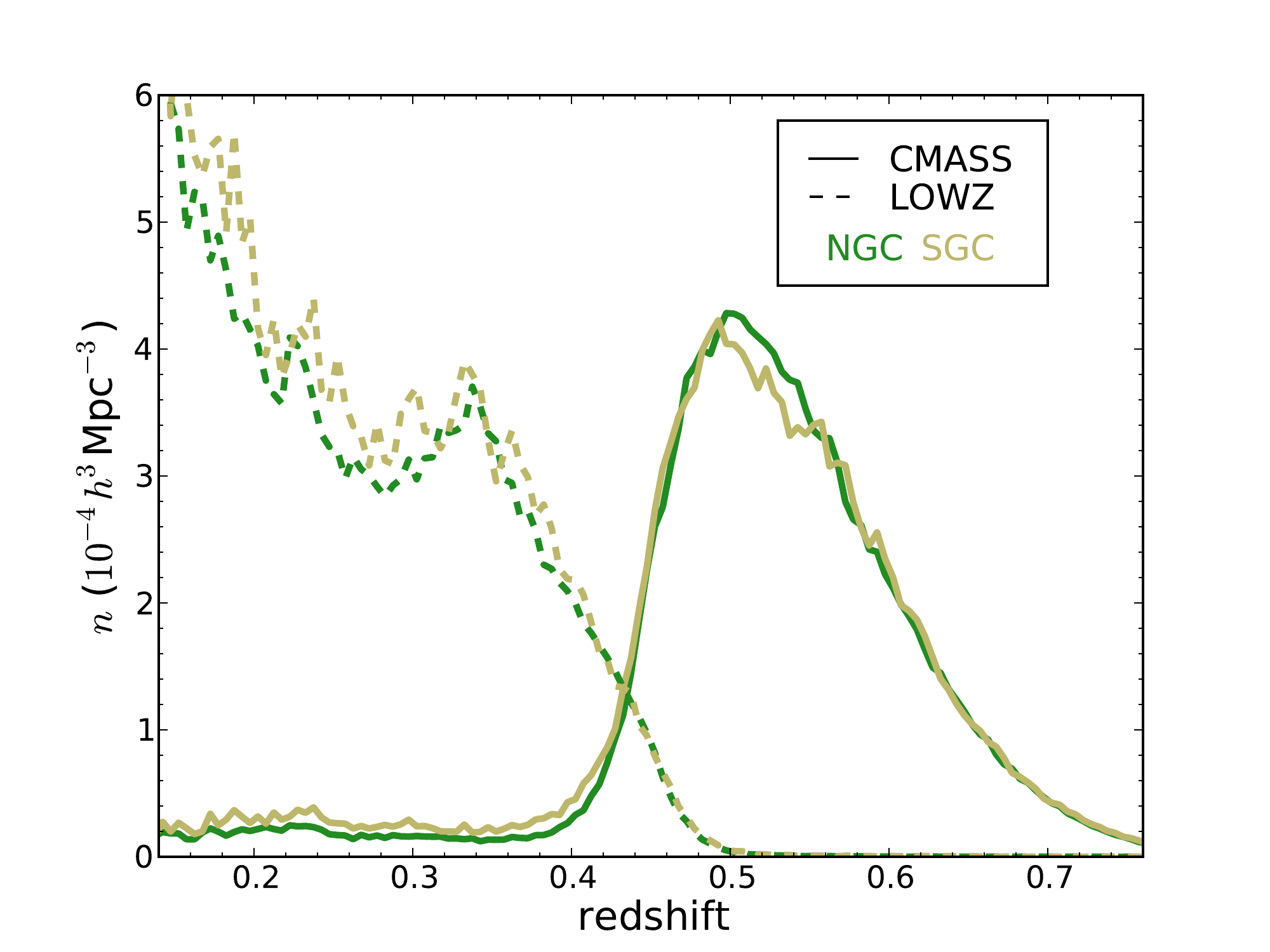}
  \caption{The number density as a function of redshift for CMASS (solid curves) and LOWZ (dashed curves) selections, in the North and South Galactic Caps (NGC, colored `forestgreen'; and SGC, colored `darkkhaki'). The overall offset between densities in the two regions is due to calibration offsets in the imaging data between the two regions.}
  \label{fig:nz}
\end{figure}

As explained in \cite{Ross11,Ross12}, we expect different number densities for BOSS galaxies in the NGC and SGC, due to the fact that \cite{Schlafly11} have shown there are measurable offsets in the DR8 \citep{DR8} photometry between the two regions. The final BOSS DR12 results are consistent with these earlier studies: Accounting for all weights, we find a 1.0 per cent larger projected density of the CMASS sample ($0.43 < z < 0.75$) in the SGC compared to the NGC. In the LOWZ sample ($0.2 < z < 0.43$), the projected density is 7.6 per cent higher in the SGC compared to the NGC. For this reason, the NGC and SGC are treated to have separate selection functions, as has been the standard practice throughout the lifetime of BOSS analyses.

Fig. \ref{fig:nz} displays the number density of the CMASS and LOWZ  samples in the NGC and SGC. One can see that the LOWZ sample in the SGC has a greater density than the NGC by a nearly constant factor. For the CMASS sample, the SGC distribution is somewhat skewed compared to the NGC selection. The number density is greater at the low redshift end, due to the fact that the offset in photometry effectively lowers $d_{\perp}$ limit (equation \ref{eq:hcut}) in the SGC compared to the NGC. These differences in $n(z)$ imply that the galaxy populations will be slightly different in the different hemispheres and should thus be considered when the results from each hemisphere are combined.

\subsection{Mock Galaxy Samples}
\label{sec:mocks}
We use two independent methods to create two samples of close to 1000 mock realizations designed to match BOSS galaxy samples\footnote{996 mocks are MD-P used for the MD-P results and 1000 for QPM}. The two methods are `QPM' \citep{QPM} and MultiDark PATCHY (MD-P)\citep{PATCHY14,Kitaura15} and each has been tuned to match the footprint, redshift distribution, and halo occupation distribution of BOSS samples. We therefore expect the clustering of the mock samples to match the BOSS measurements. We use both sets of these mock samples to generate covariance matrices and to test methodology. Each uses its own cosmology; the differences between these cosmologies aid in assessing the robustness of our results\footnote{Both include no neutrino mass, rather than the minimal allowed mass adopted for our fiducial cosmology, but as shown by \cite{ThepLewis}, this is expected to have minimal impact on BAO analyses.}.  The cosmology used for each mock and the BAO measurements we expect to find for them when analyzing them using our fiducial cosmology are listed in Table \ref{tab:baoexp}.

The tests we performed on the LOWZ and CMASS samples were completed using the QPM mocks; this work was completed (as a pre-requisite) prior to the definition of the BOSS combined sample.\footnote{We have found no indications that any conclusions would be altered if the tests are repeated with the final MD-P mocks.} This same work allowed the combined sample MD-P and QPM mocks to be created. \cite{Kitaura15} demonstrate that the MD-P mocks are a better match to the combined sample, with some improvement over QPM due to the treatment of the lightcone (see \citealt{Kitaura15} for full details). Thus, in what follows we exclusively use the QPM mocks in tests of the LOWZ and CMASS samples, use the MD-P mocks as the primary sample for tests of the combined sample, and use the QPM mocks as a robustness check on the combined sample results.

\begin{table}
\centering
\caption{Cosmology and expected values for BAO parameters for QPM and MultiDark-PATCHY (MD-P) mocks, given we have analyzed them using our fiducial cosmology and each set of mocks has their own cosmology. Each uses a flat geometry and has a density of neutrinos $\Omega_{\nu} = 0$. The exact values used for MD-P are $\Omega_{m} = 0.307115$ and $h=0.6777$, which have been rounded to 3 significant figures below.}
\begin{tabular}{lcccc}
\hline
\hline
QPM & $\Omega_{m} = 0.29$ & $h=0.7$ & $\Omega_bh^2 = 0.02247$ & $\Omega_{\nu} = 0$\\
\hline
redshift & $\alpha_{||}$ & $\alpha_{\perp}$ & $\alpha$ & $\epsilon$\\
\hline
0.38 & 0.9808 & 0.9755 & 0.9773 & 0.0018\\
0.51 & 0.9840 & 0.9770 & 0.9793 & 0.0024\\
0.61 & 0.9861 & 0.9782 & 0.9808 & 0.0027\\
\hline
\hline
MD-P & $\Omega_{m} = 0.307$ & $h=0.678$ & $\Omega_bh^2 = 0.02214$ & $\Omega_{\nu} = 0$\\
\hline
redshift & $\alpha_{||}$ & $\alpha_{\perp}$ & $\alpha$ & $\epsilon$\\
\hline
0.38 & 0.9999 & 0.9991 & 0.9993 & 0.0003\\
0.51 & 1.0003 & 0.9993 & 0.9996 & 0.0003\\
0.61 & 1.0006 & 0.9995 & 0.9999 & 0.0004\\
\hline
\label{tab:baoexp}
\end{tabular}
\end{table}

\section{Weighting Galaxies Based on Survey Properties}
\label{sec:weights}
The methods used to account for various reasons for incompleteness in observations of the BOSS spectroscopic sample are defined and justified in \cite{Reid15}. These include close pair weights, $w_{\rm cp}$, that are applied to account for fiber collisions and weights, $w_{\rm noz}$, that account for redshift failures. We include these weights as $w_z = w_{\rm cp}+w_{\rm noz}-1$ in all analyses, unless otherwise noted. In the following subsections, we test the projected BOSS galaxy density against observational parameters that affect the imaging data, and define weights to correct for systematic relationships, where identified. 

Our results require determining the uncertainty in the relationships between galaxy density and observational parameters, often for samples that are divided in ways that are not possible for our mock samples. Thus, we require some manner of estimating uncertainties that balances cosmic variance and shot-noise but does not rely on the variance of mock realizations. To do so, we weight all galaxy counts by the $w_{\rm FKP}$ weights and treat the resulting counts like Poisson statistics. Such a scheme balances shot-noise and cosmic variance, at the scale used to define the FKP weights. For example, if the FKP weight is 0.5 for all galaxies in the sample, the expected variance in the number of galaxies is twice the number of galaxies (instead of the number of galaxies in the case where the FKP weights are 1). The variance on the FKP-weighted sample would be $0.5N$, while the variance in the pure Poisson case would be $0.25N$ (as the variance of $xN$ is $x^2N$ when $N$ is drawn from a Poisson distribution). In this example, the variance is twice as large as the shot-noise contribution, because there are equal contributions from cosmic variance and shot-noise. We have compared this scheme to the variance of statistics obtained from the CMASS mock samples and found good agreement. Applying this scheme allows uncertainties to be estimated for samples that do not have matching suites of mock catalogs.

\subsection{Stellar Density}

\begin{figure}
\includegraphics[width=83mm]{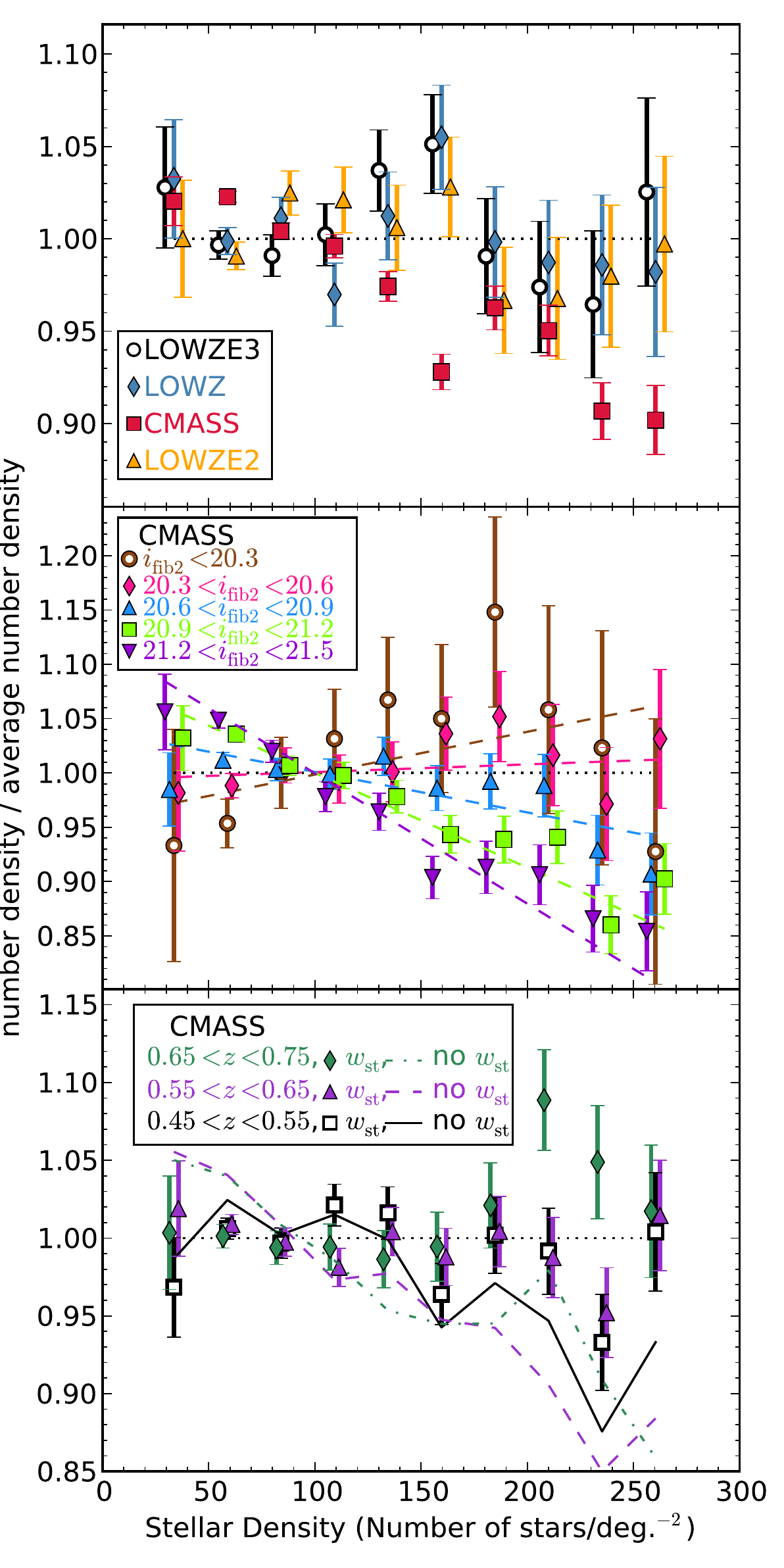}
  \caption{Projected BOSS galaxy density versus stellar density, measured as the number of $17.5 < i < 19.9$ stars in Healpix pixels with Nside=128. Top panel: the relationships for CMASS and the three LOWZ selections. Middle panel: The relationships for CMASS, split into bins of $i_{\rm fib2}$ magnitude. These are the measurements used to define the stellar density weights applied to clustering measurements. Bottom panel: The relationships for CMASS, split by redshift, before (curves) and after (points with error-bars) stellar density weights are applied. The relationships before any weighting is applied are slightly dependent on redshift, due to a weak correlation between $i_{\rm fib2}$ and redshift. 
  Weighting based on $i_{\rm fib2}$ (illustrated in the middle panel) removes this dependency.
  }
  \label{fig:nstall}
\end{figure}

The projected density of CMASS was found to depend on the local stellar density in \cite{Ross11}. This finding was confirmed in all subsequent BOSS data sets. We use SDSS DR8 stars with $17.5 < i < 19.9$ to map the stellar density at Healpix resolution Nside$=128$ (0.21 square degrees per pixel). This is the same set of stars used in \cite{Ross11, Ross12}. The systematic dependency with stellar density affects only the CMASS sample; as shown in the top panel of Fig. \ref{fig:nstall}, none of the LOWZ selections exhibit any trend; this is as expected given it is a brighter selection than the CMASS sample (see \citealt{Tojeiro14} for further details). Assuming a diagonal covariance matrix, we find the $\chi^2$ of the null test of $n/\langle n\rangle =1 $ to be 9.6, 11.1, and 9.8 for the LOWZ, LOWZE2, and LOWZE3 samples (to be compared to 10 measurement bins). Comparatively, the $\chi^2$ for the CMASS sample is 211. We therefore do not include any stellar density weights for any of the LOWZ samples.

In \cite{Ross11,Ross12}, it was shown that the relationship with stellar density also depends on the surface brightness of the galaxy. The $i_{\rm fib2}$ magnitude of the galaxy is a convenient measure of the surface brightness, as it represents the total flux within a given aperture (convolved with the seeing). The middle panel of Fig. \ref{fig:nstall} shows the relationship between the CMASS number density and the stellar density, divided into five ranges of $i_{\rm fib2}$ magnitudes ($i_{\rm fib2} < 20.3; 20.3 < i_{\rm fib2} < 20.6; 20.6 < i_{\rm fib2} < 20.9; 20.9 < i_{\rm fib2} < 21.2; 21.2 < i_{\rm fib2}$). In each bin, we find the best-fit linear relationship $n_{\rm gal} = A(i_{\rm fib2})+B(i_{\rm fib2})n_{\rm star}$. The dashed lines display the best-fit linear relationship in each $i_{\rm fib2}$ bin; the $\chi^2$ of the fits range between 4 and 8, for 8 degrees of freedom. With increasing $i_{\rm fib2}$, the best-fit $A$ and $B$ are $A(i_{\rm fib2}) =$ [0.959, 0.994, 1.038, 1.087, 1.120] and $B(i_{\rm fib2}) = [0.826, 0.149, -0.782,-1.83, -2.52] \times 10^{-4}$.

The linear fits to the relationship between galaxy and stellar density in each of the $i_{\rm fib2}$ bins are used to define weights to apply to CMASS galaxies to correct for the systematic dependency on stellar density.  To obtain the expected relationship at any $i_{\rm fib2}$, we interpolate between the results in the neighboring $i_{\rm fib2}$ bins, i.e., to find the expected relationship at $i_{\rm fib2} = 20.8$, we interpolate between the results in the $20.3 < i_{\rm fib2} < 20.6$ and $20.6 < i_{\rm fib2} < 20.9$ bins to obtain the slope, $B(i_{\rm fib2})$, and intercept, $A(i_{\rm fib2})$, of the relationship. The weight we apply to the galaxy is then
\begin{equation}
w_{\rm star}(n_{\rm star},i_{\rm fib2}) = \left(B(i_{\rm fib2})n_{\rm star}+A(i_{\rm fib2})\right)^{-1},
\label{eq:wstar}
\end{equation}
i.e., we simply weight by the inverse of the expected systematic relationship. 

The surface brightness dependence of the stellar density relationship must be accounted for in order to account for the redshift dependence of the systematic effect. The bottom panel of Fig. \ref{fig:nstall} shows the CMASS number density vs. stellar density, after applying $w_{\rm star}$. In each redshift bin, the systematic relationship is removed. After applying the systematic weights, the $\chi^2$ for the null test are 13.5, 8.4, and 11.2 (for 10 degrees of freedom), with increasing redshift; prior to applying the weights, they are 47, 117, and 65. The impact of the stellar density weights on the measured clustering is presented in Section \ref{sec:xiweights}.

\subsection{Seeing}
\label{sec:see}
There is a relationship between the observed density of BOSS CMASS galaxies and the local seeing due to the star galaxy separation cuts, as explained in \cite{Ross11}. Weights were previously defined and applied to the DR10 and DR11 CMASS samples to remove this trend, and we repeat such a procedure for DR12, while further investigating any relationship in the LOWZ samples.

\begin{figure}
\includegraphics[width=84mm]{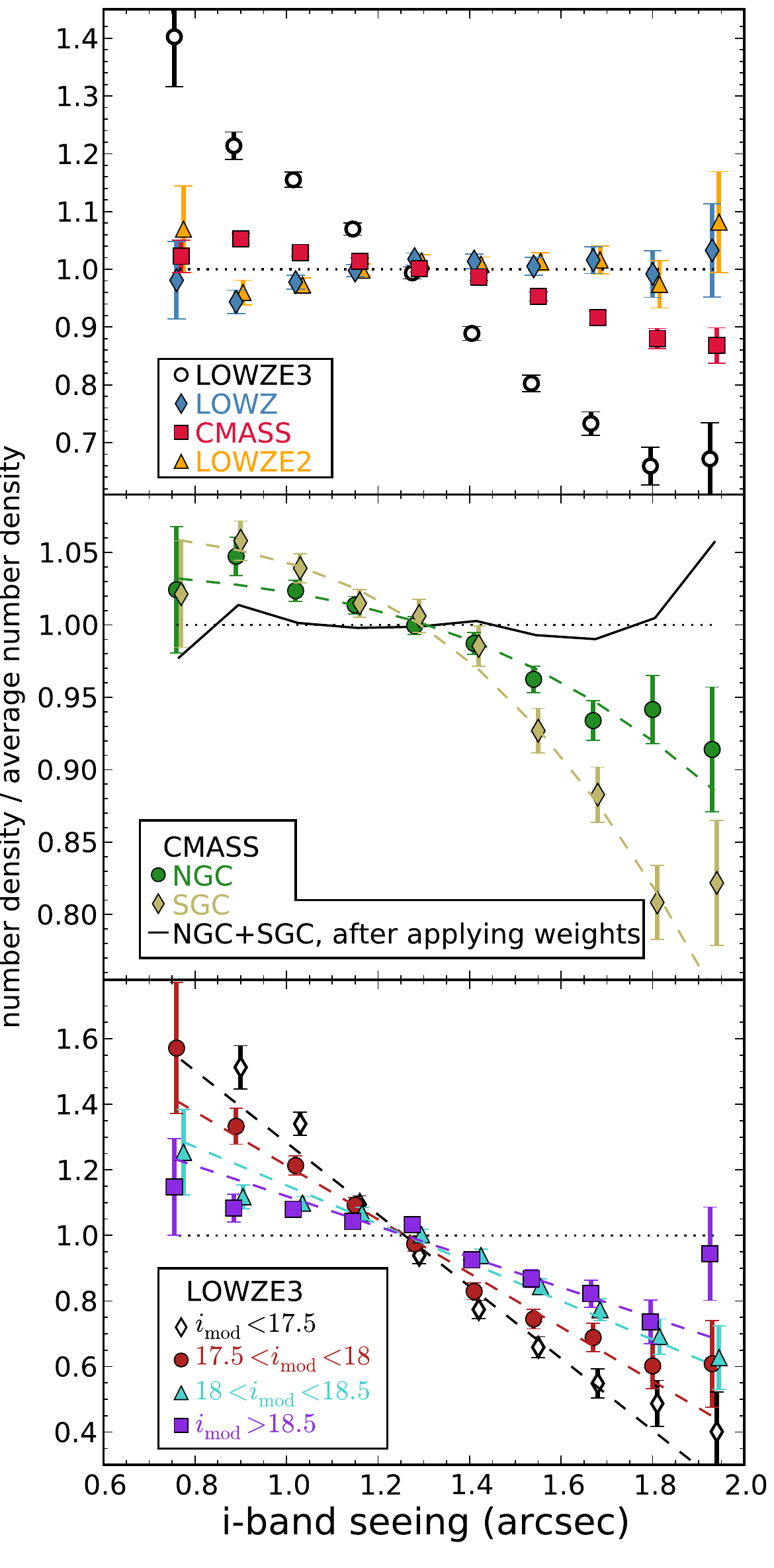}
  \caption{The relationship between observed density of BOSS galaxies and $i$-band seeing. Top panel: The relationships for CMASS and the three LOWZ selections. Middle panel: The relationships for CMASS NGC and SGC. The dashed curves display the best-fit relationship used to define the weights that correct for the observed trends. The solid curve displays the measured relationship for the combined NGC+SGC sample, after the weights have been applied. Bottom panel: The relationships for the LOWZE3 sample, split into four bins by $i_{\rm mod}$ magnitude. These relationships are used to define the weights applied the LOWZE3 sample. }
  \label{fig:nvsee}
\end{figure}

The top panel of Fig. \ref{fig:nvsee} displays the relationship between observed projected density and seeing for different BOSS selections. For the standard LOWZ selection and the LOWZE2 selection, no strong relationship is observed; the $\chi^2$ values of the null tests are 16.2 and 14.2, respectively, for 10 degrees of freedom. However, for CMASS and especially LOWZE3, clear relationships exist where the galaxy density decreases as the seeing gets worse (the $\chi^2$ values of the null tests are 225 and 877). For each sample, we will define systematic weights to correct for these relationships, and we describe this process throughout the rest of this section..

For CMASS, we define weights in a manner similar to that applied in \cite{alph}. We find the relationship with seeing is more severe in the SGC compared to the NGC, and we therefore determine the weights separately in each region\footnote{The difference in this dependency with seeing between the two regions must be related to another variable that differs considerably between the two regions, but a thorough investigation was unable to determine this variable.}. We find the best-fit parameters to the following model
\begin{equation}
n_g = A_{\rm see}\left[1-{\rm erf}\left(\frac{S_i-B_{\rm see}}{\sigma_{\rm see}}\right)\right],
\label{eq:seemod}
\end{equation}
where $S_i$ denotes the $i$-band seeing. The middle panel of Fig. \ref{fig:nvsee} displays the observed relationships for the data in each hemisphere and the best-fit model. For the NGC (SGC), the best-fit parameters are $A_{\rm see} = 0.5205 (0.5344)$, $B_{\rm see} =2.844 (2.267)$,and $\sigma_{\rm see} = 1.236 (0.906)$. The $\chi^2$ of these best-fit are 5.4 and 6.9 for the NGC and SGC, to be compared to 7 degrees of freedom. The seeing-dependent weights are simply given by the inverses of the best-fit relationships. The combined SGC+NGC relationship, after applying the seeing-dependent weights, is displayed using a solid black curve. The error-bars are suppressed, but the $\chi^2$ of the null test is 7.7 for 10 data points.

For LOWZE3, the inclusion of the $z$-band star/galaxy separation cut introduces a strong relationship between the galaxy density and the seeing. We find the effect is strongly magnitude dependent (we do not find this to be the case for the dependence of the CMASS sample with seeing). We therefore divide the sample by $i_{\rm mod}$ magnitude ($i$- and $z$-band magnitudes are strongly correlated at these redshifts and the SDSS $i$-band is less prone to zero-point fluctuations) and define weights in a manner analogous to how we defined the CMASS stellar density weights as a function of $i_{\rm fib2}$. We divide the LOWZE3 sample into four bins based on the galaxies' $i_{\rm mod}$ magnitude, $i_{\rm mod} < 17.5$, $17.5 < i_{\rm mod} < 18$, $18 < i_{\rm mod} < 18.5$, and $i_{\rm mod} > 18.5$, and fit a linear relationship to each and then interpolate to obtain the weight as a function of the local $i$-band seeing and the galaxy's $i_{\rm mod}$ magnitude. The measurement in these four magnitude bins is displayed by the points with error-bars in the bottom panel of Fig. \ref{fig:nvsee}. The dashed curves display the best-fit linear relationship to each. We find the slope of the best-fits, $\ell$, is well-approximated by
\begin{equation}
\ell = b+m(i_{\rm mod}-16)^{\frac{1}{2}},
\end{equation}
with $b=0.875$ and $m=-2.226$. Thus, given that the mean seeing over the footprint is 1.25, the relationship between $i$ band seeing, LOWZE3 density ($n_{LE3}$), and $i_{\rm mod}$ is given by
\begin{equation}
n_{LE3}(S_i,i_{\rm mod}) = 1+(S_i-1.25)\ell(i_{\rm mod}).
\label{eq:nl3si}
\end{equation}
We set any $\ell < -2$ to $\ell_{\rm min} = -2$ and take the inverse of equation (\ref{eq:nl3si}) in order to apply weights to the LOWZE3 sample, setting any weights greater than 5 to 5.

The total systematic weight (e.g., $w_{\rm star}\times w_{\rm see}$ for CMASS) is normalized such that the weights sum to the total number of galaxies in the sample they are defined for. The impact of the seeing weights we apply on the measured clustering of the CMASS and LOWZE3 samples is presented in Section \ref{sec:xiweights}.

\subsection{Sky background, Airmass, Extinction}

\begin{figure}
\includegraphics[width=84mm]{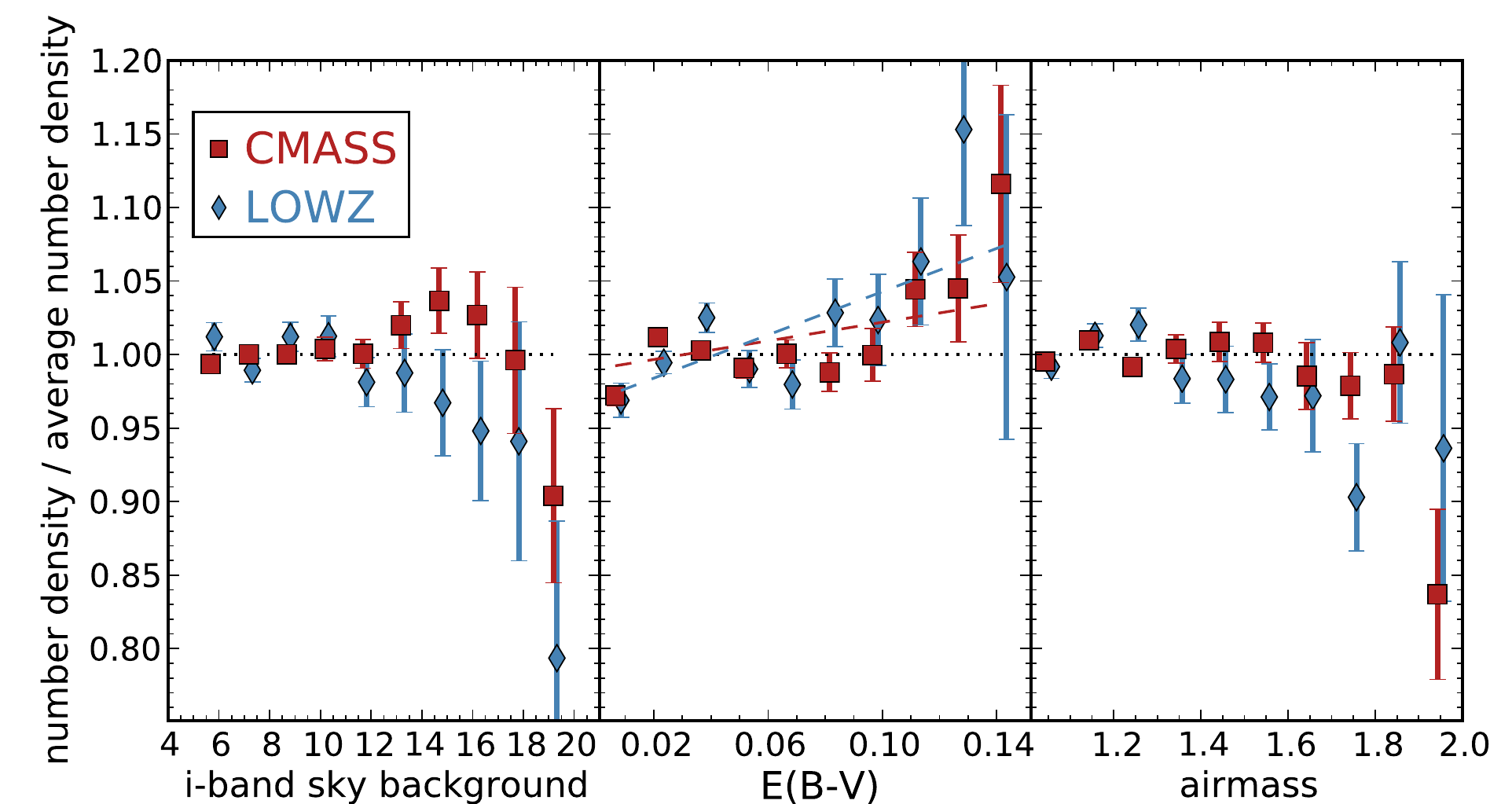}
  \caption{The relationship between galaxy density observed density and sky background (in nanomaggies per square arc second), Galactic extinction (in E(B-V)), and airmass, for CMASS and LOWZ. The dashed lines display the predicted relationship with Galactic extinction, based on the difference between the extinction coefficients applied to BOSS imaging data and those found in Schlafly \& Finkbeiner (2011).}
  \label{fig:sea}
\end{figure}

As for previous BOSS data releases, we test against three additional potential systematic quantities, each of which affects the depth of the imaging data: sky background, airmass, and Galactic extinction. These are shown for the CMASS and LOWZ samples in Fig. \ref{fig:sea}. For sky-background and airmass, the $\chi^2$ values of the null tests range between 9 (for CMASS against sky background) and 18 (for LOWZ against airmass), to be compared to the 10 data points in each case. 

For Galactic extinction, the $\chi^2$ are somewhat larger than expected: 35 for the CMASS sample and 26 for LOWZ (compared to 10 data points). However, these large $\chi^2$ are dominated by the value at the lowest extinction, which is low by 3 per cent for both LOWZ and CMASS\footnote{Masking the data at the lowest extinction values does not cause any significant change in the clustering results.}. \cite{Schlafly11} suggest somewhat different extinction coefficients than those used to target BOSS galaxies. Such a change implies extinction-dependent shifts in the color of the BOSS selection and these shifts can be translated into an expected change in target density as a function of extinction. The expected trend is shown with dashed lines and agrees with the overall trend observed for both LOWZ and CMASS. In terms of $\chi^2$, the LOWZ value is 19 when using this prediction and the CMASS value remains 35 (improvement at the extrema of the range is countered by disagreement at E(B-V)$\sim$0.08). This implies any effect on the measured clustering found when correcting for this predicted relationship would be marginal, and, indeed, we find no significant changes in the measured clustering when applying and extinction-dependent weights. We thus choose not to include any weights to correct for these trends with Galactic extinction.

Overall, we do not find any clear trends, given the uncertainty, between the density of BOSS galaxies and sky background, Galactic extinction, or airmass. Therefore, like in previous BOSS analyses, we do not weight BOSS galaxies according to any of these quantities. In the tests that follow, it will become clear that the systematic effects we correct for via weights (stellar density and seeing) would have minimal impact on the final BOSS BAO and RSD results even if they had not been corrected for. Attempts to correct for additional potential systematic effects of marginal significance are thus ill-advised. However, each individual analysis will be affected differently, and it would therefore be prudent for any future studies of the clustering of BOSS galaxies (e.g., primordial non-Gaussianity; \citealt{Ross13}) at the largest scales to reconsider this choice.

\section{BOSS Galaxy Clustering}
\label{sec:clus}
In this section, we present the configuration-space clustering of BOSS galaxies. We determine the relative importance of the systematic weights we apply, in terms of the impact on the measured correlation functions. We then show BOSS clustering results when the samples are divided by hemisphere (NGC and SGC) and by targeting selection (LOWZ, LOWZE2, LOWZE3, and CMASS). We conclude by showing the clustering of the combined BOSS sample, split by redshift.

\subsection{Effect of weights}
\label{sec:xiweights}
\begin{figure}
\includegraphics[width=84mm]{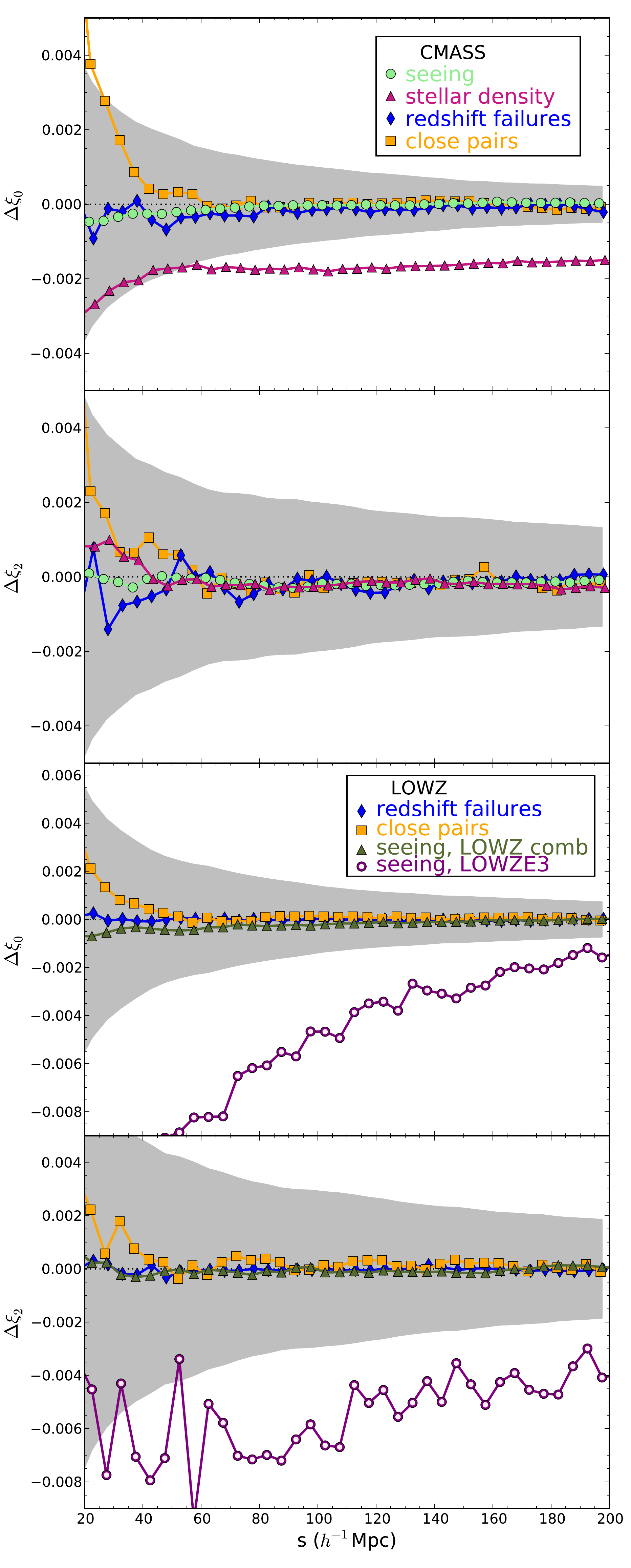}
  \caption{The change in the measured monopole and quadrupole of the BOSS CMASS (top panels) and LOWZ (bottom panels) correlation functions, when the given systematic weight is applied. `LOWZ comb' refers to the combination of the LOWZ, LOWZE2, and LOWZE3 selections. The grey shaded region displays the 1$\sigma$ uncertainty obtained from mock samples.}
  \label{fig:xi0sys}
\end{figure}

The CMASS sample contains the most signal-to-noise of any particular BOSS selection, has a significant percentage of unobserved close-pairs and redshift failures (5.4 and 1.8 per cent), and uses weights for both stellar density and seeing to correct for systematic dependencies in the observed number density. We test the impact of these weights by comparing the clustering measured with the weights applied to that without. For the monopole, these differences are displayed in the top panel of Fig. \ref{fig:xi0sys}. In order to assess the total potential impact of the weights, we find the total $\chi^2$ difference between the clustering measured with and without the weights. The relative importance of each weight is as one would expect visually: the $\chi^2$ are 13.1, 3.7, 2.1, and 0.1 for stellar density, close pair, redshift failure, and seeing weights. 

The importance of the weights is smaller for CMASS $\xi_2$ than $\xi_0$, as one can see in the 2nd to the top panel in Fig. \ref{fig:xi0sys}. The $\chi^2$ are 0.5, 2.5, 2.3, and 0.1 for stellar density, close pair, redshift failure, and seeing weights. Unsurprisingly, the weights that affect the radial distribution are most important for $\xi_2$, and the redshift failure weights are slightly more important for $\xi_2$ than for $\xi_0$. For both $\xi_0$ and $\xi_2$, the seeing weights have negligible impact. The $\chi^2$ difference is only 0.1 for both, implying that the {\it greatest} difference it could cause in the determination of a model parameter is 0.3$\sigma$ (whereas for stellar density, it is potentially a 3.6$\sigma$ effect) . 

For the nominal LOWZ sample, the only systematic weights applied are for close pairs and redshift failures, and these represent only 2.9 and 0.5 per cent of LOWZ targets. Similar to CMASS, the close-pair weights increase the small-scale clustering amplitudes. However, the effect is much smaller, compared to the uncertainty on the measurements, and the $\chi^2$ are only 0.8 and 1.4 for $\xi_0$ and $\xi_2$. For redshift failures, the $\chi^2$ are only 0.2 and 0.1 for $\xi_0$ and $\xi_2$.

For the LOWZE3 sample, selected over the full NGC footprint, we defined a weight based on seeing, in order to reverse a strong effect on the observed number density of the sample. The effect of this weight on the measured clustering of the LOWZE3 selection over the full NGC footprint is shown using circles in the bottom two panel of Fig. \ref{fig:xi0sys} (of note, the size of the uncertainty band for LOWZE3 should be larger than for the displayed LOWZ uncertainty, due to the number density being approximately half of LOWZ and the fact that the SGC footprint is not used). It has the strongest effect of any weight we apply. 

While the effect of the seeing weights is strong for the LOWZE3 sample over the full NGC footprint, our final sample will only use this selection over 755 deg$^2$. Further, when these data are used, we combine the LOWZ sample with CMASS and use data in the range $0.2 < z < 0.5$. When we consider the impact of the weights on the clustering of this combined sample (denoted `LOWZ comb'), we find a $\chi^2$ difference of only 0.2 between the $\xi_{0,2}$ measured with and without the weights applied, this comparison is plotted using triangles in the bottom two panels Fig. \ref{fig:xi0sys}. The reason for the sharp decrease in significance is two-fold: 1) the LOWZE3 sample accounts for approximately five per cent of the statistical power of the combined sample with $0.2 < z < 0.5$ and 2) the effect of the weights when restricting to only the 755 deg$^2$ of unique LOWZE3 data is considerably smaller than over the full NGC (presumably due to the particular pattern of seeing in this area). Thus, while its effect is dramatic on the LOWZE3 sample within the full NGC area, the effect of the weights on the combined sample is minor for the combined sample that we use for BOSS science. Notably, the inclusion of the LOWZE3 area allows us to include the CMASS data occupying the same footprint with $0.2 < z < 0.5$ into the combined sample, which increases the statistical power of the region to eight per cent of the total. Our tests suggest that even in the (catastrophic) event that residual systematic effects in the LOWZE3 sample are equal to those we have treated with weights for seeing, the most any derived parameter could be biased is 0.45$\sigma$ (and this is in the specific case that the signal being searched for is exactly mimicked by the systematic effect). The expected variation (assuming Gaussian statistics) when increasing a sample from 92 per cent complete to 100 per cent is 0.4$\sigma$; in this sense the expected gain is approximately equal to the worst-case scenario for the inclusion of the LOWZE3 data. We thus include the 755 deg$^2$ of unique LOWZE3 data in the BOSS combined sample.

\subsection{Hemisphere}
\begin{figure}
\includegraphics[width=84mm]{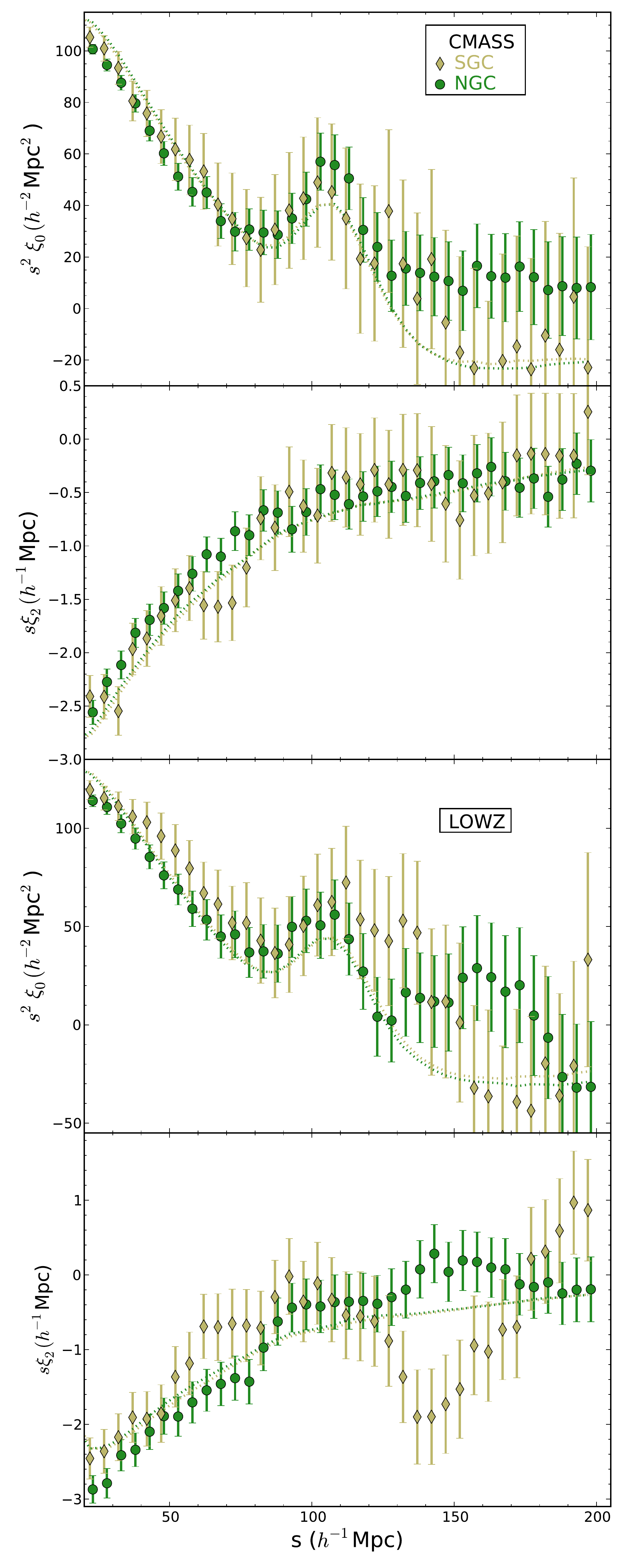}
  \caption{The clustering of BOSS CMASS (top two panels) and LOWZ (bottom two panels) galaxies, for the two contiguous regions within the SGC and NGC hemispheres. The dotted lines denote the mean of the QPM mock samples.}
  \label{fig:xicmasslowzNS}
\end{figure}

As described in Section \ref{sec:NSdata}, the selection functions for the NGC and SGC BOSS galaxy data are slightly different. Here, we compare the clustering in the two regions. This comparison is shown for CMASS in the top two panels of Fig. \ref{fig:xicmasslowzNS} for $\xi_0$ (top panel) and $\xi_2$ (2nd to top panel). In the range $20 < s < 200h^{-1}$, the $\chi^2$ obtained when testing the NGC $\xi_0$ against the SGC $\xi_0$ (determined by summing the two QPM covariance matrices) is 42 for the 36 data points. Restricting to the range $50 < s < 150h^{-1}$, the $\chi^2$ is 25 for 20 points. The CMASS clustering in the two regions agrees to an similar extent as it did for the DR9 data \citep{Ross12}. The agreement is somewhat worse for $\xi_2$, as we find a $\chi^2$ of 48 for $20 < s < 200h^{-1}$Mpc and 29 for $50 < s < 150h^{-1}$Mpc.

The comparison between NGC and SGC for the LOWZ sample is shown in the bottom panels of Fig. \ref{fig:xicmasslowzNS}. The agreement between the $\xi_0$ is quite good; the $\chi^2$ is 28 for the 36 data points with $20 < s < 200 h^{-1}$Mpc. For $\xi_2$, the agreement is worse; the $\chi^2 $ is 50 for the same 36 $s$ bins. The discrepancy is dominated by large-scales, as for the 22 data points with $s < 130h^{-1}$Mpc, the $\chi^2$ is 19, while for the 14 with $s > 130h^{-1}$Mpc, the $\chi^2$ is 29. The difference is such that it serendipitously cancels for the combined NGC+SGC sample. While unusual, no effect studied in this paper has a significant impact on the shape of the LOWZ quadrupole at $s > 130h^{-1}$Mpc and we can offer no explanation beyond a statistical fluctuation (which would be at $\sim2\sigma$ for $\chi^2$/dof$=50/36$). We note that scales $s > 130h^{-1}$Mpc have a negligible impact on RSD structure growth measurements and only a small impact on BAO measurements (see Appendix \ref{app:rob}).

We do not find any strong discrepancies between the NGC and SGC configuration-space clustering of BOSS galaxies at scales relevant to BAO or RSD studies. We therefore combine the two hemispheres in our standard analysis, but demonstrate in subsequent sections that the results applied to each hemisphere individually are consistent with the combined constraints and that the BAO results are thus robust to any concerns about combining the NGC and SGC results. \cite{Acacia} show discrepancies between the two hemispheres are more apparent at small scales when studying the power spectrum. The differences are shown to be a consequence of the color offsets between the two regions, as discussed in Section \ref{sec:NSdata}. These differences are not apparent in the correlation function analysis because they are isolated to $s < 20h^{-1}$ in configuration space.

\subsection{Targeting selection}
\begin{figure}
\includegraphics[width=84mm]{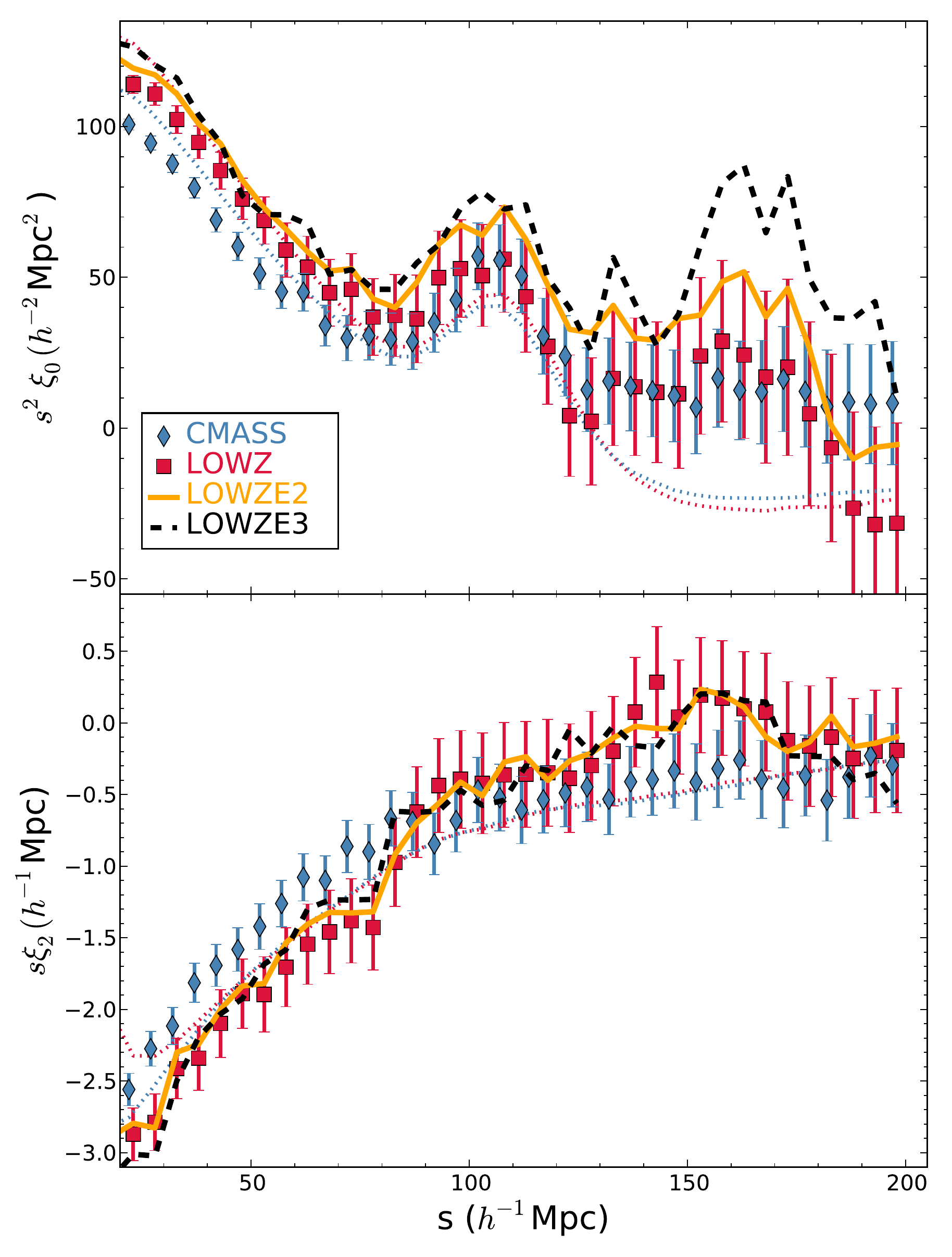}
  \caption{The clustering of BOSS galaxies, using the four different targeting specifications. The CMASS and LOWZ samples occupy different redshift regimes (see Fig. \ref{fig:nz}) and thus some difference in clustering amplitude is to be expected. The dotted lines denote the mean of the QPM mock samples.}
  \label{fig:xilowzearlycmass}
\end{figure}

Here, we compare the clustering in the nominal LOWZ selection to the clustering obtained using the LOWZE2 selection (which is the full LOWZ footprint plus the 131 deg$^2$ area where the LOWZE2 selection was applied to targeting) and to the clustering obtained using the LOWZE3 selection (which is the full LOWZ area plus the 755deg$^2$ where the LOWZE3 selection was applied to targeting.) We use the full area available, within the NGC, in order to obtain the best statistics on the galaxies that comprise each selection. 

We show this comparison in Fig. \ref{fig:xilowzearlycmass}, where the CMASS clustering is also shown. The LOWZE2 selection covers the same area as the LOWZ selection, with 131deg$^2$ more area and a lower number density. We should thus expect consistent clustering measurements. Its correlation function is displayed using a solid curve in Fig. \ref{fig:xilowzearlycmass}. For both $\xi_0$ and $\xi_2$, LOWZE2 appears consistent with the LOWZ measurements, but with a slightly higher clustering amplitude. Indeed, using the LOWZ covariance matrix, we find a $\chi^2$ of 23 for the monopole and 19 for the quadrupole when testing the range $20 < s < 200$ (36 data points). Multiplying the LOWZ $\xi_0$ by 1.04 reduces the $\chi^2$ to 20. An increase in clustering amplitude is expected, as the LOWZE2 sample applies brighter limits to the selection compared to the nominal LOWZ selection. Applying a factor to the quadrupole does not significantly reduce the $\chi^2$. These $\chi^2$/dof are much less than one, as expected for measurements that are highly correlated.

The LOWZE3 sample covers the same area as the LOWZ footprint, with an additional 755deg$^2$, a lower number density, and large weights that account for variations in target density with seeing. As detailed in Section \ref{sec:data}, its mean number density is just greater than half that of the nominal LOWZ selection. The LOWZE3 correlation functions are displayed using dashed curves in Fig. \ref{fig:xilowzearlycmass}. The measurements appear qualitatively similar to the LOWZ measurements, especially for the quadrupole, but with a slightly greater clustering amplitude for $\xi_0$. However, when repeating the test we applied to the LOWZE2 sample, using the LOWZ covariance matrix to evaluate a $\chi^2$ value for the difference between the LOWZ and LOWZE3 samples, we find the $\chi^2$ is 83 for the monopole, when multiplying the amplitudes by a factor of 1.10, in the range $20 < s < 200h^{-1}$Mpc (36 data points), and that this $\chi^2$ is not significantly better or worse for a particular range of scales (e.g., it is 31 for the 16 data points with $s > 120h^{-1}$Mpc). Similar to LOWZE2, we expect an increase in the clustering amplitude of the LOWZE3 sample compared to LOWZ, as the cuts applied to LOWZ to produce the LOWZE3 sample preferentially remove fainter galaxies. The quadrupole gives somewhat better agreement, as the $\chi^2$ is 50 for the range $20 < s < 200h^{-1}$Mpc (applying a constant factor does not significantly improve the $\chi^2$). 

If we increase the diagonal elements of the LOWZ covariance matrix by 10 per cent and repeat the test, we find the $\chi^2$ reduce to 36 for $\xi_0$ and 24 for $\xi_2$ (for the same 1.10 factor for $\xi_0$). Changing the covariance matrix in this manner represents the addition of a pure shot-noise contribution to the covariance matrix that has a variance which is 10 per cent of the LOWZ variance. This is likely conservative, as the LOWZE3 number density is approximately half of the LOWZ number density. When using the value of $P_0 = 10^4h^{3}$Mpc$^{-3}$ adopted to define the FKP weights, a number density of $3\times10^{-4}h^{3}$Mpc$^{-3}$ for the LOWZ sample, and a number density $1.5\times10^{-4}h^{3}$Mpc$^{-3}$ for the LOWZE3 sample, we find the expected increase in the variance is 56 per cent. We therefore conclude that the clustering of the LOWZ and LOWZE3 samples is consistent, when allowing for a 10 per cent increase in clustering amplitude and the extra shot noise imparted by the lower LOWZE3 number density.

The clustering amplitude of the CMASS sample is clearly lower than that of the LOWZ sample on scales $s < 80h^{-1}$Mpc. Again, using the covariance matrix of the LOWZ sample, we find the $\chi^2$ between two measurements, scaling the CMASS result by a constant factor. We find a minimum $\chi^2$ of 34 for a factor 1.12 for the monopole and 41 for the quadrupole, applying a factor of 1.27. This implies the shapes of the measured monopole and quadrupole are consistent between the CMASS and LOWZ samples.

\subsection{Combined BOSS sample}
\begin{figure}
\includegraphics[width=84mm]{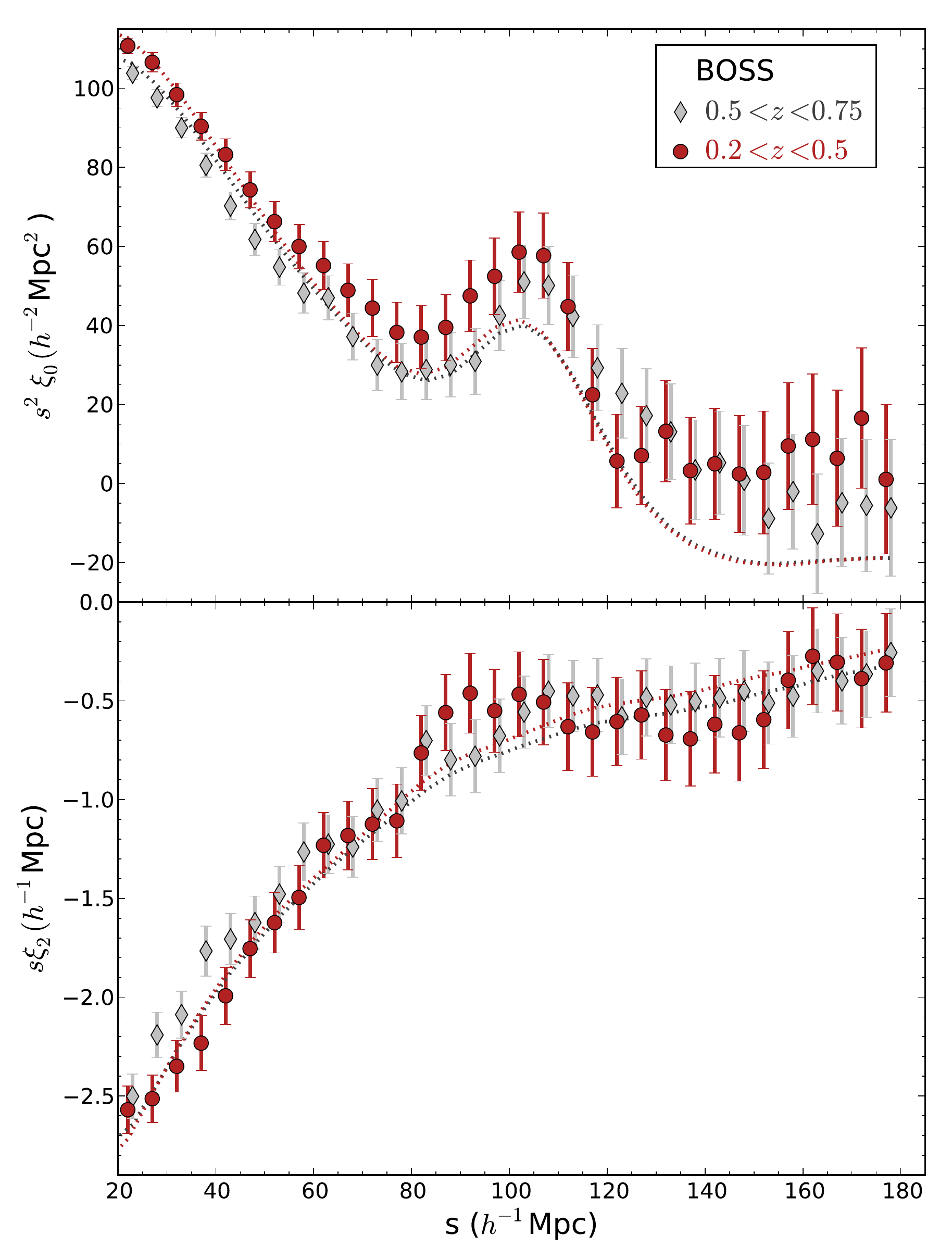}
  \caption{The measured monopole and quadrupole of the BOSS galaxy correlation function, split into two redshift shells. The dotted lines display the mean of the MultiDark-Patchy samples with the same redshift selections.}
  \label{fig:xicom}
\end{figure}

Finally, we present the clustering of the BOSS galaxy sample, i.e., the combined sample of LOWZ, LOWZE2, LOWZE3, and CMASS, applying all of the weights defined in the previous section. The clustering amplitudes of the individual BOSS samples differ by less than 20 per cent for the CMASS/LOWZ samples and less than 10 per cent for the individual LOWZ samples. The scales we are interested in are less than 150$h^{-1}$ Mpc. Thus any cross-sample pairs of galaxies will be a small percentage of the total entering any particular measurement and we do not expect any significant shift in the amplitude as a function of scale within the scales of interest. Further, we have tested weighting the individual samples such that their density field has the same amplitude in the regions of overlap. We find this weighting has no significant impact on the measured clustering, and we therefore simply add the catalogs (both the galaxy and the random ones, in the correct proportion) to produce the combined sample.
The clustering measurements for the combined BOSS sample with $0.2 < z < 0.75$, split into two redshift bins at $z=0.5$, are displayed in Fig. \ref{fig:xicom}. One can see that the clustering is similar in the two redshift regimes, with a slightly greater clustering amplitude in the lower redshift sample.

The dotted curves in Fig. \ref{fig:xicom} display the mean of the PATCHY mock samples, which are a better match to the BOSS combined sample properties than QPM (one of the biggest differences is the treatment of the lightcone in PATCHY, see \citealt{Kitaura15} for full details)\footnote{The QPM mocks are a good match }. The covariance between the $s$ bins makes the statistical match between the mean of the mocks and the measured clustering better than might be guessed by eye. For the monopole and $0.2 < z < 0.5$ it is 38 for the 32 measurement bins with $20 < s < 180h^{-1}$Mpc, while for $0.5 < z < 0.75$, it is 31 for the same range of scales. For the quadrupole, it is 35 for $0.2 < z < 0.5$ and 30 for $0.5 < z < 0.75$. Allowing the mean of the mocks to be scaled by a constant value, the $\chi^2$ decreases to 36 for the $0.2 < z < 0.5$ monopole when applying a factor of 0.98. No significant improvement is found for the $0.5 < z < 0.75$ monopole. For the quadrupole, the $\chi^2$ cannot be significantly improved by applying any factor to the mean of the $0.2 < z < 0.5$ mocks and is reduced to 27 when applying a factor of 0.93 to the $0.5 < z < 0.75$ mocks. 

For the monopole, the clustering at large scales shows an apparent excess, however it is of marginal statistical significance : for the $0.2 < z < 0.5$ bin the $\chi^2$ is 20 for the 12 data points with $s > 120h^{-1}$ and 17 for the 20 points with $s < 120h^{-1}$, but for $z>0.5$, the $\chi^2$/dof is slightly smaller for $s > 120h^{-1}$ (10/12) than for $s < 120h^{-1}$ (22/20). While all of the data points are greater than the mean of the mocks at large-scales, the large degree of covariance between the measurements makes this fact unremarkable. In Fourier space, \cite{BeutlerDR12RSD,GriebDR12RSD} find no apparent excess for $k > 0.01h$Mpc$^{-1}$.

\section{Robustness of BAO Measurements to Observational Treatment}
\label{sec:BAOrob}
\begin{table*}
\begin{minipage}{7in}
\centering
\caption{Statistics of anisotropic BAO fits on either 600 (`Sub' and `Sub Star, weighted') or 200 (`Fid.' and `Sub Star, not weighted') pre-reconstruction NGC mocks with and without stellar density systematics. Numerals match the list in the text. For these, we have assumed the same cosmology as used to construct the QPM mocks (given in Table \ref{tab:baoexp}). Thus, the expected $\alpha$ and $\epsilon$ values are 1 and 0. `Fid.' denotes the case where the fiducial mock samples have been used; `Sub Star, not weighted' denotes the case where each mock was subsampled based on the stellar density relationship observed in the data; `Sub' refers to the case where the mocks have been randomly sub-sampled so that the number density is the same as the `sub Star, not weighted' case; `Sub Star, weighted' denotes the case where the Star, not weighted mocks have stellar density weights assigned in a manner matching the procedure applied to the data. $S$ denotes standard deviation and $\sigma$ the uncertainty recovered from a likelihood.}
\begin{tabular}{lccccccccccccccccc}
\hline
\hline
Case  & $\langle\alpha_{||}\rangle$ & $S_{||}$ &$\langle\sigma_{||}\rangle$& $\langle\alpha_{\perp}\rangle$ & $S_{\perp}$ &$\langle\sigma_{\perp}\rangle$ &    $\langle\alpha\rangle$ & $S_{\alpha}$ &$\langle\epsilon\rangle$ & $S_{\epsilon}$\\
\hline
600 mocks used: & \\
(iii) Sub Star, weighted & 1.0011 & 0.0534 & 0.0567  & 1.0045 & 0.0253 & 0.0266 & 1.0029 & 0.0181 & -0.0013 & 0.0220\\
(iv) Sub  & 1.0016 & 0.0532 & 0.0564  & 1.0043 & 0.0247 & 0.0266 & 1.0029 & 0.0180 & -0.0010 & 0.0217\\
\hline
200 mocks used: & \\
(i) Fid. & 1.0011 & 0.0510 & 0.0554  & 1.0053 & 0.0241 & 0.0259 & 1.0034 & 0.0165 & -0.0015 & 0.0213\\
(ii) Sub Star, not weighted & 1.0009 & 0.0520 & 0.0550  & 1.0055 & 0.0250 & 0.0257 & 1.0035 & 0.0171 & -0.0016 & 0.0217\\
\hline
\label{tab:baoresultsmock}
\end{tabular}
\end{minipage}
\end{table*}

In this section, we measure the BAO scale for each of the BOSS target samples, and test the robustness of the measurements to our treatment of the selection function. We first test the effect of the stellar density weights by simulating the stellar density systematic in mock samples and then comparing the BAO results to those without any simulation of the stellar density systematic. We then test the BOSS BAO measurements by determining their dependency on the application of the various weights and examining the results we obtain for each Galactic hemisphere.

\subsection{Tests on mocks}
\label{sec:mocksys}
We test for the systematic impact the stellar density relationship has on the measured BAO position by simulating the effect in mock CMASS samples and thus determine an observational systematic uncertainty on BOSS BAO measurements. We take the stellar density field observed by SDSS and assume the distribution of stars is the same for each of the mocks. In order to simulate the systematic effect of stellar density observed in the BOSS data, we also must assign $i_{\rm fib2}$ magnitudes to each mock galaxy. We accomplish this by taking the observed distribution of $i_{\rm fib2}$ magnitude as a function of redshift and sampling from this for each mock galaxy redshift, i.e., we estimate $P(i_{\rm fib2}|z)$ based on the BOSS data and use this to assign the $i_{\rm fib2}$ values to each mock galaxy. This allows us to analyze the statistics of the distributions of BAO scale measurements obtained from the following four cases that include different levels of systematic contamination and correction:\begin{enumerate}
  \item Fiducial mocks; BAO fits are presented for 200 of these, in order to match the number used in case ii.
\item Mocks that have been randomly sub-sampled in a manner matching the observed stellar density systematic\footnote{E.g., if the density is expected to be 0.95 that of the nominal density, each mock galaxy is tested and kept in the sample if a randomly generated number between 0 and 1 is less than 0.95.}; the clustering of these has the spurious large-scale power similar to the unweighted data sample; BAO fits have been performed for 200 of these.
\item Mocks that first have the sub-sampling procedure applied in case (ii) and then have stellar density weights calculated and used for their clustering; the stellar density systematic is thus removed, but the weights are calculated on a per-mock basis; BAO fits have been performed for 600 of these.
\item Mocks that have been uniformly sub-sampled by 4\% to have the same number density as those sub-sampled according to the stellar density systematic; these are a more-fair comparison to cases (ii) and (iii) than the fiducial mocks: BAO fits have been performed for 600 of these.

\end{enumerate}
Cases iii) and iv) are the most realistic and will be used to determine any additional scatter from the weighting process. We therefore concentrate on performing fits for these tests, while for other tests we simply perform a number sufficient to detect any significant issues.

\begin{figure}
\includegraphics[width=84mm]{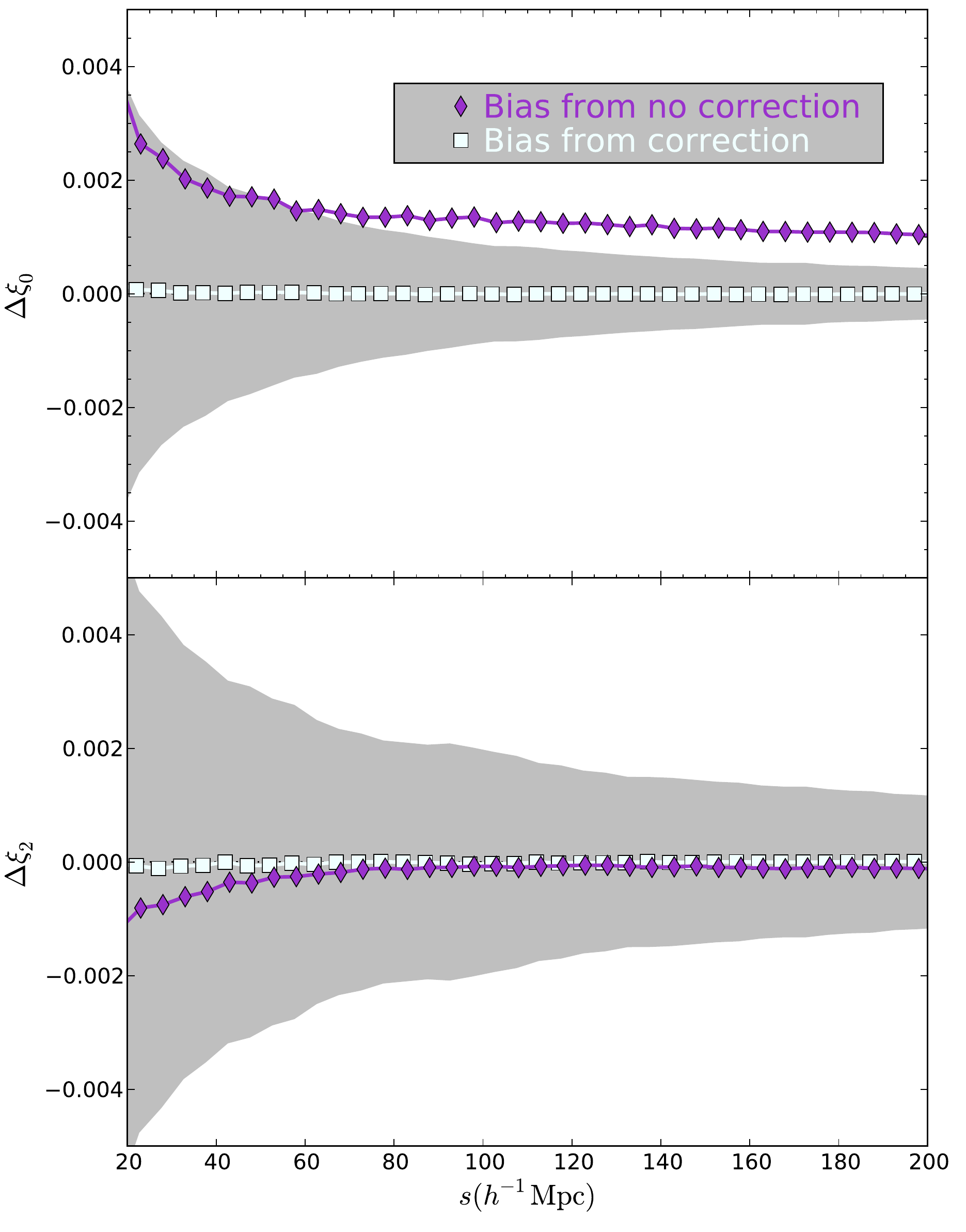}
  \caption{The change in the mean measured monopole (top) and quadrupole (bottom) of the correlation function of mock samples, when comparing the fiducial case (without any simulation of observational systematics) to the case where the stellar density systematic has been simulated (`darkorchid' diamonds) and when comparing the fiducial case to the case where the stellar density systematic has been simulated and corrected for (azure squares). The grey shaded region displays the 1$\sigma$ uncertainty obtained from mock samples.}
  \label{fig:xi0mocksys}
\end{figure}

We use the QPM CMASS NGC mocks and for all tests we use the $\xi_{0,2}$ covariance matrix determined from 1000 realizations of the fiducial case (i). For these, we have assumed the same cosmology as used to construct the QPM mocks (given in Table \ref{tab:baoexp}) both when measuring $\xi_{0,2}$ and in the BAO template. These choices match those of \cite{CuestaDR12}. Thus, the expected $\alpha$ and $\epsilon$ values are 1 and 0. 

The results of anisotropic BAO fits are shown in Table \ref{tab:baoresultsmock} (`S' denotes a standard deviation and $\sigma$ an uncertainty recovered from a likelihood). Compared to the cases with no stellar density systematic, introducing the stellar density systematic shifts the mean recovered value of $\alpha_x$ by at most 0.0005, equivalent to 0.01$\sigma$. This suggests that any potential systematic bias due to stellar density is negligibly small; i.e., if we applied no correction for stellar density systematics, we would still recover un-biased BAO measurements. All of the mean $\sigma$ are very similar (for cases using the same set of mocks), as one might expect given that the same covariance matrix is used in all cases. 

\begin{table*}
\begin{minipage}{7in}
\centering
\caption{Isotropic and anisotropic BAO fits on pre- and post-reconstruction CMASS data when different weights are applied. The fiducial case is `all' (without specification of hemisphere) and we are interested in the variations (or lack thereof) between the results in different cases. The cases are as follows: `none' denotes no systematic weights were applied; `cp' denotes close pair weights were applied; `zf' denotes redshift failure weights were applied in addition to cp; `st' denotes stellar density weights were applied in addition to cp and zf weights; `all' additionally includes the seeing weights, and thus all weights were applied. `C16' denotes that the $\xi_{0,2}$ measurements using the reconstruction applied to Cuesta et al. (2016) were applied (whereas the rest of the cases used reconstruction similar to that of Burden et al. 2014). All covariance matrices used were determined from the appropriate sample of 1000 QPM mocks. The reconstruction applied to these mocks matches what was used in Cuesta et al. (2016).}
\begin{tabular}{lcccccc}
\hline
\hline
Sample & Weights  &   $\alpha$ & $\chi^2$/dof & $\alpha_{||}$ & $\alpha_{\perp}$ & $\chi^2$/dof \\
\hline
Pre-reconstruction: & & & & &\\
\hline
CMASS & none & $0.985\pm0.013$ & 26/15 & 0.965$\pm$0.035 & 0.996$\pm$0.020 & 42/30\\
CMASS & cp & $0.986\pm0.012$ & 23/15 & 0.966$\pm$0.034 & 0.996$\pm$0.020 & 41/30\\
CMASS & zf & $0.985\pm0.012$ & 30/15 & 0.972$\pm$0.034 & 0.992$\pm$0.020 & 47/30\\
CMASS & st & $0.987\pm0.012$ & 24/15 & 0.971$\pm$0.034 & 0.996$\pm$0.020 & 41/30\\
CMASS & all & $0.987\pm0.012$ & 24/15 & 0.970$\pm$0.034 & 0.997$\pm$0.021 & 40/30\\
CMASS NGC & all & $0.985\pm0.013$ & 19/15 & 0.965$\pm$0.037 & 0.994$\pm$0.026 & 41/30\\
CMASS SGC & all & $1.020\pm0.028$ & 27/15 & 1.020$\pm$0.095 & 1.014$\pm$0.057 & 38/30\\
\hline
LOWZ & none & $0.992\pm0.026$ & 18/15 & x & x & x\\
LOWZ & zf & $0.993\pm0.026$ & 18/15 & x & x & x\\
LOWZ & all & $0.993\pm0.025$ & 17/15 & x & x & x\\
LOWZE3 NGC & all & $1.007\pm0.025$ & 38/15 & x & x & x\\
LOWZE2 NGC & all & $1.010\pm0.029$ & 14/15 & x & x & x\\
LOWZ NGC & all & $1.009\pm0.029$ & 18/15 & x & x & x\\
LOWZ SGC & all & $0.949\pm0.042$ & 10/15 & x & x & x\\
\hline
Post-reconstruction: & & & & &\\
\hline
Sample & Weights  &   $\alpha$ & $\chi^2$/dof & $\alpha_{||}$ & $\alpha_{\perp}$ & $\chi^2$/dof \\
\hline
CMASS & none & $0.9843\pm0.0093$ & 16/15 & 0.962$\pm$0.023 & 0.997$\pm$0.014 & 30/30\\
CMASS & cp & $0.9850\pm0.0083$ & 27/15 & 0.961$\pm$0.022 & 0.996$\pm$0.013 & 43/30\\
CMASS & zf & $0.9856\pm0.0087$ & 33/15 & 0.962$\pm$0.022 & 0.998$\pm$0.013 & 63/30\\
CMASS & st & $0.9859\pm0.0086$ & 18/15 & 0.957$\pm$0.021 & 1.001$\pm$0.013 & 37/30\\
CMASS & all & $0.9832\pm0.0085$ & 19/15 & 0.952$\pm$0.021 & 1.000$\pm$0.013 & 46/30\\
CMASS & C16 & $0.9849\pm0.0092$ & 14/15 & 0.949$\pm$0.024 & 1.003$\pm$0.014 & 30/30\\
CMASS NGC & all & $0.975\pm0.010$ & 15/15 & 0.942$\pm$0.022 & 0.999$\pm$0.016 & 39/30\\
CMASS SGC & all & $1.016\pm0.020$ & 15/15 & 1.005$\pm$0.044 & 1.013$\pm$0.029 & 50/30\\
\hline
\label{tab:baoresultspr}
\end{tabular}
\end{minipage}
\end{table*}

In order to assess whether the weighting process introduces any additional scatter, we have compared the standard deviations recovered from Cases iii) and iv). For both $\alpha_{||}$ and $\alpha_{\perp}$, the standard deviations increase very slightly when the mocks go through the weighting process. We determine the systematic scatter as $S^2_{sys} = S^2_{iii}-S^2_{iv}$ and estimate an uncertainty via a jackknife-like method; we omit blocks of 20 mocks and recalculate $S_{\rm sys}$. The uncertainty on $S$ is then $\sigma^2_{S} = \frac{29}{30}\sum_i (S_{{\rm sys},i}-S_{\rm sys,full} )^2$, with $i$ denoting the sample with 20 mock results removed. We find $S_{sys} = 0.005\pm0.005$ for $\alpha_{||}$ and $S_{sys} = 0.005\pm0.002$ for $\alpha_{\perp}$. The increase in the variance is thus significant for $\alpha_{\perp}$.

The variance on the recovered BAO positions is slightly larger when the mocks have the stellar density systematic applied and corrected for, compared to the case where a uniform sub-sampling has been applied. This not surprising, as the correction procedure has essentially removed the clustering modes that align with stellar density (c.f. \citealt{Elsner15}). The application of the weights has a larger (relative) effect on the $\alpha_{\perp}$ measurements; this is consistent with the fact that the weighting procedure should largely remove transverse modes that correlate with the distribution of stars in the Galaxy. The results from our mocks tests suggest that uncertainties on $\alpha_{\perp}$ using the CMASS data will be under-estimated by 2 per cent ($\sqrt{0.025^2+0.005^2}/0.025-1$) and that uncertainties on $\alpha_{||}$ by half a per cent ($\sqrt{0.05^2+0.005^2}/0.05-1$). Based on the mode-removal argument, we expect the percentages to stay constant with signal-to-noise (e.g., for post-reconstruction results)\footnote{We have focused the mock tests on pre-reconstruction results due to the computational demands of analyzing the post-reconstruction samples}.

 As demonstrated in the Appendix of \cite{Ross12}, the correction procedure we apply for observational systematics is expected to produce slightly biased clustering measurements.\footnote{See \cite{Elsner15} for analytic descriptions of similar effects in spherical harmonic space.} We test this by comparing the mean $\xi_{0,2}$ for each of the mock cases and we plot the results in Fig. \ref{fig:xi0mocksys}. We find the correction procedure produces a nearly indistinguishable change in the mean $\xi_{0,2}$ when compared to the fiducial case (squares); clearly any bias is negligible in comparison to the statistical uncertainty (denoted by the grey shaded regions). In contrast, the mean effect of simulating the stellar density systematic is of clear significance to $\xi_0$ but exhibits a difference that is well within the statistical uncertainty for $\xi_2$ (see the diamonds in Fig. \ref{fig:xi0mocksys}). This is similar to the difference between the clustering observed in the CMASS data with and without corrective weights for the stellar density systematic (the triangles in the upper two panels of Fig. \ref{fig:xi0sys}).

The conclusion of this subsection is that, as best we can measure, observational systematics impart no bias on BOSS BAO measurements. However, we do find that the known observational systematics slightly reduce the statistical power of the measurements, implying our uncertainties on $\alpha_{\perp}$ are under-estimated by 2 per cent and those on $\alpha_{||}$ by 0.5 per cent. We apply these additional errors to our final results as systematic uncertainties.

\subsection{Robustness of BOSS data}

\begin{table*}
\begin{minipage}{7in}
\centering
\caption{Statistics of anisotropic BAO fits obtained from two sets 1000 mocks, for each of the combined sample redshift bins. `MD-P' denotes the Multidark PATCHY mocks were used. $S$ denotes standard deviation and $\sigma$ the uncertainty recovered from a likelihood.  $\Delta$ represents the difference from the expected values (given in Table 1). }
\begin{tabular}{cccccccccccccccccccccc}
\hline
\hline
$z$ bin & $\Delta\langle\alpha_{||}\rangle$ & $S_{||}$ &$\langle\sigma_{||}\rangle$& $\Delta\langle\alpha_{\perp}\rangle$ & $S_{\perp}$ &$\langle\sigma_{\perp}\rangle$ &    $\Delta\langle\alpha\rangle$ & $S_{\alpha}$ &$\Delta\langle\epsilon\rangle$ & $S_{\epsilon}$ &$\langle \chi^2\rangle$/dof \\
\hline
pre-reconstruction:\\
{\bf QPM} & & & & & & & & & \\
$0.2 < z < 0.5$  & 0.003 & 0.048 & 0.049 & 0.005 & 0.025 & 0.027 & 0.004 & 0.018 & -0.001 & 0.024 & 29.4/30 \\
$0.4 < z < 0.6$ &  0.001 & 0.045 & 0.045 & 0.007 & 0.023 & 0.025 & 0.005 & 0.015 & -0.002 & 0.021 & 29.3/30 \\
$0.5 < z < 0.75$& -0.002 & 0.042 & 0.043 & 0.007 & 0.023 & 0.025 & 0.004 & 0.015 & -0.003 & 0.020 & 29.3/30 \\
{\bf MD-P} & & & & & & & & & & & &\\
$0.2 < z < 0.5$  & 0.001 & 0.057 & 0.057 & 0.008 & 0.031 & 0.032 & 0.005 & 0.021 & -0.002 & 0.025 & 29.4/30 \\
$0.4 < z < 0.6$ &  0.004 & 0.056 & 0.053 & 0.008 & 0.028 & 0.028 & 0.005 & 0.018 & -0.001 & 0.025 & 29.3/30 \\
$0.5 < z < 0.75$& -0.001 & 0.052 & 0.050 & 0.010 & 0.029 & 0.028 & 0.006 & 0.018 & -0.004 & 0.024 & 29.3/30 \\
\hline
post-reconstruction:\\
{\bf QPM} &  & & &  & & & & & \\
$0.2 < z < 0.5$  & 0.002 & 0.030 & 0.031 & 0.003 & 0.017 & 0.017 & 0.0024 & 0.0113 & -0.0003 & 0.0138 & 29.4/30 \\
$0.4 < z < 0.6$ &  0.003 & 0.027 & 0.029 & 0.001 & 0.015 & 0.016 & 0.0016 & 0.0105 & 0.0005 & 0.0125 & 29.7/30 \\
$0.5 < z < 0.75$& 0.002 & 0.029 & 0.031 & 0.002 & 0.016 & 0.017 & 0.0013 & 0.0112 & -0.0001 & 0.0130 & 29.7/30 \\
{\bf MD-P} &  & & & & & & & & \\
$0.2 < z < 0.5$  & 0.002 & 0.034 & 0.035 & -0.001 & 0.019 & 0.020 & 0.0002 & 0.0128 & 0.0009 & 0.0152 & 29.3/30 \\
$0.4 < z < 0.6$ &  0.004 & 0.031 & 0.032 & 0.001 & 0.017 & 0.017 & 0.0014 & 0.0114 & 0.0011 & 0.0140 & 29.3/30 \\
$0.5 < z < 0.75$& 0.000 & 0.031 & 0.033 & 0.002 & 0.018 & 0.019 & 0.0015 & 0.0118 & -0.0008 & 0.0145 & 29.4/30 \\
\hline
\label{tab:baoresultsmockcomb}
\end{tabular}
\end{minipage}
\end{table*}

The results of the previous section (\ref{sec:mocksys}) imply that the stellar density systematic, the most dominant systematic (in terms of greatest potential significance), has, at most, a minor effect on the resulting BAO measurements. In this section, we apply similar tests to the BOSS data, and expand them to consider all of the weights applied to BOSS galaxies that are meant to provide the correct selection function. We also compare the results from the NGC and SGC regions separately. All of the measurements in this section use the covariance matrix constructed from 1000 QPM mocks. The results are summarized in Table \ref{tab:baoresultspr} and we discuss them below.

The pre-reconstruction CMASS results are shown in the top rows of Table \ref{tab:baoresultspr}. We measure both isotropic and anisotropic BAO. Moving down by row, we add weights to the galaxy catalog (the $n(z)$ is re-created for each case). The results are stable; the biggest absolute difference is  0.007 in $\alpha_{||}$ between the cases where no weights are applied and the case where close-pair and redshift-failure weights are applied. The biggest difference in terms of fraction of the uncertainty is 0.25$\sigma$ in $\alpha_{\perp}$ between the cases the close-pair and redshift failure weights have been applied and all weights have been applied. These size changes are consistent with the scatter expected due to statistical fluctuations. For example, the level of scatter we find when applying stellar density weights in the previous section is 0.2$\sigma$ between the weighted and un-weighted data; i.e., the statistical results are consistent with the level to which we expect the weights to alter the relative importance of each given survey mode and thus cause small differences in the recovered measurements. The isotropic NGC/SGC measurements differ by 1.1$\sigma$, and therefore are consistent to this level, given they represent independent volumes. The combined result is slightly closer (by 0.004) to the NGC measurement than one would expect from Gaussian likelihoods. 

For LOWZ, pre-reconstruction, we only measure the isotropic BAO scale, due to the relatively low signal to noise.  Measurements of the isotropic BAO scale use only the monopole, $\xi_0$. The results are shown in the middle rows of Table \ref{tab:baoresultspr}. As expected, the application of close-pair or redshift failure weights has very little impact on the measurements (at most $0.04\sigma$). The difference between the NGC and SGC measurements is 1$\sigma$, but in the opposite direction as the difference found for CMASS. The combined LOWZ measurement is closer to the NGC measurement by 0.003 compared to what would be expected from Gaussian statistics. We find that the BAO measurements obtained from the LOWZE3 and LOWZE2 selections are very similar (within 0.1$\sigma$) to what we find for the nominal LOWZ sample. This agreement helps validate that the LOWZE3 and LOWZE2 samples are indeed faithful tracers of the BAO signal and that their unique areas should be added to the nominal LOWZ footprint in order to obtain the best BAO measurements using low redshift BOSS data.

Finally, we investigate the robustness of the post-reconstruction results, shown in the bottom panels of Table \ref{tab:baoresultspr}. We focus on the CMASS sample. The agreement remains quite good, but the differences are larger relative to the uncertainty than they were for the pre-reconstruction results. The biggest difference is 0.5$\sigma$ in $\alpha_{||}$, between the case where close-pair and redshift failure weights are applied and all weights are applied (with the change being shared equally between the addition of the stellar density weights and the seeing weights). A potential explanation is that there is more stochasticity in the reconstruction process; the weighted galaxy field is first used to determine the displacement field and then the weighted galaxy and random positions are displaced. This increases the chance of fluctuations in the resulting measurements. Given that the largest fluctuation we find is 0.5$\sigma$ out of 30 possible comparisons, we find no evidence for concern.

There is a 1.8$\sigma$ discrepancy between the post-reconstruction CMASS isotropic BAO measurement in the NGC and SGC. Such a discrepancy has been observed at similar significance in each BOSS data release. When decomposed, the discrepancy is largest in $\alpha_{||}$, where the difference is 1.3$\sigma$ (it is only $0.4\sigma$, and thus consistent, for $\alpha_{\perp}$). Despite the slight tension, the results recovered when combining the pair-counts of NGC and SGC samples match the expectation for Gaussian likelihoods one obtains when taking the weighted mean of the NGC and SGC results.

\section{Combined Sample BAO Measurements}
\label{sec:BAOres}
The previous subsection demonstrates that the BAO measurements are consistent between the components of BOSS, splitting by targeting algorithm, after correcting for effects due to technical issues in BOSS observations. Here, we present BAO measurements determined using the combined sample data, both for the mock and data samples. We use both the QPM and MD-P mocks, and the covariance matrix determined using them, to analyze this sample. In addition to the $0.2 < z < 0.5$ and $0.5 < z < 0.75$ redshift bins, we present results for a $0.4 < z < 0.6$ redshift bin, which we expect to be largely covariant with the two distinct redshift bins but to provide additional information when assessing the robustness of our results.

\subsection{Results from mock samples}
Table \ref{tab:baoresultsmockcomb} displays the results of our BAO fits to both sets of mock correlation functions. For the mean values, we indicate the difference from the expected value, given the cosmology used for the mocks and our fiducial cosmology. These expected values are given in Table \ref{tab:baoexp}.

All of the results are biased relative to the uncertainty on the ensembles of the 1000 mock realizations (one should divide the $S$ and $\sigma$ numbers by $\sqrt{1000}$ to obtain the uncertainty on the of average of results of 1000 mocks), but by a relatively small amount when compared to the uncertainty expected for one realization. For the pre-reconstruction results, some bias is expected due to mode-coupling from non-linear structure formation (c.f. \citealt{PadWhite09}). The bias we find is greatest in $\alpha_{\perp}$, where it is 0.006 for QPM and 0.009 for MD-P (averaged across the three redshift bins). These are 0.25$\sigma$ and 0.31$\sigma$. For $\alpha_{||}$, the bias is only 0.001 on average, making it $\ll 0.1\sigma$. The biases are of the order predicted by \cite{PadWhite09}. Studies (e.g., \citealt{BeutlerDR12RSD,SanchezDR12RSD}) that use the pre-reconstruction data to measure $f\sigma_8$, $\alpha_{||}$, and $\alpha_{\perp}$ employ modeling that takes the predicted shifts into account and are expected to obtain somewhat more accurate results for the pre-reconstruction data. We use the pre-reconstruction results primarily as a basis for comparison to the post-reconstruction results.

Post-reconstruction, as expected, the bias in $\alpha_{\perp}$ is decreased. Considering the mean results across the redshift bins, for QPM, it is 0.002 (i.e., $\sigma/8$) and for MD-P it is 0.001 (i.e., 0.06$\sigma$). For $\alpha_{||}$, it is 0.002 (i.e., $\sim 0.07 \sigma$) for both sets of mocks. In terms of $\alpha/\epsilon$, the biases are 0.16$\sigma$ for the QPM $\alpha$ and 0.08$\sigma$ for the MD-P $\alpha$, while for $\epsilon$ they are both $\ll 0.1\sigma$. The biases vary with redshift bin to a level that is significantly larger than the uncertainty on the ensemble average; for example, in MD-P the difference in $\alpha$ between the low and high redshift bins is 0.0014, while the expected 1$\sigma$ deviation is 0.0006; similarly the difference for $\epsilon$ is 0.0017 compared to an expected 1$\sigma$ deviation of 0.0007. For QPM, the differences are smaller. In terms of the expected deviations, they are $\sim 2\sigma$ for $\alpha$ and less than 1$\sigma$ for $\epsilon$ (though it is 1.5$\sigma$ comparing the $0.2 < z < 0.5$ and $0.4 < z < 0.6$ bins, which should be correlated). The biases thus appear specific to redshift bin, implying they are either related the creation of the mocks and any redshift evolution they include or choices in the reconstruction algorithms related to the expected evolution of the density field. Overall, any bias in our measured BAO parameters is less than 0.16$\sigma$ (using the expected uncertainty for a single realization) and should not impact our conclusions. See \cite{VargasDR12BAO} for further study of related issues.

In general, the uncertainties recovered from the MD-P mocks are larger than those of the QPM mocks. The differences are in the uncertainties are $\sim$ 10 per cent in $\alpha_{||}$ and are slightly larger ($<$ 13 per cent) in $\alpha_{\perp}$. The differences in the uncertainties are thus at a similar level to the biases we find in the recovered BAO parameters. These biases are absorbed by the theoretical systematic uncertainty budget derived in \cite{VargasDR12BAO} and applied in \citet{Acacia}.

\subsection{Results from data}

\begin{figure}
\includegraphics[width=84mm]{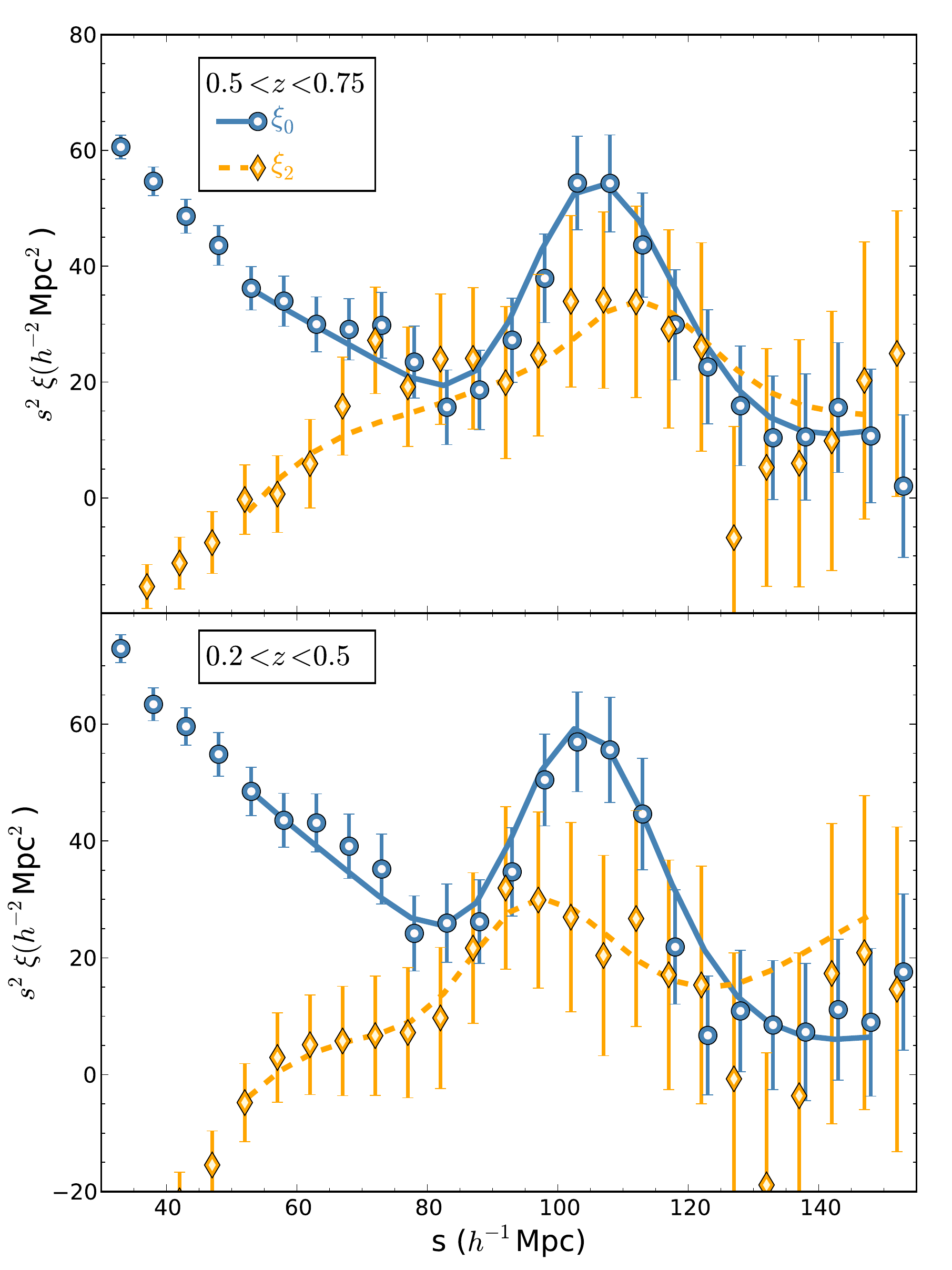}
  \caption{The measured post-reconstruction $\xi_0$ and $\xi_2$ and corresponding best-fit BAO models for BOSS galaxies. These best-fit models encode the BAO distance measurements determined in this work and are displayed for the range of scales that have been fit ($50< s < 150h^{-1}$Mpc).}
  \label{fig:xiBAOfit}
\end{figure}

\begin{table}
\centering
\caption{BAO fits on the BOSS combined sample data, using both the Multidark PATCHY (MD-P) and QPM covariance matrices.}
\begin{tabular}{cccc}
\hline
\hline
$z$ bin &  $\alpha_{||}$ & $\alpha_{\perp}$ & $\chi^2$/dof\\
\hline
pre-reconstruction: \\
{\bf QPM}  & & & \\
$0.2 < z < 0.5$  & 1.068$\pm$0.035 & 0.982$\pm$0.020 & 45/30\\
$0.4 < z < 0.6$  & 1.037$\pm$0.038 & 1.014$\pm$0.021 & 46/30\\
$0.5 < z < 0.75$&  0.963$\pm$0.035 & 0.999$\pm$0.024 & 30/30\\
{\bf MD-P}  & & & \\
$0.2 < z < 0.5$  & 1.051$\pm$0.036 & 0.983$\pm$0.022 & 37/30\\
$0.4 < z < 0.6$ & 1.024$\pm$0.042& 1.008$\pm$0.022 & 42/30\\
$0.5 < z < 0.75$&  0.953$\pm$0.034 & 1.001$\pm$0.024 & 28/30\\
\hline
post-reconstruction: \\
{\bf QPM}  &  & & \\
$0.2 < z < 0.5$  & 1.024$\pm$0.024 & 0.986$\pm$0.013 & 48/30\\
$0.4 < z < 0.6$ &  0.989$\pm$0.020 & 0.993$\pm$0.012 & 27/30\\
$0.5 < z < 0.75$& 0.962$\pm$0.024 & 0.991$\pm$0.015 & 33/30\\
{\bf MD-P}  & & & \\
$0.2 < z < 0.5$  & 1.025$\pm$0.027 & 0.988$\pm$0.015 & 39/30\\
$0.4 < z < 0.6$ & 0.986$\pm$0.024 & 0.994$\pm$0.014 & 23/30\\
$0.5 < z < 0.75$&  0.962$\pm$0.023 & 0.991$\pm$0.015 & 32/30\\
\hline
\label{tab:baoresultsdatacomb}
\end{tabular}
\end{table}

Results for BAO fits on BOSS data, using both the QPM and the MD-P covariance matrices, are displayed in Table \ref{tab:baoresultsdatacomb}. The results are similar using the two covariance matrices, but there are notable differences. In general, the uncertainties are smaller when the QPM covariance matrices are used, matching the results on the mocks. Correspondingly, the $\chi^2$ values are consistently higher for the QPM mocks (in five of the six cases to compare). None of the six QPM cases recover a $\chi^2$/dof that is less than 1, while this is the case for two of the MD-P cases. Considering the total $\chi^2$ for the two independent redshift bins, the $\chi^2$/dof for QPM is 75/60 pre-reconstruction and 81/60 post-reconstruction. This can be compared to 65/60 and 71/60 for MD-P. This is suggestive that the MD-P covariance matrix is doing the better job of characterizing the noise in the BOSS combined sample $\xi_{0,2}$ measurements. 

Pre-reconstruction, the $\alpha_{||}$ results are consistently greater for the QPM covariance matrix compared to the MD-P covariance matrix. The difference varies between 0.017 and 0.010 and is a 0.5$\sigma$ shift in the most extreme case (the $0.2 < z < 0.5$ redshift bin); given the same data is used and only the covariance matrix is altered this is a fairly large change. The differences are much smaller for $\alpha_{\perp}$, where it is at most 0.006 (0.3$\sigma$) in the $0.4 < z < 0.6$ redshift bin.

Post-reconstruction, the BAO measurements are robust to the choice of covariance matrix. The biggest difference is 0.003 (0.15$\sigma$) in $\alpha_{||}$ for the data in the $0.4 < z < 0.6$ redshift bin; the difference in the uncertainty between the results in this bin is the same. The level of agreement is consistent with the results found from the mock realizations and suggests that the choice of covariance matrix is not a major systematic uncertainty in our analysis. Given the slightly larger uncertainties for the data using the MD-P covariance matrix, we believe they represent the more conservative choice and are what we use for our final results. We use the MD-P results in all comparisons that follow unless otherwise noted. 

 Fig. \ref{fig:xiBAOfit} displays the measured post-reconstruction $\xi_{0,2}$ and the associated best-fit BAO model, using the MD-P covariance matrix. At each redshift, one can observe the strong BAO feature in the monopole, which has been enhanced by the reconstruction process, compared to previous plots. For the quadrupole, reconstruction removes most of the large-scale RSD effects and the overall amplitude is thus decreased. BAO features appear in the quadrupole to the right and left of the peak in the monopole. Such BAO features appear in the quadrupole when $\alpha_{||} \neq \alpha_{\perp}$ (and thus do not present themselves in the mocks as the two $\alpha$ parameters are expected to be nearly equal in our mock analysis). The feature appears to the right in the $0.5 < z < 0.75$ redshift bin, which yields a measurement of $\alpha_{||}$ that is lower than $\alpha_{\perp}$; the reverse is true for the $0.2 < z < 0.5$ bin. See \cite{Acacia} for further exploration and visualization of these features in the same data.

\begin{table*}
\begin{minipage}{7in}
\centering
\caption{Post-reconstruction combined sample 2D BAO fits, obtained using the covariance matrix constructed from the MultiDark-PATCHY mock catalogs. The `combined+sys' results represent the likelihoods that are used in Alam et al. (2016) and are the average of the results listed as a function of bin-center. }
\begin{tabular}{lccccccc}
\hline
\hline
sample & bin center shift & $\alpha_{||}$ & $\alpha_{\perp}$ & $r$ & $\chi^2$/dof & $\alpha$ &$\chi^2$/dof \\
\hline
{\bf $0.2 < z < 0.5$:}\\
post-recon & {\bf combined +sys} & 1.022$\pm$0.027$\pm$0.003 & 0.988$\pm$0.015$\pm$0.003 & - & - \\
& combined & 1.022$\pm$0.027 & 0.988$\pm$0.015 & -0.39 & 42/30\\
& 0 $h^{-1}$Mpc  & 1.025$\pm$0.027 & 0.988$\pm$0.015 & -0.39 & 39/30 & 0.998$\pm$0.010 & 25/15\\
& 1 $h^{-1}$Mpc  & 1.017$\pm$0.027 & 0.992$\pm$0.015 & -0.39 & 35/30 & 1.000$\pm$0.010 & 20/15\\
& 2 $h^{-1}$Mpc  & 1.022$\pm$0.028 & 0.990$\pm$0.015 & -0.39 & 40/30 & 0.999$\pm$0.010 & 19/15\\
& 3 $h^{-1}$Mpc  & 1.024$\pm$0.028 & 0.985$\pm$0.015 & -0.40 & 51/30 & 1.000$\pm$0.010 & 28/15\\
& 4 $h^{-1}$Mpc  & 1.023$\pm$0.026 & 0.986$\pm$0.015 & -0.40 & 44/30 & 1.000$\pm$0.010 & 26/15\\
pre-recon & 0 $h^{-1}$Mpc  & 1.051$\pm$0.037 & 0.983$\pm$0.022 & -0.37 & 37/30 & 1.004$\pm$0.015 & 18/15\\
\hline
{\bf $0.4 < z < 0.6$:}\\
post-recon & {\bf combined +sys} & 0.984$\pm$0.023$\pm$0.002 & 0.994$\pm$0.014$\pm$0.003 & - & - \\
& combined & 0.984$\pm$0.023 & 0.994$\pm$0.014 & -0.39 & 30/30\\
& 0 $h^{-1}$Mpc  & 0.986$\pm$0.024 & 0.994$\pm$0.014 & -0.39 & 23/30 & 0.991$\pm$0.009 & 16/15\\
& 1 $h^{-1}$Mpc  & 0.981$\pm$0.022 & 0.996$\pm$0.014 & -0.39 & 22/30 & 0.992$\pm$0.009 & 14/15\\
& 2 $h^{-1}$Mpc  & 0.981$\pm$0.023 & 0.996$\pm$0.015 & -0.39 & 37/30 & 0.993$\pm$0.009 & 19/15\\
& 3 $h^{-1}$Mpc  & 0.988$\pm$0.023 & 0.994$\pm$0.014 & -0.39 & 38/30 & 0.993$\pm$0.009 & 24/15\\
& 4 $h^{-1}$Mpc  & 0.987$\pm$0.024 & 0.992$\pm$0.014 & -0.40 & 29/30 & 0.992$\pm$0.009 & 18/15\\
pre-recon & 0 $h^{-1}$Mpc  & 1.024$\pm$0.042 & 1.008$\pm$0.022 & -0.49 & 42/30 & 1.012$\pm$0.015 & 22/15\\
\hline
{\bf $0.5 < z < 0.75$:}\\
post-recon & {\bf combined +sys} & 0.958$\pm$0.023$\pm$0.002 & 0.995$\pm$0.016$\pm$0.003 & - & - \\
& combined & 0.958$\pm$0.023 & 0.995$\pm$0.016 & -0.41 & 32/30\\
& 0 $h^{-1}$Mpc  & 0.962$\pm$0.023 & 0.991$\pm$0.015  & -0.42 & 32/30 & 0.981$\pm$0.010 & 14/15\\
& 1 $h^{-1}$Mpc  & 0.957$\pm$0.023 & 0.999$\pm$0.016 & -0.42 & 26/30 & 0.982$\pm$0.010 & 13/15\\ 
& 2 $h^{-1}$Mpc  & 0.957$\pm$0.023 & 0.994$\pm$0.016 & -0.41 & 34/30 & 0.982$\pm$0.010 & 18/15\\
& 3 $h^{-1}$Mpc  & 0.954$\pm$0.024 & 0.996$\pm$0.015 & -0.41 & 40/30 & 0.983$\pm$0.010 & 18/15\\
& 4 $h^{-1}$Mpc  & 0.957$\pm$0.024 & 0.994$\pm$0.015 & -0.41 & 29/30 & 0.982$\pm$0.010 & 14/15\\
pre-recon & 0 $h^{-1}$Mpc  & 0.953$\pm$0.035 & 1.001$\pm$0.024 & -0.49 & 28/30 & 0.984$\pm$0.015 & 14/15\\
\hline
\label{tab:bincenter}
\end{tabular}
\end{minipage}
\end{table*}

\begin{figure}
\includegraphics[width=84mm]{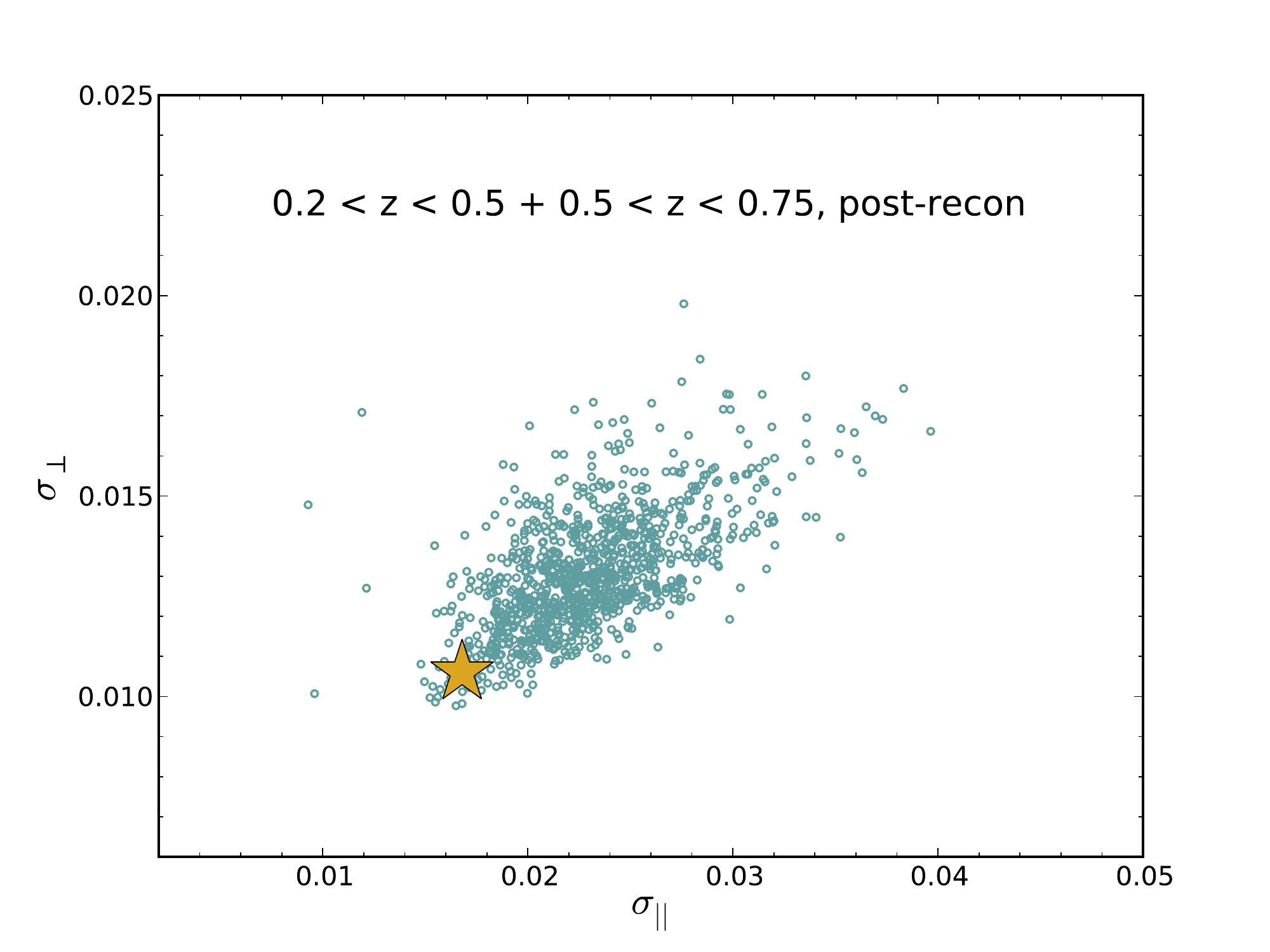}
  \caption{The uncertainty in $\alpha_{||}$ compared to the uncertainty in $\alpha_{\perp}$ for each MultiDark-PATCHY mock realization (open  `cadetblue' circles) and the DR12 data (large goldenrod star). We have combined the data in the $0.2 < z < 0.5$ and $0.5 < z < 0.75$ redshift bins, assuming Gaussian likelihoods. The DR12 uncertainties are on the low side, but are within the locus of points representing the mock realizations. 
  }
  \label{fig:errcompmocks}
\end{figure}

The uncertainties we obtain are significantly smaller than the mean uncertainties recovered from the mock realizations, by $\sim$ 25 per cent in each redshift bin. {This implies more pronounced BAO features in the data than are present in the typical mock.} In order to determine how unusual this is, we combine the results from the $0.2 < z < 0.5$ and $0.5 < z < 0.75$ redshift bins, as they are independent and the expected $\alpha$ values are nearly identical. Fig. \ref{fig:errcompmocks} displays the uncertainty in $\alpha_{\perp}$ ($\sigma_{\perp}$) vs. the uncertainty in $\alpha_{||}$ ($\sigma_{||}$) recovered for each mock realization when combining the results of the two redshift bins (blue circles) and the DR12 data (orange star). One can see that the DR12 result is within the locus of points, but at the lower edge. We can quantify the results further by comparing the area of the 1$\sigma$ confidence region in the data to the ensemble of mocks. We find 45 mocks ($\sim$5 per cent), when once more combining the results of the $0.2 < z < 0.5$ and $0.5 < z < 0.75$ redshift bins, have a smaller area contained in their 1$\sigma$ confidence region than we find for the data. Thus, we determine that we have been somewhat lucky in the region of the Universe we have observed with BOSS, but not grossly so. In this sense, these results are similar to those obtained with the previous data set \citep{alph}. To some degree, the fact that we find better results than the majority of the mock realizations is due to the fact that the grid-scales involved in the creation of the mocks effectively increase the damping of the BAO signal. This is discussed further in \cite{BeutlerDR12BAO}.

In order to produce our final measurements, we combine results across five choices of bin center, each separated by 1$h^{-1}$Mpc. This is similar to what was done in \cite{alph}. However, given our fiducial bin size is now 5$h^{-1}$Mpc (compared to 8$h^{-1}$Mpc), the variance between the results in each bin center is smaller and to obtain the combined results we simply average the likelihood surfaces for each bin center (rather attempt to determine the optimal combination with a slightly improved uncertainty, as was done for the isotropic results in \citealt{alph}).

The results for each bin center choice are presented in Table \ref{tab:bincenter}. The results from averaging each likelihood are labeled `combined'. The difference between the combined results and the fiducial bin center choice (0 $h^{-1}$Mpc) is at most 0.004 in $\alpha_{\perp}$ (0.25$\sigma$) for the $0.5 < z < 0.75$ redshift bin.

We add an observational systematic uncertainty to the combined result to obtain our final results, quoted as {\bf `combined+ sys'} in Table \ref{tab:bincenter}. Our tests on the mock samples do not suggest any systematic bias is imparted into the measurements due to observational systematic effects. However, we do find that the procedure we apply to correct for a systematic dependence with stellar density removes a small amount of the BAO information from the survey volume. The mocks we used to determine the covariance used for our BAO results do not include this small reduction in information. Thus, to account for this we add to the results a systematic uncertainty. In Section \ref{sec:mocksys}, the weighting process was found to impart a 2 per cent dilation into the standard deviation on $\alpha_{\perp}$ and a 0.5 per cent dilation on $\alpha_{||}$. We decompose these dilations into individual systematic uncertainties, so that they can be combined with any other systematic uncertainties. For the given dilations, these are $0.1\sigma_{\rm stat}$ for $\alpha_{||}$ and $0.2\sigma_{\rm stat}$ for $\alpha_{\perp}$ (e.g., solving $1.02^2\sigma_{\rm stat}^2 = \sigma_{\rm stat}^2 + \sigma_{\rm sys}^2$). This systematic uncertainty is added in a similar manner to all of the BAO distance measurements that are used to obtain cosmological constraints in \cite{Acacia}. We emphasize that these systematic uncertainties are purely observational; \cite{Acacia} presents a full accounting of potential systematic uncertainties affecting BOSS BAO measurements, incorporating theoretical systematic uncertainties (e.g., those relating to the methodology used for BAO fits and to construct the covariance matrix) that are estimated in \cite{VargasDR12BAO}.

Our final measurements determine the radial distance scale to than 2.7 per cent precision (or better) and the transverse distance to 1.6 precision (or better) in each redshift bin. If we consider the two independent redshift bins, we can add the inverse variance on each $\alpha$ parameter to determine an effective combined precision. This yields 1.8 and 1.1 per cent for the radial and transverse distance scales. These measurements are further improved in \cite{Acacia}, where results from the middle redshift bin, power spectrum BAO, and full-shape measurements are optimally combined.

Additional robustness checks are presented in Appendix \ref{app:rob}, where we find no significant concerns.

\section{Discussion}
\label{sec:disc}

\begin{figure}
\includegraphics[width=84mm]{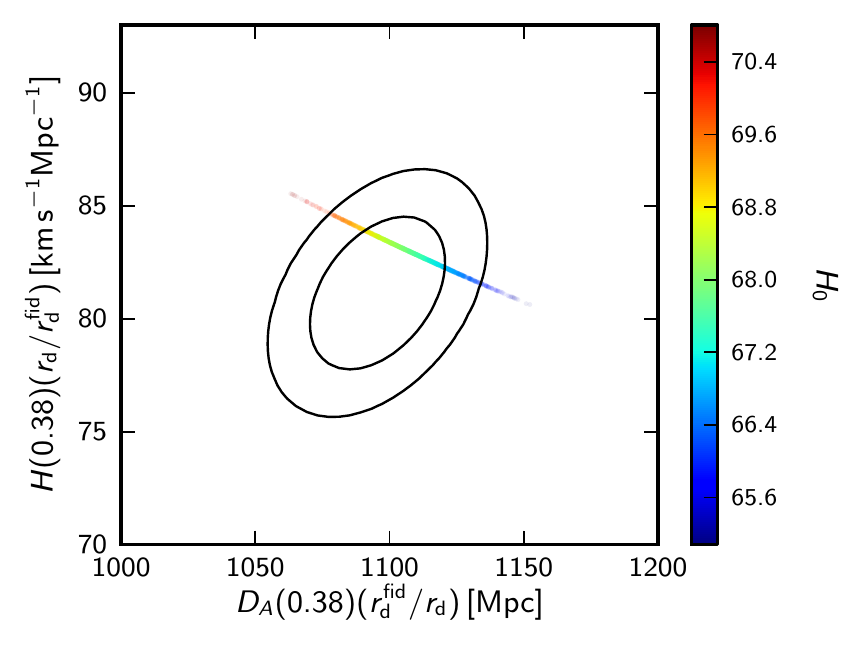}\vspace*{-0.3em}
\includegraphics[width=84mm]{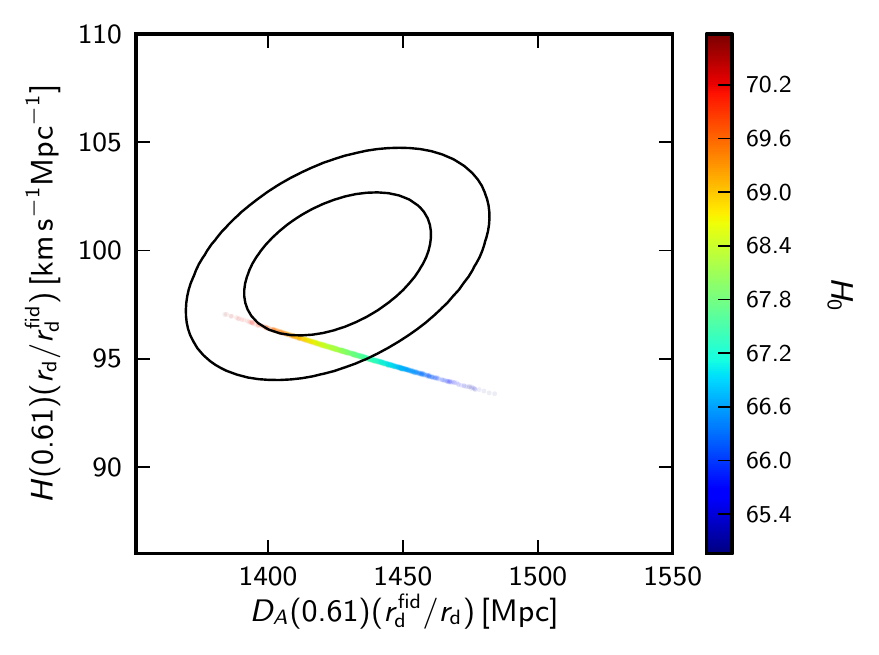}
  \caption{The allowed 1 and 2$\sigma$ regions (black ellipses) in the Hubble parameter, $H$, and the angular diameter distance, $D_A$, determined from our post-reconstruction anisotropic BAO scale measurements using BOSS galaxies with $0.2 < z < 0.5$ (top panel) and with $0.5 < z < 0.75$ (bottom panel). The colored points represent the 2$\sigma$ allowed region when assuming a flat $\Lambda$CDM cosmology and the the Planck 2015 results, with different colors representing the value of $H$ at $z=0$ (as indicated by the color bar on the right).
  }
  \label{fig:DAH}
\end{figure}

\subsection{Comparison to other DR12 BAO measurements}
The final output of this work is the BAO measurements using the post-reconstruction, anisotropic correlation function measurements of the BOSS DR12 galaxy sample in redshift bins $0.2 < z < 0.5$, $0.4 < z < 0.6$, and $0.5 < z < 0.75$. Other studies have made similar measurements using DR12 data. \cite{CuestaDR12} obtained BAO measurements using the post-reconstruction anisotropic correlation function of the DR12 CMASS and LOWZ samples. In our robustness checks, we made the same measurements for the CMASS sample. Accounting for the difference in the fiducial cosmologies assumed by each analysis, the differences between \cite{CuestaDR12} and ours are 0.018 for $\alpha_{||}$ and -0.011 for $\alpha_{\perp}$. However, once we adjust to use the same bin size (8$h^{-1}$Mpc) as \cite{CuestaDR12}, the differences reduce to 0.011 for $\alpha_{||}$ and -0.004 for $\alpha_{\perp}$. Each of these represent a difference of less than 0.5$\sigma$ and are likely due to small methodological differences in the BAO fitting. We find smaller uncertainties on $\alpha_{||}$ (for both the data and the mocks) due to these differences.

Both \cite{BeutlerDR12BAO} and \cite{VargasDR12BAO} obtain BAO measurements for the same post-reconstruction data set and redshift bins as we use. \cite{BeutlerDR12BAO} is a Fourier space analysis. Analyzing the same set of mocks, we find our results are correlated with a factor 0.9 and that the differences we obtain on the BOSS data are consistent with this high level of correlation. Both recover nearly identical uncertainties on the anisotropic BAO parameters, for both the data and the mock samples. \cite{VargasDR12BAO} uses the same configuration space data as presented in this study, but apply slightly different methodology to obtain their BAO measurements; they recover results that are consistent with ours. A more detailed comparison of these results is presented in \cite{Acacia}, where consensus sets of BOSS DR12 BAO and BOSS DR12 BAO + RSD measurements, combined as described in \cite{SanchezDR12comb}, are presented. 

\subsection{Comparison with $\Lambda$CDM}
Our measurements of $\alpha_{||}$ and $\alpha_{\perp}$ can be translated into constraints on $D_A(z)(r_{\rm d}^{\rm fid}/r_{\rm d})$ and $H(z)(r_{\rm d}/r_{\rm d}^{\rm fid})$ and thereby test cosmological models. Here, we simply compare our measurements with the allowed parameter space in $\Lambda$CDM as determined by \cite{Planck2015}\footnote{Specifically, the results from the 
'base\_plikHM\_TT\_lowTEB\_lensing' chains.}. This is show in Fig. \ref{fig:DAH} for the $0.2 < z < 0.5$ and $0.5 < z < 0.75$ redshift bins. Our low redshift result is fully consistent with the Planck $\Lambda$CDM prediction. Our high redshift result is in slight tension, as the 1$\sigma$ contours just barely overlap; this is mostly driven by the $H(z)$ measurement. This is similar to what was found in \cite{alph} for the DR11 CMASS data; the agreement is slightly better in \cite{BeutlerDR12BAO} and significantly better (to the level there is no tension) when these two post reconstruction results are optimally combined with pre-reconstruction full-shape results in \cite{Acacia}. Our results for the $0.4 < z < 0.6$ redshift slice (not plotted) are consistent with the Planck $\Lambda$CDM prediction, as one would predict based on the mean of the $0.2 < z < 0.5$ and $0.5 < z < 0.75$ results. The full cosmological context of our measurements, when combined with other BOSS DR12 results, is explored in detail in \cite{Acacia}. 

\section{Summary}
In this work, we have \begin{itemize}
\item Described and motivated the construction of the selection function for BOSS galaxies;
\item Shown how the treatment of the selection function affects the measured clustering;
\item Shown that the individual BOSS target samples can be trivially combined into one BOSS sample, allowing arbitrary splitting in redshift;
\item Demonstrated that BOSS BAO measurements are robust to the treatment of the selection function and the details of how the BOSS samples are combined;
\item Measured the BAO scale transverse to and along the line of sight from the BOSS galaxy correlation function in two independent redshift slices, $0.2 < z < 0.5$ and $0.5 < z < 0.75$, and one overlapping redshift slice. $0.4 < z < 0.6$. \end{itemize} The results of our work on the selection function are included in the BOSS galaxy catalogs described in \cite{Reid15}. The results of our BAO scale measurements are used in \cite{Acacia}, where they are combined with other BOSS DR12 results and used to evaluate cosmological models.

The main, non-standard, components to the BOSS selection function are the weights that we apply to account for fluctuations in the angular selection function. The angular selection function has been demonstrated to depend on the stellar density and the seeing conditions of the BOSS imaging data that targets are selected from. The weights we have defined correct for these variations in the selection function.

We have assessed the impact of these weights by comparing the clustering of BOSS samples with and without the weights. The stellar density weights have by far the greatest impact. The impact can be quantified by determining the $\chi^2$ difference between the two measurements (using a model that assumes the difference is zero); for the stellar density weights it was 13.1, implying the possibility of parameter estimation being biased by 3.6$\sigma$ when not accounting for the effect of stellar density on the angular selection function. However, we find both for mocks and for the data that BAO measurements are robust to whether or not any weights are included to account for the fluctuations in the selection function. We conclude that our treatment of the BOSS selection function imparts no bias into the resulting BAO measurements.

We note that our conclusions on the lack of any bias are specific to BAO measurements. We recommend that any other kind of measurement conduct a similar analysis as presented here, in order to assess any potential of systematic bias. At the least, we suggest that any configuration space analysis includes a constant term with a free amplitude to be marginalized over (like there is in the BAO model). An analysis demonstrating the robustness of structure growth measurements determined by modeling RSD under such treatment is presented in Appendix \ref{sec:rsdrobust}. Our analysis does not attempt to assess the size of possible fluctuations due to calibration uncertainties, like discussed in \cite{Huterer13}, which would need to be accounted for in any analysis where broad-band large-scale power is important (e.g., primordial non-Gaussianity).

While the location of the measured BAO position is robust to the treatment of the selection function, our treatment does add a small degree of statistical uncertainty that is not accounted for in our covariance matrices. The reason is that our methods essentially null clustering modes that are aligned with fluctuations in stellar density. A small fraction of these modes contain BAO information. We find that when approximating our procedure for correcting for the stellar density systematic the standard deviation of mock samples increases by 2 per cent for the transverse BAO measurement and 0.5 per cent for the radial BAO measurement.  In terms of the statistical uncertainty, these are 0.14$\sigma_{\rm stat}$ and 0.07$\sigma_{\rm stat}$, respectively. 

 Fundamentally, the robustness of BAO measurements is due to the fact that the BAO are a localized feature in configuration space and it is difficult for any observational feature to have such a localized effect, especially when angular and radial components are combined. Indeed, it was noted in the review of \cite{WeinbergDERev} that this nature of BAO studies makes it an especially robust probe of the expansion history of the Universe. The work we have presented shows this to be true in detail. Our results suggest this will remain fact for the next generation of BAO experiments.

\section*{acknowledgements}
AJR is grateful for support from the Ohio State University Center for Cosmology and ParticlePhysics.
Nearly all heavy computer processing made use of the facilities and staff of the UK Sciama High Performance Computing cluster supported by the ICG, SEPNet and the University of Portsmouth.
Colors made possible by 
\url{http://matplotlib.org/examples/color/named_colors.html}; figures made colorblind-friendly (hopefully) by use of Color Oracle software\\
\noindent C. C. acknowledges support as a MultiDark Fellow and from the Spanish MICINNs Consolider-Ingenio 2010 Programme under grant MultiDark CSD2009-00064, MINECO Centro de Excelencia Severo Ochoa Programme under grant SEV-2012-0249, and grant AYA2014-60641-C2-1-P\\ 
\noindent MPI acknowledges support from MINECO under the grant AYA2012-39702-C02-01.\\
\noindent Hee-Jong Seo's work is supported by the U.S. Department of Energy, Office of Science, Office of High Energy Physics under Award Number DE-SC0014329.\\
MV is partially supported by Programa de Apoyo a Proyectos de Investigaci\'on e Innovaci\'on Tecnol\'ogica (PAPITT) No IA102516 and Proyecto Conacyt Fronteras No 281\\

Funding for SDSS-III has been provided by the Alfred P. Sloan
Foundation, the Participating Institutions, the National Science
Foundation, and the U.S. Department of Energy Office of Science.
The SDSS-III web site is http://www.sdss3.org/.

SDSS-III is managed by the Astrophysical Research Consortium for the
Participating Institutions of the SDSS-III Collaboration including the
University of Arizona,
the Brazilian Participation Group,
Brookhaven National Laboratory,
Cambridge University ,
Carnegie Mellon University,
Case Western University,
University of Florida,
Fermilab,
the French Participation Group,
the German Participation Group,
Harvard University,
UC Irvine,
Instituto de Astrofisica de Andalucia,
Instituto de Astrofisica de Canarias,
Institucio Catalana de Recerca y Estudis Avancat, Barcelona,
Instituto de Fisica Corpuscular,
the Michigan State/Notre Dame/JINA Participation Group,
Johns Hopkins University,
Korean Institute for Advanced Study,
Lawrence Berkeley National Laboratory,
Max Planck Institute for Astrophysics,
Max Planck Institute for Extraterrestrial Physics,
New Mexico State University,
New York University,
Ohio State University,
Pennsylvania State University,
University of Pittsburgh,
University of Portsmouth,
Princeton University,
UC Santa Cruz,
the Spanish Participation Group,
Texas Christian University,
Trieste Astrophysical Observatory
University of Tokyo/IPMU,
University of Utah,
Vanderbilt University,
University of Virginia,
University of Washington,
University of Wisconsin
and Yale University.

\appendix

\section{Choosing a bin size and range of scales}
\label{app:binsize}
\begin{figure}
\includegraphics[width=84mm]{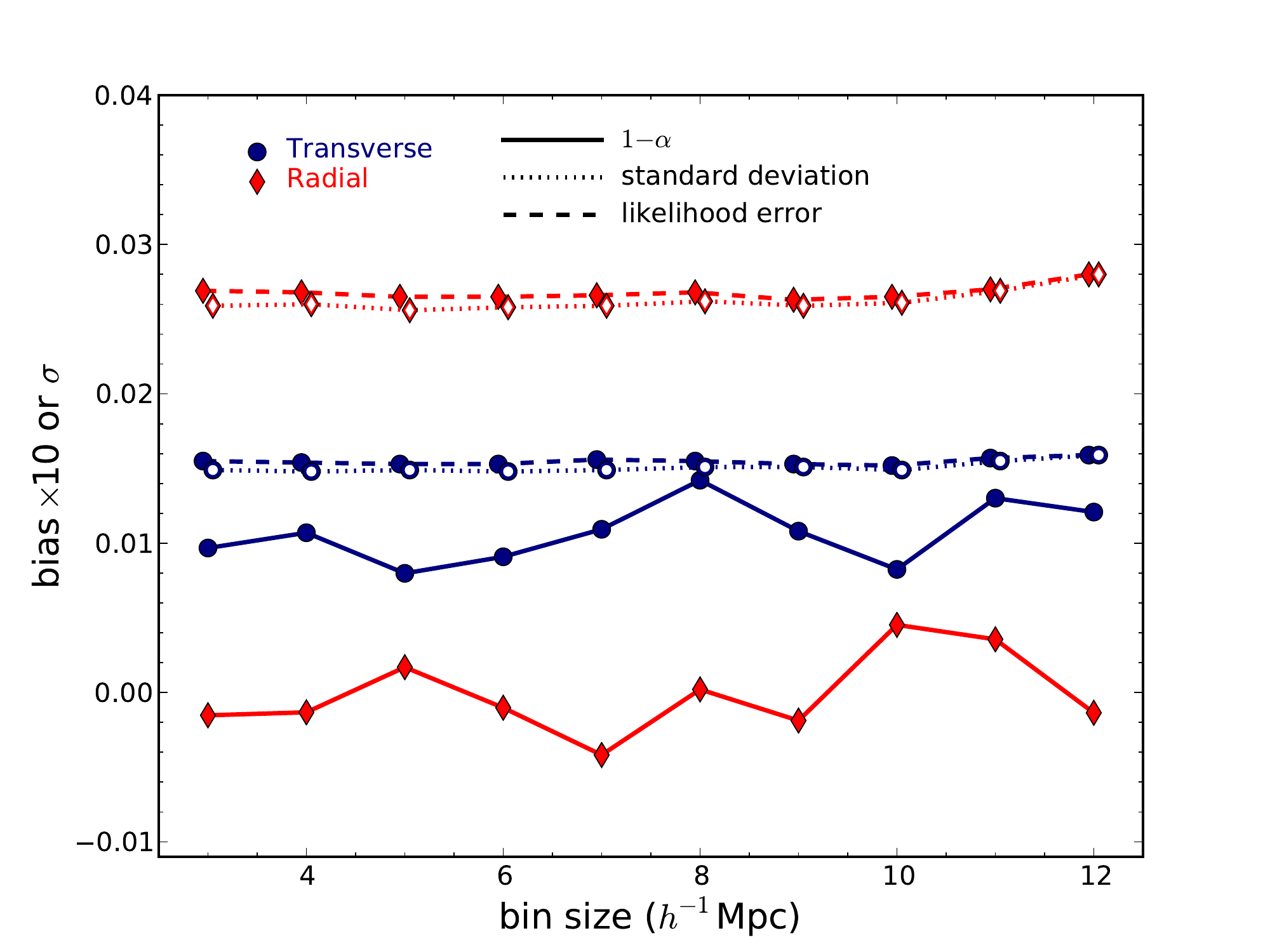}
  \caption{Statistics of 2D BAO fits on 1000 QPM CMASS post-reconstruction mocks, as a function of the bin size. Red diamonds show results for $\alpha_{\perp}$ and blue circles show the results for $\alpha_{||}$. The bias of the mean alpha, multiplied by 10, is shown with solid lines; one can see it is never greater that 0.1$\sigma$. The standard deviation of the mock results is shown with dotted lines (and open symbols) and the mean likelihood error with dashed lines.}
  \label{fig:baomockbinsize}
\end{figure}

In this appendix, we motivate the choices for the bin-size and range of scales used to obtain our BAO measurements. We thus present the results of BAO constraints obtained from the post-reconstruction CMASS sample as a function of the bin-size and the range of scales used in the analysis. All statistics are derived from the mean and variance of fits to $\alpha_{||},\alpha_{\perp}$ obtained from the QPM mocks. See \cite{VargasDR12BAO} for a more detailed study on similar tests.

We have tested the BAO constraints obtained from the post-reconstruction CMASS sample as a function of bin-size (holding the fitting range fixed to $50 < s < 150$). This is a repeat of the tests done in \cite{Per14}; naively, the results would only improve as the bin-size is decreased, but this decrease increases the size of the data vector and thus the noise in the inverse covariance matrix. The results are summarized by Fig. \ref{fig:baomockbinsize}. One can see that the trends are not strong, so any choice of bin size in the range $4-8h^{-1}$Mpc would be reasonable. The tests on the mocks suggest the correlations between results from different bin sizes are $\sim$0.95 for both $\alpha_{\perp}$ and $\alpha_{||}$. Based on these results, we choose to use a bin size of 5$h^{-1}$Mpc. Such a bin choice requires combining across less bin centers than was the case for BOSS DR11 analyses, which used a bin size of 8$h^{-1}$Mpc \citep{alph}.

\begin{figure}
\includegraphics[width=84mm]{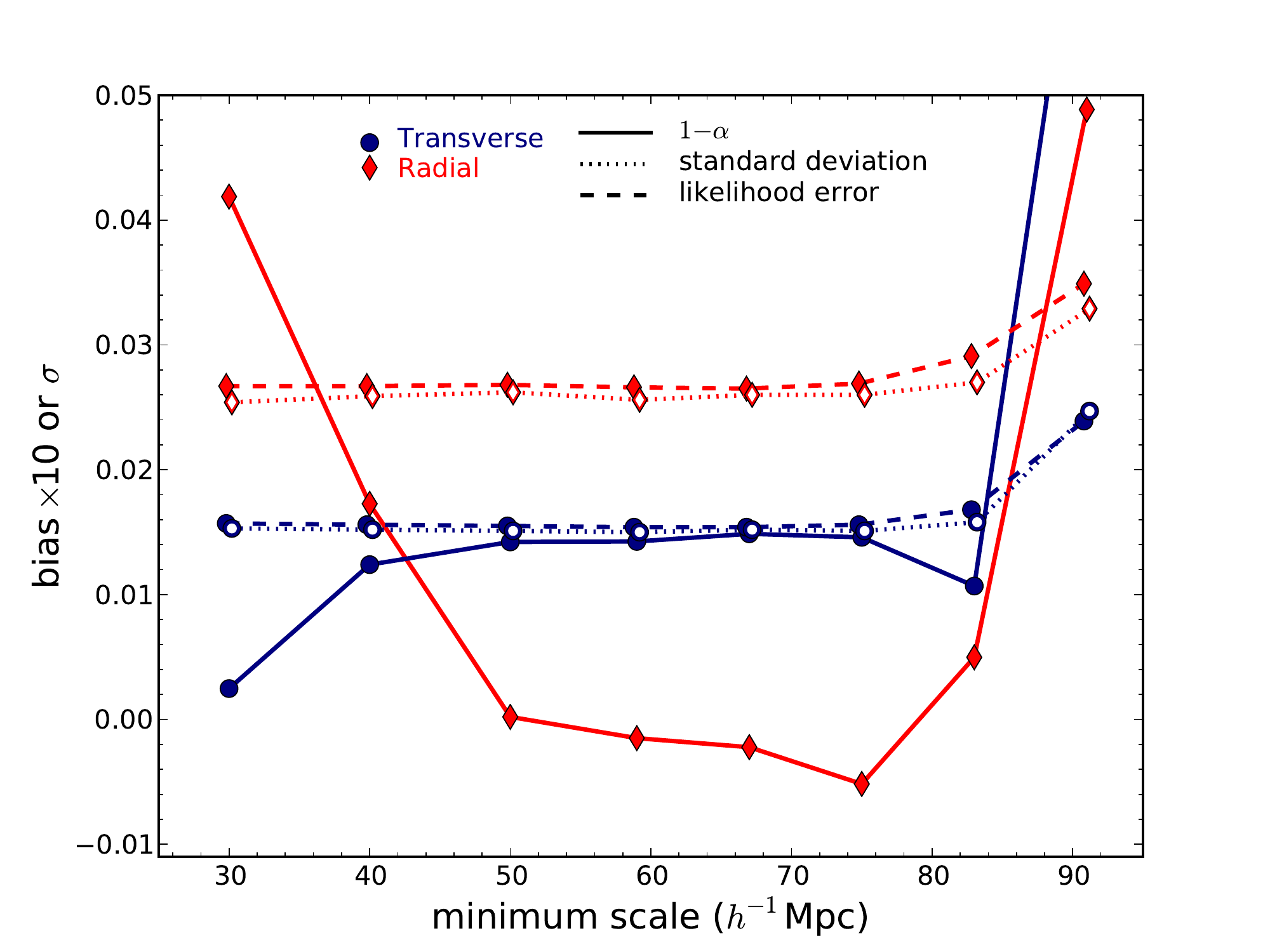}\vspace*{-0.3em}
\includegraphics[width=84mm]{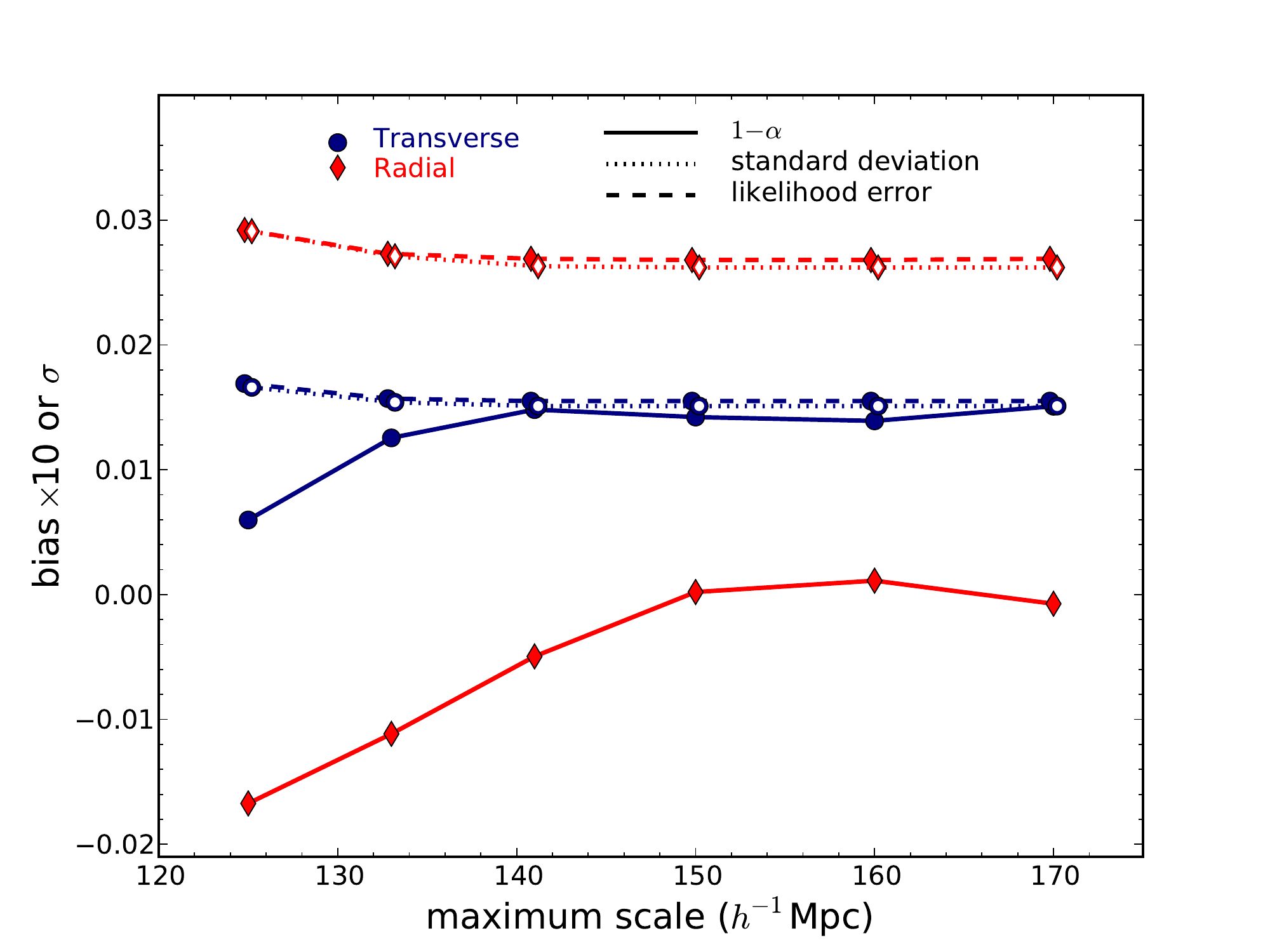}
  \caption{Statistics of 2D BAO fits on 1000 QPM CMASS post-reconstruction mocks, as a function of the minimum (top) and maximum (bottom) scale used.. Red diamonds show results for $\alpha_{\perp}$ and blue circles show the results for $\alpha_{||}$. The bias of the mean alpha, multiplied by 10, is shown with solid lines. The standard deviation of the mock results is shown with dotted lines (and open symbols) and the mean likelihood error with dashed lines.
  }
  \label{fig:scalerange}
\end{figure}

Similarly, we have tested the minimum and maximum scale used in the BAO fits. The results are summarized in Fig. \ref{fig:scalerange}. These results motivate our choice of using the range $50 < s < 150h^{-1}$Mpc. At scales $s < 50h^{-1}$Mpc, we do not recover un-biased measurements of $\alpha_{||}$. This is due to our ability to model the post-reconstruction quadrupole at such scales. A minimum scale $r > 70h^{-1}$Mpc causes a decrease in the statistical power of the measurements. Likewise, a maximum scale $r < 150h^{-1}$ increases both the statistical uncertainty and the bias of the results.

\section{Robustness tests on combined sample}
\label{app:rob}
\begin{table}
\centering
\caption{Post-reconstruction combined sample 2D BAO fits as a function of bin-size, choice of fitting range, choices for nuisance parameters, and Galactic hemisphere.}
\begin{tabular}{cccc}
\hline
\hline
 test & $\alpha_{||}$ & $\alpha_{\perp}$  & $\chi^2$/dof\\
\hline
{\bf $0.2 < z < 0.5$:}\\
bin size:\\
3 $h^{-1}$Mpc  & 1.022$\pm$0.028 & 0.987$\pm$0.015 & 51/54\\
4 $h^{-1}$Mpc  & 1.026$\pm$0.029 & 0.985$\pm$0.015 & 55/38\\
5 $h^{-1}$Mpc  & 1.025$\pm$0.027 & 0.988$\pm$0.015 & 39/30\\
6 $h^{-1}$Mpc  & 1.020$\pm$0.028 & 0.988$\pm$0.015 & 34/24\\
7 $h^{-1}$Mpc  & 1.026$\pm$0.028 & 0.985$\pm$0.015 & 26/18\\
8 $h^{-1}$Mpc  & 1.027$\pm$0.028 & 0.982$\pm$0.015 & 28/16\\
10 $h^{-1}$Mpc  & 1.027$\pm$0.029 & 0.986$\pm$0.016 & 14/10\\
$s > 70h^{-1}$Mpc & 1.023$\pm$0.026 & 0.990$\pm$0.014 & 32/26\\
$s < 170h^{-1}$Mpc & 1.027$\pm$0.027 & 0.987$\pm$0.015 & 44/34\\
$A_0 = 0$ & 1.029$\pm$0.028 & 0.986$\pm$0.015 & 46/33\\
$A_2 = 0$ & 1.021$\pm$0.028 & 0.991$\pm$0.015 & 43/33\\
$A_{\ell} = 0$ & 1.023$\pm$0.028 & 0.990$\pm$0.015 & 50/36\\
$B_0$ free & 1.025$\pm$0.027 & 0.988$\pm$0.015 & 39/30\\
$B_2$ free & 1.025$\pm$0.027 & 0.988$\pm$0.015 & 39/30\\
NGC & 1.035$\pm$0.031 & 0.997$\pm$0.016 & 36/30\\
SGC & 0.999$\pm$0.043 & 0.942$\pm$0.034 & 38/30\\
\hline
{\bf $0.4 < z < 0.6$:}\\
bin size:\\
3 $h^{-1}$Mpc  & 0.991$\pm$0.024 & 0.995$\pm$0.014 & 54/54\\
4 $h^{-1}$Mpc  & 0.991$\pm$0.024 & 0.991$\pm$0.014 & 38/38\\
5 $h^{-1}$Mpc  & 0.986$\pm$0.024 & 0.994$\pm$0.014 & 23/30\\
6 $h^{-1}$Mpc  & 0.984$\pm$0.023 & 0.995$\pm$0.014 & 23/24\\
7 $h^{-1}$Mpc  & 0.985$\pm$0.023 & 0.992$\pm$0.013 & 16/18\\
8 $h^{-1}$Mpc  & 0.989$\pm$0.024 & 0.993$\pm$0.013 & 13/16\\
10 $h^{-1}$Mpc  & 0.982$\pm$0.024 & 0.995$\pm$0.014 & 8/10\\
$s > 70h^{-1}$Mpc & 0.985$\pm$0.022 & 0.994$\pm$0.013 & 17/26\\
$s < 170h^{-1}$Mpc & 0.984$\pm$0.025 & 0.995$\pm$0.014 & 37/34\\
$A_0 = 0$ & 0.986$\pm$0.026 & 0.993$\pm$0.015 & 30/33\\
$A_2 = 0$ & 0.982$\pm$0.024 & 0.996$\pm$0.014 & 25/33\\
$A_{\ell} = 0$ & 0.982$\pm$0.025 & 0.995$\pm$0.015 & 31/36\\
$B_0$ free & 0.986$\pm$0.024 & 0.994$\pm$0.014 & 23/30\\
$B_2$ free & 0.986$\pm$0.024 & 0.994$\pm$0.014 & 22/30\\
NGC & 0.972$\pm$0.028 & 0.995$\pm$0.016 & 21/30\\
SGC & 1.025$\pm$0.057 & 0.990$\pm$0.036 & 30/30\\
\hline
{\bf $0.5 < z < 0.75$:}\\
bin size:\\
3 $h^{-1}$Mpc  & 0.962$\pm$0.023 & 0.993$\pm$0.015 & 55/54\\
4 $h^{-1}$Mpc  & 0.957$\pm$0.023 & 0.995$\pm$0.015 & 37/38\\
5 $h^{-1}$Mpc  & 0.962$\pm$0.023 & 0.991$\pm$0.015 & 32/30\\
6 $h^{-1}$Mpc  & 0.961$\pm$0.023 & 0.995$\pm$0.016 & 25/24\\
7 $h^{-1}$Mpc  & 0.963$\pm$0.025 & 0.990$\pm$0.015 & 13/18\\
8 $h^{-1}$Mpc  & 0.955$\pm$0.023 & 0.995$\pm$0.015 & 16/16\\
10 $h^{-1}$Mpc  & 0.963$\pm$0.023 & 0.989$\pm$0.015 & 12/10\\
$s > 70h^{-1}$Mpc & 0.964$\pm$0.022 & 0.990$\pm$0.014 & 23/26\\
$s < 170h^{-1}$Mpc & 0.963$\pm$0.023 & 0.989$\pm$0.015 & 41/34\\
$A_0 = 0$ & 0.963$\pm$0.027 & 0.992$\pm$0.017 & 43/33\\
$A_2 = 0$ & 0.955$\pm$0.022 & 0.994$\pm$0.015 & 35/33\\
$A_{\ell} = 0$ & 0.954$\pm$0.025 & 0.996$\pm$0.017 & 49/36\\
$B_0$ free & 0.962$\pm$0.024 & 0.991$\pm$0.015 & 31/30\\
$B_2$ free & 0.962$\pm$0.023 & 0.990$\pm$0.015 & 31/30\\
NGC & 0.944$\pm$0.025 & 0.986$\pm$0.017 & 31/30\\
SGC & 1.020$\pm$0.048 & 1.010$\pm$0.035 & 32/30\\
\hline
\label{tab:binsize}
\end{tabular}
\end{table}

\begin{table}
\centering
\caption{Post-reconstruction combined sample 2D BAO fits, varying the choice of damping parameters that enter the template.}
\begin{tabular}{lccc}
\hline
\hline
 test & $\alpha_{||}$ & $\alpha_{\perp}$  & $\chi^2$/dof\\
\hline
{\bf $0.2 < z < 0.5$:}\\
fiducial  & 1.025$\pm$0.027 & 0.988$\pm$0.015 & 39/30\\
$\Sigma_{\perp} = 0$ & 1.026$\pm$0.027 & 0.987$\pm$0.014 & 39/30\\
$\Sigma_{\perp} = 5.0h^{-1}$Mpc & 1.024$\pm$0.027 & 0.991$\pm$0.016 & 42/30\\
$\Sigma_{||} = 0$ & 1.024$\pm$0.026 & 0.988$\pm$0.015 & 39/30\\
$\Sigma_{||} = 8.0h^{-1}$Mpc & 1.029$\pm$0.028 & 0.988$\pm$0.015 & 42/30\\
$\Sigma_{s} = 0$ & 1.024$\pm$0.027 & 0.988$\pm$0.015 & 39/30\\
$\Sigma_{s} = 8.0h^{-1}$Mpc & 1.035$\pm$0.030 & 0.988$\pm$0.015 & 44/30\\
\hline
{\bf $0.4 < z < 0.6$:}\\
fiducial  & 0.986$\pm$0.024 & 0.994$\pm$0.014 & 23/30\\
$\Sigma_{\perp} = 0$ & 0.986$\pm$0.023 & 0.994$\pm$0.014 & 21/30\\
$\Sigma_{\perp} = 5.0h^{-1}$Mpc & 0.985$\pm$0.024 & 0.995$\pm$0.016 & 27/30\\
$\Sigma_{||} = 0$ & 0.985$\pm$0.022 & 0.995$\pm$0.014 & 21/30\\
$\Sigma_{||} = 8.0h^{-1}$Mpc & 0.991$\pm$0.027 & 0.992$\pm$0.014 & 27/30\\
$\Sigma_{s} = 0$ & 0.987$\pm$0.024 & 0.992$\pm$0.014 & 28/30\\
$\Sigma_{s} = 8.0h^{-1}$Mpc & 0.994$\pm$0.029 & 0.994$\pm$0.014 & 24/30\\
\hline
{\bf $0.5 < z < 0.75$:}\\
fiducial  & 0.962$\pm$0.023 & 0.991$\pm$0.015 & 32/30\\
$\Sigma_{\perp} = 0$ & 0.962$\pm$0.023 & 0.990$\pm$0.015 & 31/30\\
$\Sigma_{\perp} = 5.0h^{-1}$Mpc & 0.962$\pm$0.024 & 0.991$\pm$0.017 & 35/30\\
$\Sigma_{||} = 0$ & 0.960$\pm$0.022 & 0.992$\pm$0.015 & 30/30\\
$\Sigma_{||} = 8.0h^{-1}$Mpc & 0.968$\pm$0.027 & 0.988$\pm$0.015 & 36/30\\
$\Sigma_{s} = 0$ & 0.963$\pm$0.024 & 0.990$\pm$0.015 & 32/30\\
$\Sigma_{s} = 8.0h^{-1}$Mpc & 0.971$\pm$0.029 & 0.987$\pm$0.015 & 38/30\\
\hline
\label{tab:damping}
\end{tabular}
\end{table}

Here, we report the results of a number of robustness checks on the BAO fits to the BOSS combined sample data. Table \ref{tab:binsize} presents measurements for different bin sizes. The variation between the results is small and consistent with that found in the mock samples in the previous section. We have also tested changing the range of scales that are fit within the region, increasing the minimum and maximum scale each by 20$h^{-1}$Mpc individually, as the mock tests suggest our results should be equally valid under this change; indeed we find no significant change.

Table \ref{tab:binsize} also presents tests where we have changed the way nuisance parameters are treated. We test allowing each of the bias terms to be completely free (i.e., with no prior on $B_{\ell}$; denoted by `$B_{\ell}$ free') and find no significant changes in the results. We have also tested removing the polynomial terms from the fits (denoted by `$A_{\ell}= 0$'); the motivation of these polynomial terms is to isolate the BAO feature and ensure broadband effects, such as incomplete modeling of the post-reconstruction quadrupole and observational systematics, do not affect the recovered results. Even without these terms, the results in the table show that we recover nearly the same results. The biggest change is in the $0.5 < z < 0.75$ redshift bin, where not including the polynomial terms shifts the results by $\sim 0.3\sigma$ for both $\alpha{||,\perp}$ values (in opposite directions). Thus, despite their (well-motivated) inclusion, the polynomial terms have only a minor effect on the recovered results.

The results presented in Table \ref{tab:binsize} are for where we individually fit the BAO scale in the NGC and SGC. In the $0.2 < z < 0.5$ redshift bin, the differences are greatest in terms of the measurement of $\alpha$, where the discrepancy is $\sim$1.5$\sigma$. A similar difference is found in the high redshift bin, except that the difference is in the opposite direction. These results are therefore consistent with those presented for the CMASS and LOWZ samples in Section 6.2.

Table \ref{tab:damping} presents tests where we have significantly altered the fiducial damping scales in the template. We have either set the damping scale to 0 or doubled its size. Setting the damping scale to zero alters the results by at most $0.13\sigma$ ($\alpha_{\perp}$ in the $0.4 < z < 0.6$ bin) and the changes are otherwise $<0.1\sigma$. Doubling the damping scale for $\Sigma_{||}$ or $\Sigma_s$ has a larger effect, mainly on $\alpha_{||}$. The most extreme change is $0.33\sigma$, when doubling $\Sigma_s$ in the $0.2 < z < 0.5$ redshift bin. The size of the change in the other redshift bins is similar; increasing $\Sigma_s$ results in an increase in $\alpha_{||}$. The same is true for increasing $\Sigma_{||}$, though changes are smaller ($<0.2\sigma$). The changes are generally coupled with small decreases in $\alpha_{\perp}$, implying that in terms of $\alpha$,$\epsilon$, the changes would be observed in $\epsilon$. These results are consistent with those of \cite{VM14,VargasDR12BAO}, where template choices are studied in detail using mock galaxy catalogs and the results of which set systematic uncertainty applied to the results in \cite{Acacia}. Notably, none of the results that cause more than a $0.1\sigma$ shift in the best-fit BAO position are preferred in terms of the minimum $\chi^2$ of the fit. 

\section{Information distribution with respect to the line of sight}
In the spherically symmetric case, with no RSD, information is expected to be divided equally as a function of the cosine of the angle to the line of sight, $\mu$. In \cite{Ross152D}, it was found that the BAO information in the BOSS DR11 mock samples was nearly constant a a function of $\mu$. A speculative argument explaining this fact is that any boost in information along the line of sight due to linear RSD is canceled by non-linear RSD and finger of God effects. Here, we test the distribution of BAO information in the MultiDark Patchy mocks, compared to the DR12 data. We do this by dividing the data into five bins by $\mu$ (or `wedges'; \citealt{Kazin12}), with $\Delta\mu = 0.2$. 

\begin{figure}
\includegraphics[width=84mm]{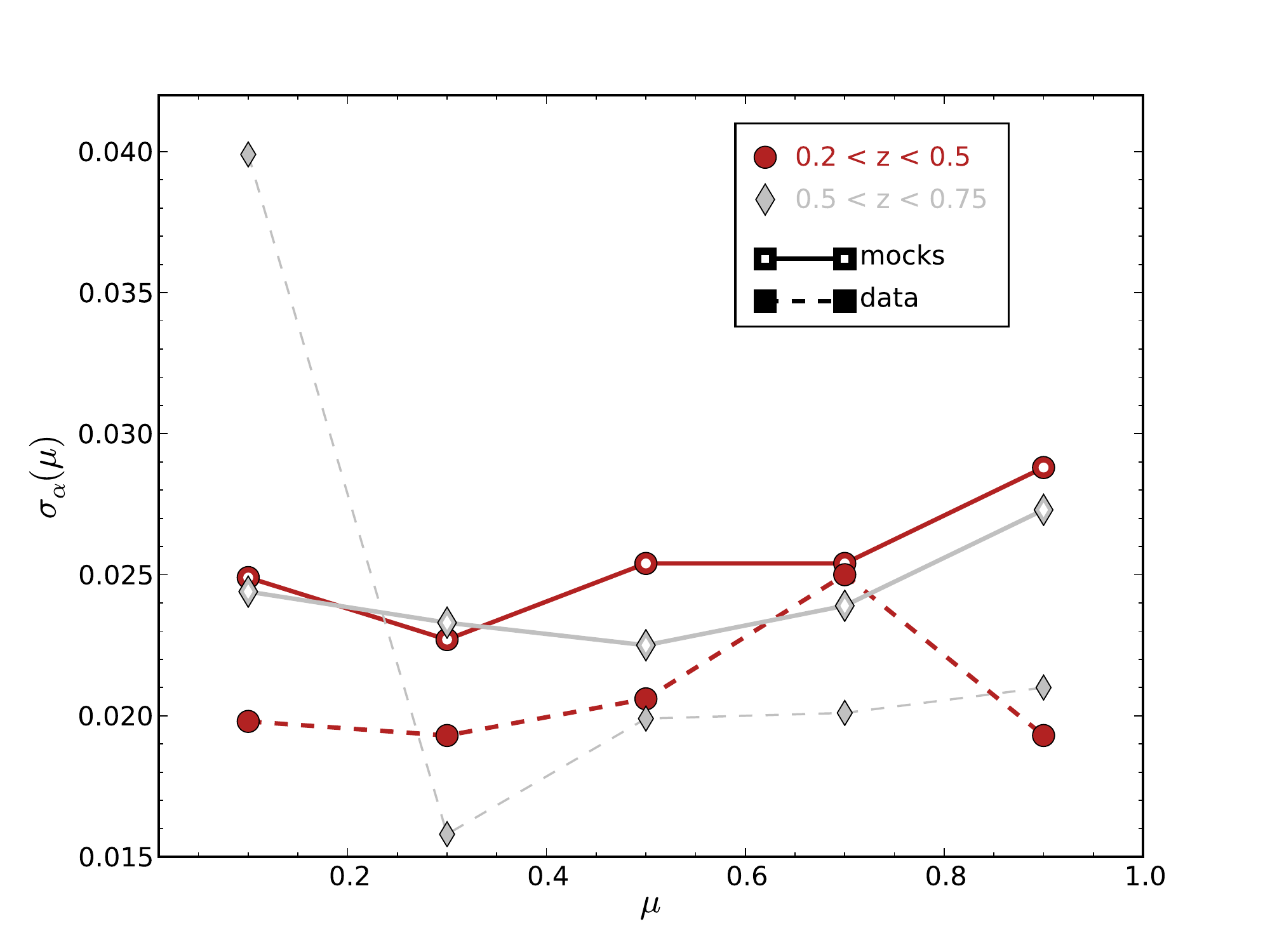}
  \caption{The mean BAO uncertainty as a function of $\mu$ for post-reconstruction MultiDark-Patchy mocks (solid lines/open symbols) compared to the results for the BOSS galaxy data (dashed lines/filled symbols).
  }
  \label{fig:baomu}
\end{figure}

In each $\mu$ bin, we apply the same BAO model described in Section 2.3, but with the template determined via integration over the particular $\mu$ range. The results for both the data and the mean results from the mocks are presented in Fig. \ref{fig:baomu}. For the mocks, the mean uncertainty is approximately constant with $\mu$, except in the $\mu > 0.8$ bin, where it is about 20 per cent greater than the $\mu$ bin with the lowest uncertainty. This is a bigger difference than was found in \cite{Ross152D}, where differences were at most 15 per cent and the uncertainties were the same in the low and high $\mu$ bins. It is possible the differences are due to differences between the MultiDark-Patchy mocks and the PTHalos \citep{Manera13} mocks used in the \cite{Ross152D} analysis. Regardless, the fundamental result that the uncertainty is approximately constant with $\mu$ remains. We find no clear trend in the uncertainty on the data. This is not overly surprising, as it is a single realization.

\begin{figure}
\includegraphics[width=84mm]{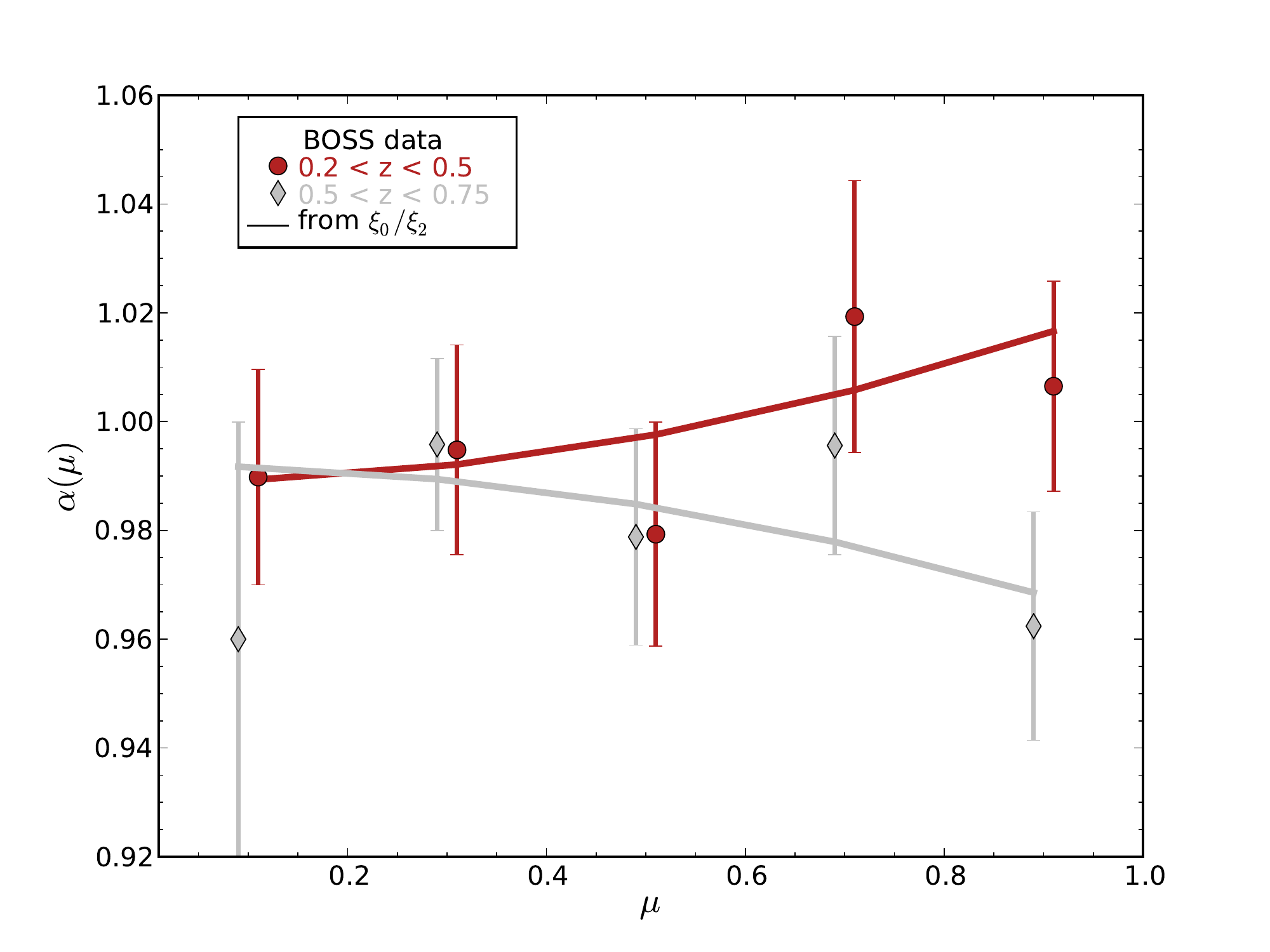}
  \caption{The measured BAO scale as a function of $\mu$, measured from the post-reconstruction BOSS galaxy correlation function, in $\mu$ bins of thickness 0.2. The solid lines represent the prediction based on the $\alpha_{||},\alpha_{\perp}$ measured from $\xi_0,\xi_2$.
  }
  \label{fig:baomud}
\end{figure}

Finally, we have looked at the measured BAO position as a function of $\mu$. These measurements can be compared a prediction based on $\alpha(\mu) = \sqrt{\mu^2\alpha_{||}^2+(1-\mu^2)\alpha_{\perp}^2}$ and our measurements of $\alpha_{||},\alpha_{\perp}$. Fig. \ref{fig:baomud} shows this comparison. The curves are consistent with the measured points, as one would expect.

\input{RSD_systematics.tex}

\label{lastpage}

\end{document}

%% file: RSD_systematics.tex
%\documentclass[a4paper,10pt,oneside]{article}
%
%\usepackage[english]{babel}
%\usepackage[pdftex]{graphicx} 
%\usepackage{amsmath}  
%\usepackage{amssymb} 
%\usepackage{bm}
%\usepackage{amsfonts}
%\usepackage{fancyhdr} 
%\usepackage{enumerate}
%\usepackage{subfig}
%\usepackage{braket}
%\usepackage{textcomp}
%
%\textwidth 14 cm
%\oddsidemargin -0.25cm 
%\evensidemargin 0cm
%\textheight 9.1in
% \parindent 0pt 
%\parskip 2ex
%
%\begin{document}
\section{Robustness of BOSS structure growth measurements to observational treatment}
\label{sec:rsdrobust}

\begin{table*}
\begin{minipage}{7in}
\label{tab:rsd}
 \begin{center}
 \caption{Measurements of $\{f(z)\sigma_8(z)$, $\alpha_{\parallel}$, $\alpha_{\perp}$, $\alpha$, $\epsilon\}$ from the mean correlation functions of the same four sets mock catalogues BAO results are presented for in Table \ref{tab:baoresultsmock}. We use 200 mock catalogues from each set. In the top four rows, we have applied MCMC analysis using the fast model described in Chuang et al. (2016). In the bottom four rows, we use a `Gaussian streaming model' like that of Reid \& White (2011). One can see that the measurements are insensitive to the systematic treatment.}
  \begin{tabular}{lccccc} 
     \hline \hline
     `fast model':\\
case&		$f(z)\sigma_8(z)$&				$\alpha_{\parallel}$&				$\alpha_{\perp}$&				$\alpha$&				$\epsilon$			\\ \hline
(i) Fid.	&$	0.507	\pm	0.067	$&$	0.995	\pm	0.045	$&$	1.005	\pm	0.021	$&$	1.001	\pm	0.015	$&$	-0.004	\pm	0.019	$\\
(ii) Sub Star, not weighted	&$	0.511	\pm	0.066	$&$	0.992	\pm	0.044	$&$	1.008	\pm	0.021	$&$	1.002	\pm	0.016	$&$	-0.005	\pm	0.018	$\\
(iii) Sub Star, weighted	&$	0.510	\pm	0.066	$&$	0.992	\pm	0.043	$&$	1.006	\pm	0.021	$&$	1.001	\pm	0.015	$&$	-0.005	\pm	0.019	$\\
(iv) Sub	&$	0.506	\pm	0.067	$&$	0.995	\pm	0.044	$&$	1.006	\pm	0.022	$&$	1.002	\pm	0.015	$&$	-0.004	\pm	0.019	$\\
    \hline
%`Gaussian streaming model':\\    
%case&		$f(z)\sigma_8(z)$&				$\alpha_{\parallel}$&				$\alpha_{\perp}$&				$\alpha$&				$\epsilon$			\\ \hline
%(i) Fid.	&$	0.494	\pm	0.049	$&$	0.998	\pm	0.030	$&$	1.004	\pm	0.016	$&$	1.002	\pm	0.011	$&$	-0.002	\pm	0.013	$\\
%(ii) Sub Star, not weighted	&$	0.496	\pm	0.050	$&$	0.996	\pm	0.030	$&$	1.005	\pm	0.015	$&$	1.002	\pm	0.011	$&$	-0.003	\pm	0.013	$\\
%(iii) Sub	&$	0.492	\pm	0.050	$&$	0.997	\pm	0.031	$&$	1.004	\pm	0.015	$&$	1.002	\pm	0.011	$&$	-0.002	\pm	0.013	$\\
%(iv) Sub Star, weighted	&$	0.497	\pm	0.050	$&$	0.995	\pm	0.030	$&$	1.004	\pm	0.015	$&$	1.001	\pm	0.011	$&$	-0.003	\pm	0.013	$\\
%    \hline
`Gaussian streaming model':\\ 
case&		$f(z)\sigma_8(z)$&				$\alpha_{\parallel}$&				$\alpha_{\perp}$&				$\alpha$&				$\epsilon$			\\ \hline
(i) Fid.	&$	0.489	\pm	0.065	$&$	0.996	\pm	0.044	$&$	1.001	\pm	0.022	$&$	0.999	\pm	0.016	$&$	-0.002	\pm	0.019	$\\
(ii) Sub Star, not weighted	&$	0.499	\pm	0.064	$&$	0.992	\pm	0.041	$&$	1.005	\pm	0.022	$&$	1.000	\pm	0.016	$&$	-0.004	\pm	0.017	$\\
(iii) Sub Star, weighted	&$	0.492	\pm	0.066	$&$	0.992	\pm	0.045	$&$	1.003	\pm	0.021	$&$	0.999	\pm	0.015	$&$	-0.004	\pm	0.020	$\\
(iv) Sub	&$	0.488	\pm	0.071	$&$	0.994	\pm	0.045	$&$	1.004	\pm	0.022	$&$	1.000	\pm	0.015	$&$	-0.003	\pm	0.020	$\\
     \hline
 \end{tabular}
 \end{center}
\end{minipage}
  \end{table*}

This work has focused on BAO scale measurements and their robustness to observational systematics. A key component of BOSS analysis has been to measure the rate of structure growth, $f\equiv d{\rm ln} D/d{\rm ln} a$, where $a$ is the scale factor and $D$ is the linear growth factor. Measurements of the clustering of galaxies are able to measure the parameter combination $f(z)\sigma_8(z)$, $\alpha_{\parallel}$, $\alpha_{\perp}$ (c.f., \citealt{Reid12,Samushia14}). Here, we investigate the extent to which the $f(z)\sigma_8(z)$ measurements are affected by observational systematic uncertainties.

We focus on the stellar density systematic, as this has the most significant effect on the clustering of BOSS galaxies. We use the mean correlation functions of the 4 sets of 200 mock catalogues with varying simulation/treatment of the BOSS stellar density systematic, as described in Section 6.1. We apply Markov Chain Monte Carlos (MCMC) analysis using the two-dimensional dewiggle model \citep{Eisenstein:2006nj} to measure $\{f(z)\sigma_8(z)$, $\alpha_{\parallel}$, $\alpha_{\perp}$, $\alpha$, $\epsilon\}$ from the mean correlation functions. The model we use for RSD tests is similar to the one for BAO tests (i.e. Eq 14 and 15) but there is some difference in detail. While the $\Sigma_\perp$ and $\Sigma_\parallel$ are fixed for the BAO model, we compute them from nonlinear perturbation theory following \cite{Crocce:2005xz,Eisenstein:2006nj,Matsubara:2007wj}. We model the systematics on monopole with a polynomial $A_0(s)=a_2/s^2+a_1/s+a_0$ but we do not apply the same for the quadrupole since the quadrupole measurements at large scales are insensitive to the observational systematics as shown in Fig. \ref{fig:xi0sys}. This methodology has been applied to DR11 CMASS data analysis to obtain the measurements of RSD + BAO \citep{Chuang:2013wga}, fit in the range $40 < s < 180 h^{-1}$Mpc. 

We also test the `Gaussian streaming model' described in
\cite{Reid11}; this model has been applied to multiple BOSS analyses (c.f., \citealt{Reid12,Samushia14}) and is fully described in these references. In this study, we consider relatively large scales (i.e. $40 < s < 180 h^{-1}$Mpc), which we do not expect to be affected by any FoG effects, so we do not include any parameter for this (as have previous analyses). Thus,  only one nuisance parameter is included in our analysis: $b_{1L} = b-1$, the first-order
Lagrangian host halo bias in {\em real} space. Further details of the model, its numerical implementation, and its accuracy
can be found in \cite{Reid11}.

The results of the tests are shown in Table \ref{tab:rsd}
%\ref{tab:rsd}
 and we find that they are insensitive to the treatment of stellar density systematics. Note that since we aim to test the impact from the stellar density systematics, the most important quantity is the differences between the four results (rather than any difference from the true value expected for the cosmology of the mocks). One can observe that the fluctuations are at a level that is $<0.1\sigma$, and that this is true for both types of modeling. The cases where the stellar density systematic is present (corrected for or not), exhibit slight ($\leq 0.1\sigma$), but coherent shifts in $f\sigma_8$ (an increase), $\alpha_{||}$ (a decrease), and $\epsilon$ (a decrease). Given the small size of the shifts, we do not believe they are of concern for BOSS DR12 analysis. However, revisiting this issue for future surveys that will have greater statistical precision is of clear importance.

%\begin{table*}
%\begin{minipage}{7in}
% \begin{center}
% \caption{Measurements of $\{f(z)\sigma_8(z)$, $\alpha_{\parallel}$, $\alpha_{\perp}$, $\alpha$, $\epsilon\}$ from the mean correlation functions of the 4 sets mock catalogues. We use 200 mock catalogues from each set. We apply MCMC analysis described in Chuang et al. in perp to include the systematics modeling with advanced RSD model. One can see that the measurements are insensitive to the systematics tested.}
%  \begin{tabular}{lccccc} 
%     \hline \hline
%case&		$f(z)\sigma_8(z)$&				$\alpha_{\parallel}$&				$\alpha_{\perp}$&				$\alpha$&				$\epsilon$			\\ \hline
%5st0	&$	0.494	\pm	0.049	$&$	0.998	\pm	0.030	$&$	1.004	\pm	0.016	$&$	1.002	\pm	0.011	$&$	-0.002	\pm	0.013	$\\
%st5st0	&$	0.496	\pm	0.050	$&$	0.996	\pm	0.030	$&$	1.005	\pm	0.015	$&$	1.002	\pm	0.011	$&$	-0.003	\pm	0.013	$\\
%sub5st0	&$	0.492	\pm	0.050	$&$	0.997	\pm	0.031	$&$	1.004	\pm	0.015	$&$	1.002	\pm	0.011	$&$	-0.002	\pm	0.013	$\\
%wst5st0	&$	0.497	\pm	0.050	$&$	0.995	\pm	0.030	$&$	1.004	\pm	0.015	$&$	1.001	\pm	0.011	$&$	-0.003	\pm	0.013	$\\
%     \hline
% \end{tabular}
% \end{center}
%\label{table:rsd2}
%\end{minipage}
%  \end{table*} 
%\end{document}